\pdfoutput=1

\documentclass[12pt]{article}
\usepackage{amsthm,amssymb,amsmath}
\usepackage{mathrsfs}
\usepackage{graphicx}
\usepackage{enumerate}
\usepackage{hyperref}
\usepackage{apacite}
\usepackage{natbib}
\usepackage{float}
\usepackage{url} % not crucial - just used below for the URL 
\usepackage[utf8]{inputenc}
\usepackage{cleveref}

\newcommand{\blind}{0}

\theoremstyle{plain}
\newtheorem{theorem}{Theorem}
\newtheorem{assumption}{Assumption}
\newtheorem{step}{Step}
\newtheorem{remark}{Remark}

\newtheorem{lemma}{Lemma}
\newtheorem{proposition}{Proposition}

\begin{document}

\def\spacingset#1{\renewcommand{\baselinestretch}%
{#1}\small\normalsize} \spacingset{1}

%%%%%%%%%%%%%%%%%%%%%%%%%%%%%%%%%%%%%%%%%%%%%%%%%%%%%%%%%%%%%%%%%%%%%%%%%%%%%%

% \if0\blind
% {
%   \title{\bf Estimating Heterogeneous Treatment Effects on Survival Outcomes Using Counterfactual Censoring Unbiased Transformations}
%   \author{Author 1 % \thanks{}\hspace{.2cm}\\
%     Department of YYY, University of XXX\\
%     and \\
%     Author 2 \\
%     Department of ZZZ, University of WWW}
%   \maketitle
% } \fi

\if0\blind
{
  \title{\bf Estimating Heterogeneous Treatment Effects on Survival Outcomes Using Counterfactual Censoring Unbiased Transformations}
  \author{Shenbo Xu$^1$ \and Raluca Cobzaru$^2$ \and Stan N. Finkelstein$^1$ \and Roy E. Welsch$^2$ \and Kenney Ng$^3$ \and Zach Shahn$^4$} % Bang Zheng$^2$ \and Ioanna Tzoulaki$^7$ \and

\date{\small %
    $^1$ Institute for Data, Systems, and Society, Massachusetts Institute of Technology. \\ % Email: \href{mailto:xushenbo@mit.edu}{xushenbo@mit.edu} \\%
    $^2$ Sloan School of Management, Massachusetts Institute of Technology. \\ % Email: \href{mailto:rcobzaru@mit.edu}{rcobzaru@mit.edu} \\%
    % $^3$ Ageing Epidemiology Research Unit, Imperial College London. \\
    % Email: \href{mailto:b.zheng17@imperial.ac.uk}{b.zheng17@imperial.ac.uk} \\%
    % $^4$ Institute for Data, Systems, and Society, Massachusetts Institute of Technology. \\ % Email: \href{mailto:snf@mit.edu}{snf@mit.edu} \\%
    % $^3$ Sloan School of Management, Massachusetts Institute of Technology. \\ % Email: \href{mailto:rwelsch@mit.edu}{rwelsch@mit.edu} \\%
    $^3$ Center for Computational Health, IBM Research. \\ % Email: \href{mailto:Kenney.Ng@us.ibm.com}{Kenney.Ng@us.ibm.com} \\%
    % $^7$ School of Public Health, Imperial College London. \\ Email: \href{mailto:i.tzoulaki@imperial.ac.uk}{i.tzoulaki@imperial.ac.uk} \\%
    $^4$ School of Public Health, The City University of New York. \\ % Email: \href{mailto:zachary.shahn@sph.cuny.edu}{zachary.shahn@sph.cuny.edu} \\ [2ex]
    \date{} % \today
}
  % \author{Shenbo Xu \thanks{
  %   The authors gratefully acknowledge \textit{please remember to list all relevant funding sources in the unblinded version}}\hspace{.2cm}\\
  %   Institute for Data, Systems, and Society, Massachusetts Institute of Technology \\
  %   Email: \href{xushenbo@mit.edu}{xushenbo@mit.edu}
  %   and \\
  %   Raluca Cobzaru \\
  %   Sloan School of Managemen, Massachusetts Institute of Technology \\
  %   and \\
  %   Bang Zheng \\
  %   Ageing Epidemiology Research Unit, Imperial College London \\
  %   Stan N. Finkelstein}
  \maketitle
} \fi

\begin{abstract}

% less than 200 words.

Methods for estimating heterogeneous treatment effects (HTE) from observational data have largely focused on continuous or binary outcomes, with less attention paid to survival outcomes and almost none to settings with competing risks. In this work, we develop censoring unbiased transformations (CUTs) for survival outcomes both with and without competing risks. %, where competing risks involve measures based on total effect and seperable effects.
 After converting time-to-event outcomes using these CUTs, direct application of HTE learners for continuous outcomes yields consistent estimates of heterogeneous cumulative incidence effects, total effects, and separable direct effects. %We prove the equivalence between the minimizer of the efficiency augmented (orthogonal) censoring loss functions and the minimizer of the corresponding loss functions based on CUTs. 
 Our CUTs enable application of a much larger set of state of the art HTE learners for censored outcomes than had previously been available, especially in competing risks settings. We provide generic model-free learner-specific oracle inequalities bounding the finite-sample excess risk. The oracle efficiency results depend on the oracle selector and estimated nuisance functions from all steps involved in the transformation. We demonstrate the empirical performance of the proposed methods in simulation studies.% and illustrate their advantages on real data by examining the treatment effect of metformin compared to sulfonylureas on cancer clinical outcomes.

\end{abstract}

\noindent %
{\it Keywords:} % 3 to 6 keywords, that do not appear in the title

Censoring unbiased transformation; competing risks; limited overlap; Neyman orthogonality; oracle inequality; separable effects % ; super learning (question: remove one)

% heterogeneous treatment effect, precision medicine, cumulative treatment effect, (cause-specific) restricted mean time lost

\vfill

\newpage
\spacingset{1.9} % DON'T change the spacing!
\section{Introduction}
\label{sec:intro}

% introduce HTE and censored data

Heterogeneous treatment effect (HTE) estimation is central to decision-making across a wide range of fields, from medicine to marketing.
%, ranging from precision medicine~\citep{robertson2021assessing} to learning mindsets~\citep{athey2019estimating}, and can be applied to learning optimal in individualized decision rules~\citep{chakraborty2013statistical,kosorok2015adaptive,tsiatis2019dynamic}. 
Accordingly, considerable effort has been devoted to developing various HTE `learners' that incorporate machine learning regression to achieve state-of-the-art empirical performance~\citep{knaus2021machine,jacob2021cate,talisa2021estimating,caron2022estimating}. However, most HTE learners are only applicable in settings with continuous or binary outcomes. Relatively fewer options are available for censored time-to-event outcomes, and even fewer options exist for time-to-event outcomes in the presence of competing risks. 

`Censoring unbiased transformations' (CUTs) convert potentially censored time-to-event outcomes into standard continuous outcomes that have the same conditional expectation functions as quantities of interest and are amenable to regression modeling approaches designed for uncensored outcomes. We use CUTs to construct transformed variables that can be plugged directly into a wide range of continuous HTE learners to produce estimates of HTEs on survival outcomes, providing a bridge from survival settings to existing HTE learners. We also derive error bounds for these CUT based HTE learners. 

While CUTs for survival outcomes in the absence of competing risks have a long history, we develop new CUTs for outcomes in the presence of competing risks. Competing risks are said to be present in time-to-event studies if it is possible for subjects to experience one of multiple mutually exclusive events. For example, interest may center on a drug's effect on death from prostate cancer, in which case death from other causes is a competing risk. In such settings, it is necessary to carefully account for competing risks when defining and estimating causal contrasts of scientific interest\citep{young2020causal}. We show that estimators of heterogeneous effects appropriate for the competing risks setting (i.e. separable direct and indirect effects \citep{stensrud2021generalized,stensrud2022conditional} and cause specific effects) obtained from plugging our CUTs into continuous HTE learners are consistent, greatly expanding the supply of available methods. %Further, we derive finite sample nonparametric oracle bounds for HTE learners applied to CUT-based variables for both the competing risks and no competing risks settings. 
%While CUTs for the no competing risk setting have been around for a long time, it had not been formally established that they can be plugged into certain modern HTE learners and bounds accounting for uncertainty in CUT estimation had not been derived.%, one major difficulty of applying these methods to clinical application is that the outcome such applications are often subject to censoring-and existing methods for HTE cannot be used in this situation directly.
% existing survival HTE learners, conditional mean difference

HTE learners for continuous outcomes can be categorized into `conditional mean difference' learners and `transformed minimization' learners. Conditional mean difference learners are more intuitive. They straightforwardly estimate the conditional expectation of the outcome given covariates (assumed sufficient to adjust for confounding) and each treatment value, then take the difference of the conditional expectations across treatment values. Conditional mean difference learners can be further broken down into `single-learners' (S-learners), which learn a single regression model for the outcome given treatment and covariates ~\citep{su2008interaction}; `two-learners' (T-learners), which learn separate regression models for the outcome given covariates in each treatment group ~\citep{zhang2017mining}; and `multitask-learners' (MT-learners) which learn separate regressions within treatment groups like T-learners but using covariate representations based on the full data ~\citep{alaa2017deep,lee2019dynamic,curth2021survite}. It is trivial to see that plugging any valid CUT into any conditional mean difference learner will produce consistent HTE estimates by virtue of the CUT sharing a conditional expectation function with the outcome of interest. By developing new CUTs appropriate for competing events effects, we enable analysts to use the well developed continuous outcome versions of these learners to investigate heterogeneous separable and cause specific effects. The bounds we derive for these learners with CUTs plugged in as the outcome (even for previously existing CUTs in the absence of competing risks) also constitute a new contribution. % ,henderson2020individualized,hu2021subgroup; chapfuwa2021enabling; lee2018deephit,schrod2022bites,tabib2020non

% recent advances in survival HTE learners, transformed minimization
Transformed minimization learners, by contrast, regress a transformed version of the outcome on covariates, possibly minimizing a weighted loss function. See Table~\ref{tab:weighted_minimization} for a survey of these learners. We show that plugging CUTs (for both the competing risks and no competing risks settings) into these continuous outcome learners also produces consistent HTE estimators. This had not been formally established even for previously developed CUTs in the no competing risks setting. The nonparametric bounds for all CUTs plugged into these learners are also a new contribution.

Our CUT based HTE learners are not the only available options. There have been several recent advances in HTE learners for censored time-to-event outcomes. For example,~\citet{yadlowsky2021estimation,boileau2024nonparametric,gao2021estimating,cui2023estimating,zhang2017mining,curth2021survite,xu2023treatment,curth2023understanding}. We do not claim sweeping theoretical advantages of CUT based estimators over these methods. Our contribution is simply to open up additional alternatives and study their statistical properties. % schrod2022bites,tabib2020non,abraich2022survcaus, often involve inverse probability censoring weights (IPCW), including learners based on solving IPCW estimating equations~\citep{yadlowsky2021estimation,liu2021estimation}, transformed regression with IPCW as sample weights~\citep{xu2023treatment}, uncentered time-to-event influence function learner (IF-learner)~\citep{diaz2018targeted,zhu2020targeted,hu2021estimating}, R-learner that belongs to the exponential family~\citep{gao2021estimating}, and causal survival forest~\citep{cui2023estimating}.

The organization of the paper is as follows. In Section 2, we define notation, causal quantities-of-interest, and corresponding identifying assumptions for both the competing events and no competing events settings. In Section 3, we introduce CUTs and show that our proposed CUTs have the desirable property that they share conditional expectation functions with quantities of interest. We then map the terrain of continuous outcome HTE learners and show that plugging our CUTs into any of the continuous outcome HTE learners discussed leads to a consistent estimator of the HTE on the time-to-event outcome of interest. Finally, we provide a sample-splitting estimation algorithm for our CUT plugin estimators.  In section 4, we discuss uncertainty quantification and  derive oracle bounds. In Section 5, we present simulation studies exploring the properties of alternative estimators with a particular emphasis on sensitivity to poor overlap. In Section 6, we present an application to real data estimating heterogeneous effects of reduced intensity conditioning prior to stem cell transplantation on relapse and treatment induced mortality. In Section 7, we conclude. 

\section{Notation and Estimands} \label{sec:setting}

\subsection{Notation}\label{notation}

Let $X \in \mathcal{X}$ be a vector of baselines covariates and $A \in \{0, 1\}$ be a binary exposure. In the absence of competing risks, we define $T^{a}>0$ as the counterfactual time-to-event under exposure $A=a$. We make the Consistency assumption that the actual time-to-event is $T=I(A=a)T^{a}$. We denote the observed censoring time by $C>0$. Then the censored time-to-event is $\widetilde{T}=\min\{T, C\}$, and the event indicator is $\Delta=I(T \leq C)$. We observe tuples $O=\{X, A, \widetilde{T}, \Delta\} \sim P_{0}$, where $P_{0}$ is the unknown true data-generating distribution. Suppose that we observe $n$ independent and identically distributed (i.i.d.) observations $O_{1}, \ldots, O_{n}$ sampled from $P_{0}$ and let $P_{n}$ be its empirical probability distribution.

When $j^{*}>1$ and there are competing events, let $T_{j}^{a}>0$ be the counterfactual time to event $j$ had intervention been set to $a$ and $T_{j}^{a}=\infty$ if $j$ would not occur under exposure $a$. Define the uncensored potential survival time as $T^{a}=\min_{j}\{T_{j}^{a}\}$ and the counterfactual cause of failure as $J^{a}=\operatorname{argmin}_{j}\{T_{j}^{a}\}$. When $A=a$, we make the consistency assumption that the actual failure time is $T=T^{A}$, and the actual cause of failure is $J=J^{A}$. The observed cause of failure in the presence of censoring can be defined as $\widetilde{J}=J \Delta$ where $\widetilde{J} \in \{0, 1, \ldots, j^{*}\}$ and the observed data unit becomes $O=\{X, A, \widetilde{T}, \widetilde{J},\Delta\} \sim P_{0}$. 

We introduce all-cause counting process $N(t)=I(\widetilde{T} \leq t)$, cause-specific counting process $N_j(t)=I(\widetilde{T} \leq t, \widetilde{J}=j)$, and censoring process $N^C(t)=I(\widetilde{T} \leq t, \Delta=0)=(1-\Delta) N(t)$. Their corresponding true conditional cumulative hazard functions are $\Lambda_{0}(t \mid a, X)$, $\Lambda_{j, 0}(t \mid a, X)$, $\Lambda_{0}^{C}(t \mid a, X)$, respectively. The counting processes can be compensated into true conditional martingales as $\mathrm{d} M_{0}(t \mid a, X)=\mathrm{d} N(t)-R(t) \mathrm{d}\Lambda_{0}(t \mid a, X)$, $\mathrm{d} M_{j, 0}(t \mid a, X)=\mathrm{d} N_j(t)-R(t) \mathrm{d}\Lambda_{j, 0}(t \mid a, X)$, and $\mathrm{d} M_{0}^C(t \mid a, X)=\mathrm{d} N^C(t)-R(t) \mathrm{d}\Lambda_{0}^C(t \mid a, X)$ where $R(t)=I(\widetilde{T} \geq t)$ is the at-risk process. % the all-cause martingale, cause-specific martingale, and censoring martingale can be defined as , respectively.
We denote the true conditional censoring survival function by $G_{0}(t \mid a, X)=P_{0}(C>t \mid A=a, X)$, the true conditional survival function by $S_{0}(t \mid a, X)=P_{0}(T>t \mid A=a, X)$, and the true conditional cumulative incidence function by $F_{j, 0}=P_{0}(T \leq t, J=j \mid A=a, X)$. We denote the set of true hazard functions (for events, competing events, and censoring) that will later be nuisance parameters for our estimators by $\eta_{ 0}^{\Lambda}=\{\Lambda_{0}^C(t \mid a, X),  \Lambda_{1, 0}(t \mid a, X),\ldots,\Lambda_{j^{*}, 0}(t \mid a, X)\}$. We let $\eta_{0}=\{\pi_{0}(a \mid X),\eta_{ 0}^{\Lambda}\}$ denote the nuisance hazard functions plus the propensity score function.  For any functional $q_0=q(P_{0})$ of the true population distribution, let $\widehat{q}$ denote an arbitrary estimator of $q_0$. We will use the subscript $\infty$ to denote probability limits of estimators.%, i.e. we will use $\eta_{\infty}^{\Lambda}$, $\eta_{j, \infty}^{\Lambda}$, $\eta_{\infty}$, $\eta_{j, \infty}$, $\pi_{\infty}$, $\Lambda_{\infty}^{C}$, $M_{\infty}^{C}$, $G_{\infty}$, $\Lambda_{\infty}$, $M_{\infty}$, $S_{\infty}$, $\mathrm{RMST}_{\infty}$, $\Lambda_{j, \infty}$, 
%$M_{j, \infty}$, $F_{j, \infty}$, and $\mathrm{RMTL}_{j, \infty}$ to denote the probability limits of estimators. $\widehat{\eta}^{\Lambda}$, $\widehat{\eta}_{j}^{\Lambda}$, $\widehat{\eta}$, $\widehat{\eta}_{j}$, $\widehat{\pi}$, $\widehat{\Lambda}^{C}$, $\widehat{M}^{C}$, $\widehat{G}$, $\widehat{\Lambda}$, $\widehat{M}$, $\widehat{S}$, $\mathrm{R\widehat{MS}T}$, $\widehat{\Lambda}_{j}$, 
%$\widehat{M}_{j}$, $\widehat{F}_{j}$, and $\mathrm{R\widehat{MT}L}_{j}$, respectively.

\subsection{Estimands of interest} \label{sec:estimands}

Absent competing events, we consider the difference between conditional survival functions under each exposure, i.e.
\begin{equation*}
\psi_{0}^{S}(t, X)=P_{0}(T^{a=1}>t \mid X)-P_{0}(T^{a=0}>t \mid X),
\end{equation*} 
or the difference between conditional restricted mean survival times (RMST) under each exposure, i.e.
\begin{equation*}  
\psi_{0}^{\mathrm{RMST}}(\tau, X)=E_{0}\{\min(T^{a=1},\tau) \mid X\}-E_{0}\{\min(T^{a=0},\tau) \mid X\}.
\end{equation*}

When there are competing events and we are interested in event $j$, we consider the difference between conditional cause-specific cumulative incidence functions (CIF), i.e.
\begin{equation*}
\psi_{j, 0}^{F}(t, X)=P_{0}(T^{a=1}\leq t, J^{a=1}=j \mid X)-P_{0}(T^{a=0}\leq t, J^{a=0}=j \mid X),
\end{equation*}
or the difference between conditional cause-specific restricted mean time lost (RMTL), i.e.
\begin{equation*}
\psi_{j, 0}^{\mathrm{RMTL}}(\tau, X)=E_{0}[\{\tau-\min(T^{a=1},\tau)\}I(J^{a=1}=j)\mid X]-E_{0}[\{\tau-\min(T^{a=0},\tau)\}I(J^{a=0}=j) \mid X].
\end{equation*}
Cause-specific effects capture the impact of exposure on the event of interest via all pathways, including through the competing event. For example, a drug that instantly causes death from heart attack would have a strong preventive cause specific effect on death from prostate cancer. Conversely, a drug that prevents death from causes other than prostate cancer would have a `harmful' cause specific effect on death from prostate cancer. Examples such as these (but more realistic) can make cause specific effects difficult to interpret.

Alternatively, one could treat death from other causes as a censoring event and proceed as if there were no competing risks. Such an analysis would estimate the controlled direct effect of exposure on death from prostate cancer under a hypothetical additional intervention that somehow prevents anyone from dying from other causes. Of course, such an intervention is infeasible. And further, even if death prevention were feasible, outcomes would likely depend on how exactly it was achieved, which cannot be known. Thus, results from analyses that treat competing events as censoring events are also often difficult to interpret. 

The difficulty in interpreting cause-specific effects and controlled direct effects (where, again, controlled direct effects are the estimands that correspond to treating competing events as censoring events) motivated the development of separable direct and indirect effects~\citep{stensrud2022separable}. To define separable effects, the treatment assignment $A$ must be decomposed into two variables, $A_{j}$ and $A_{\overline{j}}$. $A_{j}$ only exerts its effect on the event of interest $j$ directly not through the other competing events $\overline{j}$, while $A_{\overline{j}}$ only exerts its effect on $j$ indirectly through $\overline{j}$. Even if $A_j$ and $A_{\overline{j}}$ are hypothetical or even outlandish, separable direct and indirect effects can still be interpretable and of scientific interest. 

The separable direct HTE is the conditional effect on the event of interest of a hypothetically modified version of exposure that has no effect on the competing event, i.e. 
\begin{equation*}
\psi_{j, 0}^{F, \mathrm{sep-D}}(t, a_{\overline{j}}, X)=P_{0}(T^{a_{j}=1, a_{\overline{j}}}\leq t, J^{a_{j}=1, a_{\overline{j}}}=j \mid X)-P_{0}(T^{a_{j}=0, a_{\overline{j}}}\leq t, J^{a_{j}=0, a_{\overline{j}}}=j \mid X)
\end{equation*}
for $a_{\overline{j}} \in \{0, 1\}$, where $T^{a_{j}, a_{\overline{j}}}$ and $J^{a_{j}, a_{\overline{j}}}$ are the counterfactual failure time and counterfactual cause of failure, respectively, had $A_{j}$ been set to $a_{j}$ and $A_{\overline{j}}$ been set to $a_{\overline{j}}$. Note that when $a_{j}=a_{\overline{j}}=a$, $(T^{a_{j}, a_{\overline{j}}}, J^{a_{j}, a_{\overline{j}}})=(T^{a}, J^{a})$. % with $\psi_{j, 0}^{\mathrm{RMTL}}(\tau,  X)$, 
The separable direct HTE on the RMTL is
\begin{align*}
&\psi_{j, 0}^{\mathrm{RMTL}, \mathrm{sep-D}}(\tau, a_{\overline{j}}, X)=\\
&E_{0}[\{\tau-\min(T^{a_{j}=1, a_{\overline{j}}},\tau)\}I(J^{a_{j}=1, a_{\overline{j}}}=j)\mid X]-E_{0}[\{\tau-\min(T^{a_{j}=0, a_{\overline{j}}},\tau)\}I(J^{a_{j}=0, a_{\overline{j}}}=j) \mid X]
\end{align*}

The separable \textit{indirect} HTE is the conditional effect on the event $j$ of interest of a hypothetically modified version of exposure that has no effect on event $j$ except through the competing risk, i.e.,
{\footnotesize
$$
\psi_{j, 0}^{F, \mathrm{sep-I}}(t, a_{j}, X)=P_{0}(T^{a_{j}, a_{\overline{j}}=1}\leq t, J^{a_{j}, a_{\overline{j}}=1}=j \mid X)-P_{0}(T^{a_{j}, a_{\overline{j}}=0}\leq t, J^{a_{j}, a_{\overline{j}}=0}=j \mid X)
$$
}
for $a_{j} \in \{0, 1\}$. The separable indirect HTE on the RMTL can be defined as
{\footnotesize
$$
\psi_{j, 0}^{\mathrm{RMTL}, \mathrm{sep-I}}(\tau, a_{j}, X)=E_{0}[\{\tau-\min(T^{a_{j}, a_{\overline{j}}=1},\tau)\}I(J^{a_{j}, a_{\overline{j}}=1}=j) \mid X]-E_{0}[\{\tau-\min(T^{a_{j}, a_{\overline{j}}=0},\tau)\}I(J^{a_{j}, a_{\overline{j}}=0}=j) \mid X]
$$
}

%Note that pairs of conditional separable direct and conditional indirect effects sum up to the conditional cause-specific effect, i.e.
%{\footnotesize
%$$
%\begin{aligned}
%&\psi_{j, 0}^{F, \mathrm{sep-D}}(t, a_{\overline{j}}=a^{*}, X)+\psi_{j, 0}^{F, \mathrm{sep-I}}(t, a_{j}=1-a^{*}, X) \\
%=&P_{0}(T^{a_{j}=1, a_{\overline{j}}=a^{*}}\leq t, J^{a_{j}=1, a_{\overline{j}}=a^{*}}=j \mid X)-P_{0}(T^{a_{j}=0, a_{\overline{j}}=a^{*}}\leq t, J^{a_{j}=0, a_{\overline{j}}=a^{*}}=j \mid X) \\
%&+P_{0}(T^{a_{j}=1-a^{*}, a_{\overline{j}}=1}\leq t, J^{a_{j}=1-a^{*}, a_{\overline{j}}=1}=j \mid X)-P_{0}(T^{a_{j}=1-a^{*}, a_{\overline{j}}=0}\leq t, J^{a_{j}=1-a^{*}, a_{\overline{j}}=0}=j \mid X) \\
%=&P_{0}(T^{a_{j}=1, a_{\overline{j}}=1}\leq t, J^{a_{j}=1, a_{\overline{j}}=1}=j \mid X)-P_{0}(T^{a_{j}=0, a_{\overline{j}}=0}\leq t, J^{a_{j}=0, a_{\overline{j}}=0}=j \mid X) \\
%=&P_{0}(T^{a=1}\leq t, J^{a=1}=j \mid X)-P_{0}(T^{a=0}\leq t, J^{a=0}=j \mid X)=\psi_{j, 0}^{F}(t, X)
%\end{aligned}
%$$
%}
%for $a^{*} \in \{0, 1\}$. Similarly, $\psi_{j, 0}^{\mathrm{RMTL}, \mathrm{sep-D}}(\tau, a_{\overline{j}}=a^{*}, X)+\psi_{j, 0}^{\mathrm{RMTL}, \mathrm{sep-I}}(\tau, a_{j}=1-a^{*}, X)=\psi_{j, 0}^{\mathrm{RMTL}}(\tau, X)$.

\subsection{Identification}\label{identification}
We require different subsets of the following assumptions to identify our various HTEs of interest. For $a \in\{0,1\}$:

\begin{assumption}\label{assumption:positivity}
(Positivity) $P_{0}(A=a \mid X) >0$ and $P_{0}(C \geq t \mid A=a, X)>0$ almost surely for all $a$ and $t \in[0, \tau]$;
\end{assumption}

\begin{assumption}\label{assumption:SUTVA_survival}
(Consistency) $T=T^{A}$;
\end{assumption}

\begin{assumption}\label{assumption:ignorability_survival}
(Conditional ignorability) $T^{a} \perp\!\!\!\perp A \mid X$;
\end{assumption}

\begin{assumption}\label{assumption:depcens_survival}
(Conditionally independent censoring) $T^{a} \perp\!\!\!\perp C \mid A, X$.
\end{assumption}

\begin{assumption}\label{assumption:SUTVA_competing}
(Competing risks consistency) $(T^{A}, J^{A})=(T, J)$;
\end{assumption}

\begin{assumption}\label{assumption:ignorability_competing}
(Competing risks conditional ignorability) $(T^{a}, J^{a}) \perp\!\!\!\perp A \mid X$;
\end{assumption}

\begin{assumption}\label{assumption:depcens_competing}
(Competing risks conditionally independent censoring) $(T^{a}, J^{a}) \perp\!\!\!\perp C \mid A, X$.
\end{assumption}

\begin{assumption}\label{assumption:dismissible}
(Dismissible component assumptions) $\lambda_{j}^{a_{j}, a_{\overline{j}}=1}(t \mid X=x)=\lambda_{j}^{a_{j}, a_{\overline{j}}=0}(t \mid X=x)$ for $a_{j} \in \{0, 1\}$ and $\lambda_{\overline{j}}^{a_{j}=1, a_{\overline{j}}}(t \mid X=x)=\lambda_{\overline{j}}^{a_{j}=0, a_{\overline{j}}}(t \mid X=x)$ for $a_{\overline{j}} \in \{0,1\}$ where $\lambda_{j}^{a_{j}, a_{\overline{j}}}(t \mid X=x)$ is the counterfactual conditional cause-specific hazard functions under joint exposure of $a_{j}$ and $a_{\overline{j}}$~\citep{martinussen2023estimation}.
\end{assumption}

Under~\Cref{assumption:positivity}-\Cref{assumption:depcens_survival}, we have
{\footnotesize
$$
\psi_{0}^{S}(t, X)=S_{0}(t \mid 1, X)-S_{0}(t \mid 0, X),
$$
}
where $S_{0}(t \mid a, X)=P_{0}(T>t \mid A=a, X)$, and
{\footnotesize
$$
\psi_{0}^{\mathrm{RMST}}(\tau, X)=\mathrm{RMST}_{0}(\tau \mid 1, X)-\mathrm{RMST}_{0}(\tau \mid 0, X),
$$
}
where $\mathrm{RMST}_{0}(\tau \mid a, X)=E_{0}\{\min(T,\tau) \mid A=a, X\}$.

Under~\Cref{assumption:positivity} and~\Cref{assumption:SUTVA_competing}-\Cref{assumption:depcens_competing}, we have
{\footnotesize
$$
\psi_{j, 0}^{F}(t, X)=F_{j, 0}(t \mid 1, X)-F_{j, 0}(t \mid 0, X),
$$
}
where $F_{j, 0}(t \mid a, X)=P_{0}(T\leq t, J=j \mid A=a, X)$, and
{\footnotesize
$$
\psi_{j, 0}^{\mathrm{RMTL}}(\tau, X)=\mathrm{RMTL}_{j, 0}(\tau \mid 1, X)-\mathrm{RMTL}_{j, 0}(\tau \mid 0, X)
$$
}
where $\mathrm{RMTL}_{j, 0}(\tau \mid a, X)=E_{0}[\{\tau-\min(T,\tau)\}I(J=j) \mid A=a, X]$.

In addition to~\Cref{assumption:positivity} and~\Cref{assumption:SUTVA_competing}-\Cref{assumption:depcens_competing}, conditional separable effects require~\Cref{assumption:dismissible}. The separable direct HTE is then identified as
{\footnotesize
$$
\psi_{j, 0}^{F, \mathrm{sep-D}}(t, a_{\overline{j}}, X)=F_{j, 0}(t \mid 1, a_{\overline{j}}, X)-F_{j, 0}(t \mid 0, a_{\overline{j}}, X)
$$
}
where $F_{j, 0}(t \mid a_{j}, a_{\overline{j}}, X)=P_{0}(T\leq t, J=j \mid A_{j}=a_{j}, A_{\overline{j}}=a_{\overline{j}}, X)=\int_0^t e^{-\Lambda_{j} (u \mid a_{j}, X)-\Lambda_{\overline{j}}(u \mid a_{\overline{j}}, X)} \mathrm{d} \Lambda_{j}(u \mid a_{j}, X)$~\citep{martinussen2023estimation}. Similarly,
{\footnotesize
$$
\psi_{j, 0}^{\mathrm{RMTL}, \mathrm{sep-D}}(\tau, a_{\overline{j}}, X)=\mathrm{RMTL}_{j, 0}(\tau \mid 1, a_{\overline{j}}, X)-\mathrm{RMTL}_{j, 0}(\tau \mid 0, a_{\overline{j}}, X)
$$
}
where $\mathrm{RMTL}(\tau \mid a_{j}, a_{\overline{j}}, X)=E_{0}[\{\tau-\min(T,\tau)\}I(J=j) \mid A_{j}=a_{j}, A_{\overline{j}}=a_{\overline{j}}, X]$. The separable indirect HTE is identified as
{\footnotesize
$$
\psi_{j, 0}^{F, \mathrm{sep-I}}(t, a_{j}, X)=F_{j, 0}(t \mid a_{j}, 1, X)-F_{j, 0}(t \mid a_{j}, 0, X).
$$
}
The separable indirect HTE on the RMTL is identified as
{\footnotesize
$$
\psi_{j, 0}^{\mathrm{RMTL}, \mathrm{sep-I}}(\tau, a_{j}, X)=\mathrm{RMTL}_{j, 0}(\tau \mid a_{j}, 1, X)-\mathrm{RMTL}_{j, 0}(\tau \mid a_{j}, 0, X).
$$
}

To facilitate general discussion in the presence of so many estimands, let $\psi(\cdot, X)$ denote a generic estimand defined above. 
%$(\psi^{S}(t, X),  \psi^{\mathrm{RMST}}(\tau, X), \psi_{j}^{F}(t, X), \psi_{j}^{\mathrm{RMTL}}(\tau, X), \psi_{j}^{F}(t, a_{\overline{j}}, X), \psi_{j}^{\mathrm{RMTL}}(\tau, a_{\overline{j}}, X), \psi_{j}^{F}(t, a_{j}, X), \psi_{j}^{\mathrm{RMTL}}(\tau, a_{j}, X))$ defined above. 
Further, let $\psi(\cdot)\equiv E_X[\psi(\cdot, X)]$ denote the marginal effect corresponding to arbitrary conditional effect $\psi(\cdot, X)$.

\section{Censoring unbiased orthogonal learning}

\subsection{(Counterfactual) Censoring unbiased transformations}

% time scale

A CUT is a function mapping a right-censored time-to-event variable to a continuous variable with the same conditional expectation as the \textit{uncensored} time-to-event variable.
%\begin{definition} \label{definition:cut}
%(Censoring unbiased transformation). A transformation $Y(\cdot; \eta_{0}^{\Lambda}): \mathcal{O} \rightarrow \mathcal{R}$ is said to be a CUT if $E_{0}\{Y(\cdot; \eta_{0}^{\Lambda}) \mid A, X\}=\mu_{0}(\cdot, A, X)$.
%\end{definition}
CUTs are useful because, by their definition, they can be substituted for right censored outcomes in regressions. Regressions for continuous outcomes (like CUTs) can be much easier to fit than regressions for right-censored outcomes that must take into account censoring. CUTs have a long history in statistics, including the Buckley-James (BJ) transformation~\citep{buckley1979linear}, %is one of the earliest CUTs and it has been popularized with statistical learning in recent years~\citep{wang2008doubly,steingrimsson2020deep,hu2021estimating,kong2023buckley}
the IPCW transformation~\citep{koul1981regression}, %and the cause-specific IPCW CUT has been taken as the response for direct binomial regression~\citep{scheike2008predicting,scheike2007direct,scheike2008flexible,azarang2017direct}. The third type of CUT 
and the augmented IPCW (AIPCW) transformation~\citep{rubin2007doubly}(which combines the BJ and IPCW transformations). ~\citep{rubin2007doubly} provides a nice overview of CUTs. 

In this work, we consider what we will dub `counterfactual CUTs'. A counterfactual CUT is a transformation of the observed outcome that, under causal assumptions, has the same conditional expectation as some function of a counterfactual value (or values) of the outcome under an intervention (or interventions) of interest. 
%\begin{definition} \label{definition:cut}
%(Counterfactual censoring unbiased transformation). A transformation $Y(\cdot, a; \eta_{0}^{\Lambda}): \mathcal{O} \rightarrow \mathcal{R}$ is said to be a counterfactual CUT if $E_{0}\{Y(\cdot, a; \eta_{0}^{\Lambda}) \mid A=a, X\}=\mu_{0}(\cdot, a, X)$, where $\eta_{0}^{\Lambda}$ represents all true distribution of conditional hazards that are required to compute the corresponding counterfactual CUT.
%\end{definition}
In~\Cref{theorem:cut}, we provide counterfactual CUTs and show that they have properties that will facilitate estimation of the estimands introduced in~\Cref{sec:estimands} under identification assumptions from ~\Cref{identification}. \\

\begin{theorem} \label{theorem:cut} Let $\eta_{\infty}^{\Lambda}$ and $\eta_{j, \infty}^{\Lambda}$ denote the element-wise $L_{2}(P_{0})$ limits of estimators $\widehat{\eta}^{\Lambda}$ and $\widehat{\eta}_{j}^{\Lambda}$ for for nuisance hazard functions $\eta^{\Lambda}_0$ and $\eta_{j, 0}^{\Lambda}$ defined in Section \ref{notation}. %We define CUTs as $Y(\cdot, A)=(1-A) Y(\cdot, 0)+A Y(\cdot, 1)$, where $Y(\cdot, a)$ denote a generic element of the set of counterfactual CUTs \sloppy $\{Y^{S}(t, a), Y^{\mathrm{RMST}}(\tau, a), Y_{j}^{F}(t, a), Y_{j}^{\mathrm{RMTL}}(\tau, a), Y_{j}^{F}(t, A_{j}=a, A_{\overline{j}}=a^{*}), Y_{j}^{\mathrm{RMTL}}(\tau, A_{j}=a, A_{\overline{j}}=a^{*}), Y_{j}^{F}(t, A_{j}=1-a^{*}, A_{\overline{j}}=a), Y_{j}^{\mathrm{RMTL}}(\tau, A_{j}=1-a^{*}, A_{\overline{j}}=a)\}$.

(i) (Counterfactual CUTs to facilitate estimation of $\psi_0^S(t,X)$) Define:
{\footnotesize
$$
\begin{aligned} \label{S_cut}
Y^{S, \mathrm{BJ}}(t, a; S)=&\frac{S(t \mid a, X)-\Delta I(\widetilde{T} \leq t) S(t \mid a, X)}{S(t \wedge \widetilde{T} \mid a, X)}. \\
% =&I(\widetilde{T}>t)+S(t \mid A, X) \int_{0}^{t \wedge \widetilde{T}} \frac{\mathrm{d} N^{C}(u \mid A, X)}{S(u \mid A, X)} \\
% =&I(\widetilde{T}>t)+\frac{(1-\Delta)I(\widetilde{T} \leq t)S(t \mid A, X)}{S(t \wedge \widetilde{T} \mid A, X)}
Y^{S, \mathrm{IPCW1}}(t, a; G)=&\frac{I(\widetilde{T}>t)}{G(t \mid a, X)} \quad \text{and} \quad Y^{S, \mathrm{IPCW2}}(t, a; G)=\frac{\Delta I(\widetilde{T}>t)}{G(\widetilde{T}- \mid a, X)} \\
Y^{S, \mathrm{AIPCW}}(t,a; \eta^{\Lambda})=&\frac{I(\widetilde{T}>t)}{G(t \mid a, X)}+S(t \mid a, X) \int_{0}^{t \wedge \widetilde{T}} \frac{\mathrm{d} M^{C}(u \mid a, X)}{S(u \mid a, X) G(u- \mid a, X)} \\
=&S(t \mid a, X)-S(t \mid a, X)\int_0^{t \wedge \widetilde{T}} \frac{\mathrm{d} M(u \mid a, X)}{S(u \mid a, X) G(u-\mid a, X)}.
\end{aligned}
$$
}
Under ~\Cref{assumption:positivity}-\Cref{assumption:depcens_survival}: if $S_{\infty}(t \mid a, X)=S_{0}(t \mid a, X)$, then $E_{0}\{Y^{S, \mathrm{BJ}}(t, A; S_{\infty}) \mid A=a, X\}=P_{0}(T^a>t \mid X)$; if $G_{\infty}(t \mid a, X)=G_{0}(t \mid a, X)$, then $E_{0}\{Y^{S, \mathrm{IPCW1}}(t, A; G_{\infty}) \mid A=a, X\}=E_{0}\{Y^{S, \mathrm{IPCW2}}(t, A; G_{\infty}) \mid A=a, X\}=P_{0}(T^a>t \mid X)$; if  either $S_{\infty}(t \mid a, X)=S_{0}(t \mid a, X)$ or $G_{\infty}(t \mid a, X)=G_{0}(t \mid a, X)$, then $E_{0}\{Y^{S, \mathrm{AIPCW}}(t, A; \eta_{\infty}^{\Lambda}) \mid A=a, X\}=P_{0}(T^a>t \mid X)$, where $P_{0}(T^a>t \mid X)$ is the true conditional counterfactual survival function.

(ii) (Counterfactual CUTs to facilitate estimation of $\psi_{j,0}^F(t,X)$) Define: % In the competing risk setting, the CUTs for cause-specific CIF are
{\footnotesize
$$
\begin{aligned} \label{Fj_cut}
Y_{j}^{F, \mathrm{BJ}}(t,a; (S, \Lambda_{j}))=&I(\widetilde{T} \leq t, \widetilde{J}=j)+\frac{(1-\Delta)I(\widetilde{T} \leq t) \{F_{j}(t \mid a, X)-F_{j}(t \wedge \widetilde{T} \mid a, X)\}}{S(t \wedge \widetilde{T} \mid a, X)} \\
Y_{j}^{F, \mathrm{IPCW}}(t,a; G)=&\frac{I(\widetilde{T} \leq t, \widetilde{J}=j)}{G(\widetilde{T}-\mid a, X)} \\
%=&I(\widetilde{T} \leq t, \widetilde{J}=j)+\int_0^{t \wedge \widetilde{T}} \frac{\{F_{j}(t \mid A, X)-F_{j}(u \mid A, X)\} \mathrm{d} N^C(u)}{S(u \mid A, X)} \\
Y_{j}^{F, \mathrm{AIPCW}}(t,a; \eta_{j}^{\Lambda})=&\frac{I(\widetilde{T} \leq t, \widetilde{J}=j)}{G(\widetilde{T}-\mid a, X)}+\int_0^{t \wedge \widetilde{T}} \frac{\{F_{j}(t \mid a, X)-F_{j}(u \mid a, X)\} \mathrm{d} M^C(u \mid a, X)}{S(u \mid a, X) G(u-\mid a, X)} \\
=&F_{j}(t \mid a, X)-\int_0^{t \wedge \widetilde{T}} \frac{\{F_{j}(t \mid a, X)-F_{j}(u \mid a, X)\}\mathrm{d} M(u \mid a, X)}{S(u \mid a, X) G(u-\mid a, X)}+\int_0^{t \wedge \widetilde{T}} \frac{\mathrm{d} M_{j}(u \mid a, X)}{G(u-\mid a, X)};
\end{aligned}
$$
}
Under ~\Cref{assumption:positivity} and~\Cref{assumption:SUTVA_competing}-\Cref{assumption:depcens_competing}: if $S_{\infty}(t \mid a, X)=S_{0}(t \mid a, X)$ and $\Lambda_{j, \infty}(t \mid a, X)=\Lambda_{j, 0}(t \mid a, X)$, then $E_{0}\{Y_{j}^{F, \mathrm{BJ}}(t, A; (S_{\infty}, \Lambda_{j, \infty})) \mid A=a, X\}=P_{0}(T^{a} \leq t, J^{a}=j \mid X)$; if $G_{\infty}(t \mid a, X)=G_{0}(t \mid a, X)$, then $E_{0}\{Y_{j}^{F, \mathrm{IPCW}}(t, A; G_{\infty}) \mid A=a, X\}=P_{0}(T^{a} \leq t, J^{a}=j \mid X)$; if  either $S_{\infty}(t \mid a, X)=S_{0}(t \mid a, X)$ and $\Lambda_{j, \infty}(t \mid a, X)=\Lambda_{j, 0}(t \mid a, X)$ or $G_{\infty}(t \mid a, X)=G_{0}(t \mid a, X)$, then $E_{0}\{Y_{j}^{F, \mathrm{AIPCW}}(t, A; \eta_{j, \infty}^{\Lambda}) \mid A=a, X\}=P_{0}(T^{a} \leq t, J^{a}=j \mid X)$, where $P_{0}(T^{a} \leq t, J^{a}=j \mid X)$ is the true conditional counterfactual cause-specific CIF. 

(iii) (Counterfactual CUTs to facilitate estimation of  $\psi^{F,\mathrm{sep-D}}_{j,0}(t,a_{\bar{j}},X)$) Define: % The CUTs for cause-specific separable direct effect,
{\footnotesize
$$
\begin{aligned} \label{Fj_sep_direct_cut}
&Y_{j}^{F,\mathrm{sep-D},\mathrm{AIPCW}}(t, a_j,a_{\bar{j}}; \eta_{j}^{\Lambda}) \\
=&F_{j}(t \mid a_j, a_{\bar{j}}, X)+\int_0^{t \wedge \widetilde{T}} \frac{S_{\overline{j}}(u \mid a_{\bar{j}}, X)\mathrm{d} M_{j}(u \mid a_j, X)}{S_{\overline{j}}(u \mid a_j, X)G(u- \mid a_j, X)} \\
&-\int_0^{t \wedge \widetilde{T}} \frac{S_{\overline{j}}(u \mid a_{\bar{j}}, X)\mathrm{d} M(u \mid a_j, X)}{S_{\overline{j}}(u \mid a_j, X)S(u \mid a_j, X)G(u- \mid a_j, X)}\{F_{j}(t \mid a_j, X)-F_{j}(u \mid a_j, X)\} \\
&%+\left\{1-\frac{I(A=a^{*}) \pi(a \mid X)}{\pi(a^{*} \mid X) I(A=a)} \right\}
+I(A=a_j)
\int_0^{t \wedge \widetilde{T}}\frac{\mathrm{d} M_{\overline{j}}(u \mid a_j, X)}{S(u \mid a_j, X)G(u- \mid a_j, X)}\{F_{j}(t \mid 1-a_{\bar{j}}, a_{\bar{j}}, X)-F_{j}(u \mid 1-a_{\bar{j}}, a_{\bar{j}}, X)\}.
% =&F_{j}(t \mid a, a^{*}, X)+\int_0^{t \wedge \widetilde{T}} \frac{S_{\overline{j}}(u \mid a^{*}, X)\mathrm{d} M_{j}(u \mid a, X)}{S_{\overline{j}}(u \mid a, X)G(u- \mid a, X)}\left[1-\frac{\{F_{j}(t \mid a, X)-F_{j}(u \mid a, X)\}}{S(u \mid a, X)}\right] \\
% &+\int_0^{t \wedge \widetilde{T}}\frac{\mathrm{d} M_{\overline{j}}(u \mid a, X)}{S(u \mid a, X)G(u- \mid a, X)}\left[F_{j}(t \mid 1-a^{*}, a^{*}, X)-F_{j}(u \mid 1-a^{*}, a^{*}, X) \right. \\
% &\left.-\frac{S_{\overline{j}}(u \mid a^{*}, X)}{S_{\overline{j}}(u \mid a, X)}\{F_{j}(t \mid a, X)-F_{j}(u \mid a, X)\}\right] \\
% =&F_{j}(t \mid a, a^{*}, X)+\int_0^{t \wedge \widetilde{T}} \frac{S_{\overline{j}}(u \mid a^{*}, X)\mathrm{d} M_{j}(u \mid a, X)}{S_{\overline{j}}(u \mid a, X)G(u- \mid a, X)} \\
% &-\int_0^{t \wedge \widetilde{T}} \frac{S_{\overline{j}}(u \mid a^{*}, X)\mathrm{d} M(u \mid a, X)}{S_{\overline{j}}(u \mid a, X)S(u \mid a, X)G(u- \mid a, X)}\{F_{j}(t \mid a, X)-F_{j}(u \mid a, X)\} \\
% &+\int_0^{t \wedge \widetilde{T}}\frac{\mathrm{d} M_{\overline{j}}(u \mid a, X)}{S(u \mid a, X)G(u- \mid a, X)}\{F_{j}(t \mid 1-a^{*}, a^{*}, X)-F_{j}(u \mid 1-a^{*}, a^{*}, X)\}.
\end{aligned}
$$
}

Under ~\Cref{assumption:positivity} and~\Cref{assumption:SUTVA_competing}-\Cref{assumption:dismissible}: if $\Lambda_{j, \infty}(t \mid a, X)=\Lambda_{j, 0}(t \mid a, X)$ and $\Lambda_{\overline{j}, \infty}(t \mid a, X)=\Lambda_{\overline{j}, 0}(t \mid a, X)$, then $E_{0}\{Y_{j}^{F, \mathrm{sep-D}, \mathrm{AIPCW}}(t,A,a_{\bar{j}}; \eta_{j, \infty}^{\Lambda}) \mid A=a_j, X\}=P_0(T^{a_j, a_{\bar{j}}}\leq t,J^{a_j, a_{\bar{j}}}=j \mid X)$, which is the true conditional counterfactual separable cause-specific CIF.

%As $\pi(\cdot \mid X)$ doesn't show up, $Y_{j}^{F, \mathrm{AIPCW}}(t, A_{j}=A, A_{\overline{j}}=a^{*}; \eta_{j}^{\Lambda})$ is not doubly robust and shares the same consistency conditions with $Y_{j}^{F, \mathrm{BJ}}(t, A_{j}=A, A_{\overline{j}}=a^{*}; \eta_{j}^{\Lambda})$ whose $G(\cdot \mid A, X)$ is defined as 1. The influence function, $\psi_{j}^{F}(t, A_{\overline{j}}=a^{*}; \eta_{j})$, requires weaker consistency conditions and is doubly robust. \\

%  or $G_{\infty}(t \mid a, X)=G_{0}(t \mid a, X)$ and $\Lambda_{\overline{j}, \infty}(t \mid a, X)=\Lambda_{\overline{j}, 0}(t \mid a, X)$

(iv) (Counterfactual CUTs to facilitate estimation of $\psi_{j,0}^{F,\mathrm{sep-I}}(t,a_j,X)$) Define: 
% the following are counterfactual CUTs for cause-specific separable indirect CIF $F_{j}(t \mid a_{j}=1-a^{*}, A, X)=P_{0}(T\leq t, J=j \mid A_{j}=1-a^{*}, A_{\overline{j}}=A, X)$ % The CUTs for cause-specific separable indirect effect are
{\footnotesize
$$
\begin{aligned} \label{Fj_sep_indirect_cut}
&Y_{j}^{F, \mathrm{sep-I}, \mathrm{AIPCW}}(t,a_j,a_{\bar{j}}; \eta_{j}^{\Lambda}) \\
=&F_{j}(t \mid a_j, a_{\bar{j}}, X)+\int_0^{t \wedge \widetilde{T}} \frac{\mathrm{d} M_{j}(u \mid a_{\bar{j}}, X)}{G(u- \mid a_{\bar{j}}, X)}\left\{1-\frac{S_{\overline{j}}(u \mid 1-a_j, X)}{S_{\overline{j}}(u \mid a_{\bar{j}}, X)}\right\} \\
&-\int_0^{t \wedge \widetilde{T}}\frac{\mathrm{d} M(u \mid a_{\bar{j}}, X)}{S(u \mid a_{\bar{j}}, X)G(u- \mid a_{\bar{j}}, X)}\left\{1-\frac{S_{\overline{j}}(u \mid 1-a_j, X)}{S_{\overline{j}}(u \mid a_{\bar{j}}, X)}\right\}\{F_{j}(t \mid a_{\bar{j}}, X)-F_{j}(u \mid a_{\bar{j}}, X)\} \\
&-I(A=a_{\bar{j}}) \int_0^{t \wedge \widetilde{T}}\frac{\mathrm{d} M_{\overline{j}}(u \mid a_{\bar{j}}, X)}{S(u \mid a_{\bar{j}}, X)G(u- \mid a_{\bar{j}}, X)}\{F_{j}(t \mid a_j, 1-a_j, X)-F_{j}(u \mid a_j, 1-a_j, X)\}. \\
% =&F_{j}(t \mid 1-a^{*}, a, X)+\int_0^{t \wedge \widetilde{T}} \frac{\mathrm{d} M_{j}(u \mid a, X)}{G(u- \mid a, X)} \left\{1-\frac{S_{\overline{j}}(u \mid a^{*}, X)}{S_{\overline{j}}(u \mid a, X)}\right\}\left[1-\frac{\{F_{j}(t \mid a, X)-F_{j}(u \mid a, X)\}}{S(u \mid a, X)}\right] \\
% &+\int_0^{t \wedge \widetilde{T}}\frac{\mathrm{d} M_{\overline{j}}(u \mid a, X)}{S(u \mid a, X)G(u- \mid a, X)}\left[-F_{j}(t \mid 1-a^{*}, a^{*}, X)+F_{j}(u \mid 1-a^{*}, a^{*}, X) \right. \\
% &\left.-\left\{1-\frac{S_{\overline{j}}(u \mid a^{*}, X)}{S_{\overline{j}}(u \mid a, X)}\right\}\{F_{j}(t \mid a, X)-F_{j}(u \mid a, X)\}\right] \\
% =&F_{j}(t \mid 1-a^{*}, a, X)+\int_0^{t \wedge \widetilde{T}} \frac{\mathrm{d} M_{j}(u \mid a, X)}{G(u- \mid a, X)}\left\{1-\frac{S_{\overline{j}}(u \mid a^{*}, X)}{S_{\overline{j}}(u \mid a, X)}\right\} \\
% &-\int_0^{t \wedge \widetilde{T}}\frac{\mathrm{d} M(u \mid a, X)}{S(u \mid a, X)G(u- \mid a, X)}\left\{1-\frac{S_{\overline{j}}(u \mid a^{*}, X)}{S_{\overline{j}}(u \mid a, X)}\right\}\{F_{j}(t \mid a, X)-F_{j}(u \mid a, X)\} \\
% &-\int_0^{t \wedge \widetilde{T}}\frac{\mathrm{d} M_{\overline{j}}(u \mid a, X)}{S(u \mid a, X)G(u- \mid a, X)}\{F_{j}(t \mid 1-a^{*}, a^{*}, X)-F_{j}(u \mid 1-a^{*}, a^{*}, X)\}.
\end{aligned}
$$
}
Under ~\Cref{assumption:positivity} and~\Cref{assumption:SUTVA_competing}-\Cref{assumption:dismissible}: if  $\Lambda_{j, \infty}(t \mid a, X)=\Lambda_{j, 0}(t \mid a, X)$ and $\Lambda_{\overline{j}, \infty}(t \mid a, X)=\Lambda_{\overline{j}, 0}(t \mid a, X)$, then $E_{0}\{Y_{j}^{F, \mathrm{sep-I}, \mathrm{AIPCW}}(t,a_j,A; \eta_{j}^{\Lambda, \infty}) \mid A=a_{\bar{j}}, X\}=P_0(T^{a_j, a_{\bar{j}}}\leq t,J^{a_j, a_{\bar{j}}}=j \mid X)$, which is the true conditional counterfactual separable cause-specific CIF.

% That is, \sloppy $E\{Y_{j}^{F, \mathrm{AIPCW}}(t, A_{j}=1-a^{*}, A_{\overline{j}}=A; \eta_{j}^{\Lambda}) \mid A=a, X\}=E\{I(T^{1-a^{*}, a}\leq t,J^{1-a^{*}, a}=j) \mid X\}=P_0(T^{1-a^{*}, a}\leq t,J^{1-a^{*}, a}=j \mid X)$, which is the separable indirect counterfactual cause-specific CIF. \\
\end{theorem}

For similar results demonstrating how CUTs facilitate the estimation of RMST and RMTL estimands, refer to \Cref{additional:cut}. The proof of Theorem 1 is provided in \Cref{theorem:cut:proof}.

\begin{remark} (Influence Function Transformation CUTs)
    For generic estimand $\psi(\cdot,X)$, define the `influence function transformation' (IF transformation) to be the nonparametric influence function $\phi(\cdot; \eta_{0})$ of the corresponding marginal effect $\psi(\cdot)$, where $\eta_{0}=\{\eta_{0}^{\Lambda}, \pi_{0}(a \mid X)\}$ denotes nuisance parameters. $\phi(\cdot; \eta_{0})$ is also a useful counterfactual CUT, in that its conditional expectation given $X$ yields the HTE, i.e. $E_{0}\{\phi(\cdot; \eta_{0}) \mid X\}=\psi_{0}(\cdot, X)$. The influence functions for most marginal effects considered here have been previously derived~\citep{hubbard2000nonparametric,diaz2018targeted,ozenne2020estimation,martinussen2023estimation}, and others have studied HTE estimation via regression of IF transformations on $X$~\citep{lee2017doubly,semenova2021debiased,kennedy2023towards,shin2023improved,knaus2022double}. However, as we will discuss later, IF transformations cannot be plugged into arbitrary HTE learners like the counterfactual CUTs defined in Theorem 1 can. Therefore, we focus on the CUTs from Theorem 1 in the remainder of the paper. %For completeness, we provide the IF transformations for each of the estimands of interest in the Appendix. % curth2020estimating, fan2022estimation, zimmert2019nonparametric,
    \end{remark}
% \begin{remark}\label{remark:consistency}
% (Consistency of CUTs) In Theorem 1, all counterfactual CUTs were functions of the true values of nuisance parameters ($\eta_0^{\Lambda}$ or $\eta_{j, 0}^{\Lambda}$) representing conditional hazard functions of event and censoring processes. Of course, in practice these nuisance parameters would not be known and estimates $\hat{\eta}^{\Lambda}$ or $\hat{\eta}_{j}^{\Lambda}$ would be substituted. All of the AIPCW transformations relies on consistent estimation of either the censoring process or all the event processes. The consistency of all non-AIPCW transformations depends on the consistent estimation of all nuisance parameters. The counterpart pseudo-observations follow the same spirit.
% \end{remark}

\begin{remark}
\label{remark:consistency}
(Consistency condition comparison) We note that the consistency conditions for some estimands are stronger for the counterfactual CUTs from~\Cref{theorem:cut} than for the IF transformation in the sense that more nuisance functions need to be consistently estimated. Thus the IF transformations have stronger robustness properties. Because we are considering machine learning nuisance models that can typically be assumed to be consistent, this is not a major consideration. Details are in the Appendix.
\end{remark}

\begin{remark}\label{remark:pseudo}
(Pseudo-observations) The proposed $Y^{S, \mathrm{IPCW}1}(t, A; \eta^{\Lambda})$ and $Y^{S, \mathrm{IPCW}2}(t, A; \eta^{\Lambda})$ transformations are numerically identical to modified versions of the jackknifed pseudo-observations introduced in~\citet{andersen2003generalised,overgaard2017asymptotic} by replacing Nelson-Aalen estimators with IPCW estimators.

% $Y^{\mathrm{BJ}}(\cdot, a; \eta^{\Lambda})$ and $Y^{\mathrm{AIPCW}}(\cdot, a; \eta^{\Lambda})$, This property also holds for $Y^{\mathrm{BJ}}(\cdot, a; \eta^{\Lambda})$ and $Y^{\mathrm{AIPCW}}(\cdot, a; \eta^{\Lambda})$.

%(Should I comment out the following paragraph)
%IPCW jacknifed pseudo-observations and asymptotically equivalent to existing pseudo-observations which uses the Nelson-Aalen estimator for cumulative hazards. Though the AIPCW and BJ pseudo-observations haven't be proposed, the proposed AIPCW CUTs are numerically identical to the AIPCW jacknifed pseudo-observations and the proposed BJ CUTs are numerically identical to the BJ jacknifed pseudo-observations.
\end{remark}

\subsection{Review of HTE learners with continuous outcomes}

In this section, we provide a review of HTE learners for continuous outcomes. HTE learners for continuous outcomes can be categorized into `conditional mean difference' learners and `transformed minimization' learners. Conditional mean difference learners estimate the conditional expectation of the outcome given covariates and each treatment value, then take the difference of the conditional expectations across treatment values. Let $\mu_{0}(a, X)=E_{0}(Y \mid A=a, X)$ and $\psi_{0}(X)=\mu_{0}(1, X)-\mu_{0}(0, X)$ where $Y$ is a continuous outcome. The S-learner estimates the HTE by $\widehat{\psi}^{\mathrm{SL}}(X)=\widehat{\mu}^{\mathrm{SL}}(1, X)-\widehat{\mu}^{\mathrm{SL}}(0, X)$, where the single regression function $\mu^{\mathrm{SL}}(A, X)$ is estimated by an arbitrary base learner. A shortcoming of S-learners is that by fitting one model, the same regularization is imposed on both treatment arms, which can be inappropriate. % Examples of proposed S-learners include parametric interaction~\citet{vanderweele2014tutorial}, Bayesian additive regression trees (BART)~\citet{hill2011bayesian}, random forests~\citet{foster2011subgroup}, and neural networks~\citet{farrell2020deep}. 

To overcome this shortcoming, the T-learner fits separate regressions in each arm, i.e.
{\footnotesize
$$
\widehat{\psi}^{\mathrm{TL}}(X)=\widehat{\mu}^{\mathrm{TL}}(1, X)-\widehat{\mu}^{\mathrm{TL}}(0, X)
$$
}
where $\mu^{\mathrm{TL}}(0, X)$ and $\mu^{\mathrm{TL}}(1, X)$ are treatment specific outcome models. % Proposed T-learners include causal trees~\citep{athey2016recursive}, random forests~\citep{lu2018estimating}, causal boosting~\citep{sugasawa2019estimating}, and causal multivariate adaptive regression splines (MARS)~\citep{powers2018some}. ~\citet{gao2020minimax} derived the minimax rates and lower bounds of smooth nonparametric T-learners. % Motivated by~\citet{steingrimsson2020deep},~\citet{hu2021estimating} applied both BJ and AIPCW CUT which can accommodate both S-learner and T-learner.
A drawback of T-learners is that common information between treatment groups is not permitted as we fit two curves on two samples. MT-learners~\citep{alaa2017bayesian} share aspects of covariate representation across the treatment specific regressions to address this issue.

Instead of contrasting conditional means, a second category of HTE estimators directly solve various transformed outcome (possibly weighted) loss minimization problems. Defining a pointwise loss function $\ell_{2}^{*}(Y, \psi(X); (\mu_{0}, \pi_{0}))$ and population loss $L_{2}^{*}(Y, \psi(X); (\mu_{0}, \pi_{0}))=E\{\ell_{2}^{*}(Y, \psi(X); (\mu_{0}, \pi_{0}))\}$, weighted minimization methods can be expressed in the form
{\footnotesize
$$
\begin{aligned}
&\min _{\psi_{0}(\cdot, X)} \; E_{0}[\ell_{2}^{*}(Y, \psi(X); (\mu_{0}, \pi_{0}))+\mathrm{RL}\{\psi(X)\} \\
=&E_{0}[w^{*}(\pi_{0}) \{Y^{*}(Y; (\mu_{0}, \pi_{0}))-\psi(X)\}^{2}+\mathrm{RL}\{\psi(X)\}]
\end{aligned}
$$
}
with choices of sample weight $w^{*}$ and transformed outcome $Y^{*}$ summarized in~\Cref{tab:weighted_minimization}. %This approach estimates the nuisance functions and the HTE separately and imposes various regularizers such that both quantities do not share similar complexity. % We generalize from~\citet{knaus2021machine} and categorize the weighted minimization method into outcome tranformation, covariate tranformation, and orthogonal learning.

\begin{table}[h]
\small
\centering
\caption{Summary of proposed HTE learners based on transformed minimization}
\begin{tabular}{lcc}
\hline Learner ($^{*}$) & $w^{*}$ & $Y^{*}$ \\
\hline
IPTW-learner~\citep{hirano2003efficient} & 1 & $Y \frac{A-\widehat{\pi}(1 \mid X)}{\widehat{\pi}(0 \mid X)\widehat{\pi}(1 \mid X)}$ \\
RA-learner~\citep{curth2021nonparametric} & 1 & $I(A=1)\{Y-\widehat{\mu}(1, X)\}$ \\
(X-learner~\citet{kunzel2019metalearners}) & & $+I(A=0)\{\widehat{\mu}(0, X)-Y\}$ \\
AIPTW-learner~\citep{robins1995semiparametric} & 1 & $\widehat{\mu}(1, X)-\widehat{\mu}(0, X)$ \\
\citep{kennedy2023towards} & & $+\frac{I(A=1)\{Y-\widehat{\mu}(1, X)\}}{\widehat{\pi}(1 \mid X)}-\frac{I(A=0)\{Y-\widehat{\mu}(0, X)\}}{\widehat{\pi}(0 \mid X)}$ \\
IF-learner~\citet{diaz2018targeted} & 1 & $\phi(\tau; \widehat{\eta})$ \\
MC-learner~\citet{tian2014simple,chen2017general} & $\frac{(2 A-1)\{A-\widehat{\pi}(1 \mid X)\}}{4\widehat{\pi}(0 \mid X)\widehat{\pi}(1 \mid X)}$ & $2(2 A-1)Y$ \\
MCEA-learner~\citep{meng2022augmented} & $\frac{(2 A-1)\{A-\widehat{\pi}(1 \mid X)\}}{4\widehat{\pi}(0 \mid X)\widehat{\pi}(1 \mid X)}$ & $2(2 A-1)\{Y-\widehat{\mu}(X)\}$ \\
R-learner~\citep{robinson1988root,nie2021quasi}
 & $\{A-\widehat{\pi}(1 \mid X)\}^2$ & $\frac{Y-\widehat{\mu}(X)}{A-\widehat{\pi}(1 \mid X)}$ \\
\citep{foster2023orthogonal} & & \\
U-learner & 1 & $\frac{Y-\widehat{\mu}(X)}{A-\widehat{\pi}(1 \mid X)}$ \\
\hline
\end{tabular}
\label{tab:weighted_minimization}
\end{table}
%Learners that require a prior distribution on the HTE are called Bayesian-learners (B-learners)~\citep{hahn2018regularization,caron2022shrinkage} and are not considered in this work.

% (question: shall I use subscript 0 or hat? shall I remove $\widehat{\eta}^{\Lambda}$ in $\widehat{\mu}$? See proofs)

% handles HTE estimation in a Bayesian fashion under orthogonal learning. The observed outcome can be parametrized as
% $$
% Y=\mu(0, X)+\psi(X) A+\varepsilon
% $$
% where the prognostic score $\mu(0, X)$ acts as the intercept and $\psi(X)$ as the slope. A prior distribution of $p(\psi(\cdot))$ can be specified under subject matter knowledge such that regularization can preserve simple heterogeneous structures. Typical Bayesian methods include BART~\citep{hahn2018regularization}, Bayesian causal forest (BCF)~\citep{hahn2020bayesian}, Shrinkage BCF~\citep{caron2022shrinkage}, Gaussian process, and Dirichlet process~\citep{shiraito2016uncovering}. B-learners obtain full predictive posterior distribution of HTE, which facilitates computing point estimates and credible intervals. This advantage is also shared by S-learner with BART as the base-learner~\citep{hill2011bayesian}.

\subsection{Plugging in counterfactual CUTs to continuous HTE learners} % Learning algorithm of the proposed estimators

Let $Y(\cdot, A;\eta^{\Lambda})$ denote an arbitrary counterfactual CUT from ~\Cref{theorem:cut} evaluated at the observed treatment value and using whichever components of $\eta^{\Lambda}$ are required, i.e. any of $\{Y^{S,\mathrm{BJ}}(t, A;S), Y^{S,\mathrm{IPCW1}}(t,A;G),Y^{S,\mathrm{IPCW2}}(t,A;G), Y^{S,\mathrm{AIPCW}}(t,A;\eta^{\Lambda}), Y^{F,\mathrm{BJ}}(t,A;S,\Lambda_j)$, \\
$Y^{F,\mathrm{IPCW}}(t,A;G),Y^{F,\mathrm{AIPCW}}(t,A;\eta^{\Lambda}),Y^{F,\mathrm{sep-D,AIPCW}}(t,A,a_{\bar{j}};\eta^{\Lambda}),Y^{F,\mathrm{sep-I,AIPCW}}(t,a_{j},A;\eta^{\Lambda})\}$. In this section, we define survival outcome HTE learners obtained by plugging any $Y(\cdot, A;\eta^{\Lambda})$ into any of the continuous outcome HTE learners defined in Section 3.2. The consistency of the resulting plug-in learners relies on the consistency of the CUTs (established in ~\Cref{theorem:cut}) and the consistency of the HTE learner. % To facilitate subsequent general discussions of all of the counterfactual CUTs we introduced in ~\Cref{theorem:cut}, we let.

\begin{theorem} \label{theorem:equivalence}
The true HTE $\psi_{0}(\cdot, X)$ minimizes relevant oracle squared error loss functions with the counterfactual CUT $Y(\cdot, A; \eta_{0}^{\Lambda})$ as the outcome, i.e.
{\footnotesize
$$
\psi_{0}(\cdot, X)=\underset{\psi(\cdot, X)}{\arg \min} \; L_{2}^{*}(Y(\cdot, A; \eta_{0}^{\Lambda}), \psi(\cdot, X); (\mu_{0}, \pi_{0}))= \underset{\psi(\cdot, X)}{\arg \min} \; E_{0}[w^{*}(\pi_{0}) \{Y^{*}(Y(\cdot, A; \eta_{0}^{\Lambda}); (\mu_{0}, \pi_{0}))-\psi(\cdot, X)\}^{2}
$$
}

% The AIPCW Brier squared orthogonal loss $L\textbf{}_{2}^{\mathrm{AIPCW}, *}(f((T, J), \cdot), \psi(\cdot, X); \eta_{0})$, where $f((T, J), \cdot)$ defines the quantity of interest using uncensored survival times (and cause-of-failure), produces the same minimzer $\psi_{0}(\cdot, X)$ as the squared loss using AIPCW CUT $L_{2}^{*}(Y^{\mathrm{AIPCW}}(\cdot; \eta_{0}^{\Lambda}), \psi(\cdot, X); (\mu_{0}, \pi_{0}))$. % HTE estimated from the full data orthogonal $L_{2}$ loss function $\ell\{\psi_{0}(\cdot; x); \eta\}$ is  the censoring unbiased learning algorithm with $L_{2}$ loss function $\widetilde{\ell}\{\psi_{0}(\cdot;x); \eta\}$.
\end{theorem}
Practically, Theorem~\ref{theorem:equivalence} implies that any of the HTE learners from Section 3.2 applied to any of the counterfactual CUT outcomes from Theorem 1 yield consistent estimates of the corresponding heterogeneous effects of interest. Furthermore, we have that the AIPCW CUTs in particular plugged into AIPTW estimators are asymptotically equivalent to the IF transformations, which we know are equal to HTEs in conditional expectation.%that constructing complex HTE learners on the censored loss functions are identical to estimating HTE from existing learners by taking AIPCW CUT as continuous outcome.

\begin{proposition} \label{proposition:augmented}

(Asymptotic equivalence) The AIPTW transformation of the AIPCW CUT is asymptotically equivalent to the time-to-event IF transformation for all estimands, i.e.
{\footnotesize
$$
\begin{aligned}
& \frac{I(A=a)Y^{S, \mathrm{AIPCW}}(t, A; \eta_{0}^{\Lambda})}{\pi_{0}(a \mid X)}+\left\{1-\frac{I(A=a)}{\pi_{0}(a \mid X)}\right\}E_{0}\{Y^{S, \mathrm{AIPCW}}(t, A; \eta_{0}^{\Lambda}) \mid A=a, X\}=\phi^{S}(t, a; \eta_{0}) \\
& \frac{I(A=a)Y_{j}^{F, \mathrm{AIPCW}}(t, A; \eta_{ 0}^{\Lambda})}{\pi_{0}(a \mid X)}+\left\{1-\frac{I(A=a)}{\pi_{0}(a \mid X)}\right\}E_{0}\{Y_{j}^{F, \mathrm{AIPCW}}(t, A; \eta_{0}^{\Lambda}) \mid A=a, X\}=\phi_{j}^{F}(t, a; \eta_{0}) \\
& \frac{I(A=a) Y_{j}^{F, \mathrm{sep-D}, \mathrm{AIPCW}}(t,A,a^{*}; \eta_{0}^{\Lambda})}{\pi_{0}(a \mid X)}+\left\{1-\frac{I(A=a)}{\pi_{0}(a \mid X)}\right\}E_{0}\{Y_{j}^{F, \mathrm{sep-D}, \mathrm{AIPCW}}(t,A,a^{*}; \eta_{j, 0}^{\Lambda}) \mid A=a, X\} \\
=&\phi^{F}_{j}(t, A_{j}=a, A_{\overline{j}}=a^{*}; \eta_{0}) \\
& \frac{I(A=a) Y_{j}^{F, \mathrm{sep-I}, \mathrm{AIPCW}}(t,a^{*},A; \eta_{0}^{\Lambda})}{\pi_{0}(a \mid X)}+\left\{1-\frac{I(A=a)}{\pi_{0}(a \mid X)}\right\}E_{0}\{Y_{j}^{F, \mathrm{sep-D}, \mathrm{AIPCW}}(t,a^{*},A; \eta_{0}^{\Lambda}) \mid A=a, X\} \\
=&\phi^{F}_{j}(t, A_{j}=a^{*}, A_{\overline{j}}=a; \eta_{0}) \\
%& \frac{I(A=a)Y^{\mathrm{RMST}, \mathrm{AIPCW}}(\tau, A; \eta_{0}^{\Lambda})}{\pi_{0}(a \mid X)}+\left\{1-\frac{I(A=a)}{\pi_{0}(a \mid X)}\right\}E_{0}\{Y^{\mathrm{RMST}, \mathrm{AIPCW}}(\tau, A; \eta_{0}^{\Lambda}) \mid A=a, X\}=\phi^{\mathrm{RMST}}(\tau, a; \eta_{0}) \\
%& \frac{I(A=a)Y_{j}^{\mathrm{RMTL}, \mathrm{AIPCW}}(\tau, A; \eta_{j, 0}^{\Lambda})}{\pi_{0}(a \mid X)}+\left\{1-\frac{I(A=a)}{\pi_{0}(a \mid X)}\right\}E_{0}\{Y_{j}^{\mathrm{RMTL}, \mathrm{AIPCW}}(\tau, A; \eta_{j, 0}^{\Lambda}) \mid A=a, X\}=\phi_{j}^{\mathrm{RMTL}}(\tau, a; \eta_{j, 0}) \\
%& \frac{I(A=a) Y_{j}^{\mathrm{RMTL}, \mathrm{sep-D}, \mathrm{AIPCW}}(\tau,A,a^{*}; \eta_{j, 0}^{\Lambda})}{\pi_{0}(a \mid X)}+\left\{1-\frac{I(A=a)}{\pi_{0}(a \mid X)}\right\}E_{0}\{Y_{j}^{\mathrm{RMTL}, \mathrm{sep-D}, \mathrm{AIPCW}}(\tau,A,a^{*}; \eta_{j, 0}^{\Lambda}) \mid A=a, X\} \\
%=&\phi^{\mathrm{RMTL}}_{j}(\tau, A_{j}=a, A_{\overline{j}}=a^{*}; \eta_{j, 0}) \\
%& \frac{I(A=a) Y_{j}^{\mathrm{RMTL}, \mathrm{sep-I}, \mathrm{AIPCW}}(\tau,a^{*},A; \eta_{j, 0}^{\Lambda})}{\pi_{0}(a \mid X)}+\left\{1-\frac{I(A=a)}{\pi_{0}(a \mid X)}\right\}E_{0}\{Y_{j}^{\mathrm{RMTL}, \mathrm{sep-D}, \mathrm{AIPCW}}(\tau,a^{*},A; \eta_{j, 0}^{\Lambda}) \mid A=a, X\} \\
%=&\phi^{\mathrm{RMTL}}_{j}(\tau, A_{j}=a^{*}, A_{\overline{j}}=a; \eta_{j, 0})
\end{aligned}
$$
}

\end{proposition}
% Note that the validity of outcome transformation lies on $\psi_{0}(X)=E_{0}(Y_{0}^{\mathrm{IPTW}}(X) \mid X=x)=E_{0}(Y_{0}^{\mathrm{RA}}(X) \mid X=x)=E_{0}(Y_{0}^{\mathrm{APITW}}(X) \mid X=x)$. In practice, as the values of $\pi_{0}(a \mid X)$ and $\mu_{0}(a, X)$ are not available from data, these nuisance parameters can be predicted by statistical learning algorithms with cross-fitting.
See~\Cref{proposition:augmented:extension} for an extension of~\Cref{proposition:augmented} to RMTL and RMST estimands. Despite being equivalent asymptotically, finite sample performance can differ between learners based on AIPTW transformed AIPCW CUTs and based on IF transformations. 

% \begin{remark}\label{remark:connection}
% (Coarsening unbiased transformation) Fundamentally, the AIPCW CUT is the uncentered IF of the censored quantity-of-interest from a RCT where intervention is randomized~\citet{reid1979influence}; the AIPTW transformation is the uncentered IF of a continuous outcome from an observational study with confounded exposure; the IF transformation is the uncentered IF of the censored outcome from an observational study. The CUTs, including IPCW, BJ, and AIPCW, and the outcome transformation $Y^{*}$ with $w^{*}=1$ in~\Cref{tab:weighted_minimization} belong to a broader class called coarsening unbiased transformation where both censoring and treatment can be considered as different coarsening processes at random (no unmeasured confounding). The empirical mean of the AIPCW, AIPTW and IF transformations are double/debiased machine learning (DML) estimators for the average treatment effect of corresponding causal estimands whose finite sample variance can be estimated by the empirical mean of the squared of the centered transformations. % The BJ CUTs are the censoring version of RA transformation; the IPCW CUTs are the censoring version of IPTW transformation; the AIPCW CUTs are the censoring version of AIPTW transformation; the IF transformation is the combination of the AIPCW CUT and the AIPTW transformation.
% \end{remark}

We will now describe a cross-fitting algorithm to estimate the HTE at $X=x$. For S-learners, T-learners, or IF-learners, the algorithm comprises ~\Cref{step1}-~\Cref{step4}. For transformed minimization learners, the algorithm comprises ~\Cref{step1}-\Cref{step3} then~\Cref{step5}-\Cref{step7}. \\

\begin{step} \label{step1}
First sample-splitting. Randomly split the data $O_1, \ldots, O_n$ into $K_{1}$ disjoint validation sets of approximately equal size $\mathcal{V}_{1}, \ldots, \mathcal{V}_{K_{1}}$ with sizes $n_1, \ldots, n_{K_{1}}$, where $K_{1} \in\{2,3, \ldots,\lfloor n / 2\rfloor\}$. For each $k = 1, \ldots, K_{1}$, define training set $\overline{\mathcal{V}}_{k}=\{O_i: i \notin \mathcal{V}_{k}\}$. % and treatment-specific training set $\overline{\mathcal{V}}_{a, k_{1}}=\{O_i: i \in \overline{\mathcal{V}}_{k}, A_i=a\}$.
\end{step}

\begin{step} \label{step2}
Nuisance model estimation. For every $k = 1, \ldots, K_{1}$, construct any estimators $\widehat{\eta}_{k}^{\Lambda}$ of $\eta^{\Lambda}_0$ and/or $\widehat{\pi}_{k}(a \mid X)$ of $\pi_0(a \mid X)$ required in later steps using $\overline{\mathcal{V}}_{k}$.
\end{step}

\begin{step} \label{step3}
CUT construction. For every $k = 1, \ldots, K_{1}$, construct transformed outcomes by plugging $\widehat{\eta}_{k}^{\Lambda}$ into $Y(\cdot, A; \widehat{\eta}_{k}^{\Lambda})$ or (for the IF transformation) $\widehat{\eta}_{k}=\{\widehat{\eta}_{k}^{\Lambda}, \widehat{\pi}_{k}(a \mid X)\}$ into $\phi(\cdot; \widehat{\eta}_{k})$ using $\mathcal{V}_{k}$. For every $k = 1, \ldots, K_{1}$, append any $Y(\cdot, A; \widehat{\eta}_{k}^{\Lambda})$ or $\phi(\cdot; \widehat{\eta}_{k})$ required in later steps to $\mathcal{V}_{k}$ and combine into one augmented dataset $O'$.
\end{step}

% \begin{step}
% Survival conditional mean difference HTE learners and CUTs construction. For every $k = 1, \ldots, K_{1}$,
% \begin{enumerate}
%   \item 

%   \item 
  
%   train models for $\eta_{k_{1}}^{\Lambda}$ or $\eta_{j, k_{1}}^{\Lambda}$ using $\overline{\mathcal{V}}_{k_{1}}$
  
%   \item  $\widehat{\eta}_{k_{1}}^{\Lambda}$ or $\widehat{\eta}_{j, k_{1}}^{\Lambda}$ using training data $\mathcal{V}_{k_{1}}$
  
%   \item evaluate at unique observed survival times from $\mathcal{V}_{k_{1}}$

%   \item when $j^{*}=1$, train time-to-event learners on $\overline{\mathcal{V}}_{k_{1}}$ and evaluate $\widehat{\Lambda}_{k_{1}}(t \mid A, X)$ using $\mathcal{V}_{k_{1}}$ at unique observed survival times $\widetilde{T}$ in $\mathcal{V}_{k_{1}}$ to capture jump information from observed counting processes $\mathrm{d} N(t)$ and $\mathrm{d} N^C(t)$; when $j^{*}>1$, train a time-to-event model with $(A, X, AX)$ or a MT-learner with $X$ on $\overline{\mathcal{V}}_{k_{1}}$ for cause $j$ and all causes, and evaluate $\widehat{\Lambda}_{j, k_{1}}(t \mid A, X)$ and $\widehat{\Lambda}_{k_{1}}(t \mid A, X)$ using $\mathcal{V}_{k_{1}}$ at unique observed survival times $\widetilde{T}$ in $\mathcal{V}_{k_{1}}$ to capture jump information from observed counting processes $\mathrm{d} N(t)$, $\mathrm{d} N^C(t)$, and $\mathrm{d} N_{j}(t)$;

%   \item train the propensity model on $\overline{\mathcal{V}}_{k_{1}}$ and predict $\widehat{\pi}_{k_{1}}(a \mid X)$ on $\mathcal{V}_{k_{1}}$

% \end{enumerate}

% \end{step}

\begin{step} \label{step4}
S-learner, T-learner, and IF-learner estimation. \begin{itemize}
    \item IF-Learner: Estimate the regression $\widehat{E}\{\phi(\cdot; \widehat{\eta}) \mid X\}$ on $O'$. Obtain the IF-learner estimates of $\psi(\cdot, x)$ as $\widehat{E}\{\phi(\cdot; \widehat{\eta}) \mid X=x\}$.
    \item S-Learner and T-Learner: Estimate the treatment specific conditional mean of $Y(\cdot, A; \widehat{\eta}^{\Lambda})$ as $\widehat{\mu}(\cdot, A, x; \widehat{\eta}^{\Lambda})$ for $A=0, 1$ using learner-appropriate regression models and obtain the HTE estimate as $\widehat{\mu}(\cdot, 1, x; \widehat{\eta}^{\Lambda})-\widehat{\mu}(\cdot, 0, x; \widehat{\eta}^{\Lambda})$.
\end{itemize}
  % Train the IF transformation $\phi(\cdot; \widehat{\eta})$ on $O'$ and predict $\widehat{E}\{\phi(\cdot; \widehat{\eta}_{k}) \mid X=x\}$ as the IF-learner;
\end{step}

\begin{step} \label{step5}

% % CUTs conditional mean difference HTE learners

Second sample-splitting. Randomly split the data $O'_1, \ldots, O'_n$ into $K_{2}$ disjoint validation sets $\mathcal{V}'_{1}, \ldots, \mathcal{V}'_{K_{2}}$ with sizes $n_1, \ldots, n_{K_{2}}$, where $n_k \in\{2,3, \ldots,\lfloor n / 2\rfloor\}$. For each $k = 1, \ldots, K_{2}$, we define training set $\overline{\mathcal{V}}'_{k}=\{O'_i: i \notin \mathcal{V}'_{k}\}$. % and treatment-specific training set $\overline{\mathcal{V}}_{a, k_{2}}=\{O_i: i \in \overline{\mathcal{V}}_{k_{2}}, A_i=a\}$.
\end{step}

\begin{step} \label{step6}
Transformed minimization weight and outcome construction. Construct estimators $\widehat{\pi}_{k}(a \mid X)$ and $\widehat{\mu}_{k}(\cdot, a, X; \widehat{\eta}^{\Lambda})$ of $\pi(a \mid X)$ and $\mu(\cdot, a, X)$ on $\overline{\mathcal{V}}_{k}'$. Construct outcome transformations by plugging $\widehat{\mu}_{k}(\cdot, a, X; \widehat{\eta}^{\Lambda})$ and $\widehat{\pi}_{k}(a \mid X)$ into $Y^{*}(Y(\cdot, A; \widehat{\eta}_{k}^{\Lambda}); (\widehat{\mu}_{k}, \widehat{\pi}_{k}))$. Construct weights by plugging $\widehat{\pi}_{k}(a \mid X)$ into $w^{*}(\widehat{\pi}_{k})$. For every $k = 1, \ldots, K_{2}$, add $Y^{*}(Y(\cdot, A; \widehat{\eta}_{k}^{\Lambda}); (\widehat{\mu}_{k}, \widehat{\pi}_{k}))$ and $w^{*}(\widehat{\pi}_{k})$ to $\mathcal{V}_{k}$ and combine into one augmented dataset $O''$. Here, we take $\widehat{\mu}_{k}(\cdot, X; \widehat{\eta}^{\Lambda})=\widehat{\pi}_{k}(0 \mid X)\widehat{\mu}_{k}(\cdot, 0, X; \widehat{\eta}^{\Lambda})+\widehat{\pi}_{k}(1 \mid X)\widehat{\mu}_{k}(\cdot, 1, X; \widehat{\eta}^{\Lambda})$. (We calculate $\widehat{\mu}_{k}(\cdot, X; \widehat{\eta}^{\Lambda})$ in this way to handle potential limited overlap in observational studies. Alternatively, one can regress $X$ on $Y(\cdot, A; \widehat{\eta}_{k}^{\Lambda})$ or use corresponding quantity of $\widehat{\eta}_{k}^{\Lambda}$ as $\widehat{\mu}_{k}(\cdot, a, X; \widehat{\eta}_{k}^{\Lambda})$ or $\widehat{\mu}_{k}(\cdot, X; \widehat{\eta}_{k}^{\Lambda})$ without regression on the CUTs.)
\end{step}

% S-learner, T-learner, and IF-learner estimation. For every $k = 1, \ldots, K_{2}$, train the IF transformation $\phi(\cdot; \widehat{\eta}_{k})$ on $\overline{\mathcal{V}}_{k}'$ and predict $\widehat{E}\{\phi(\cdot; \widehat{\eta}_{k}) \mid X\}$ on $\mathcal{V}_{k}'$ as the IF-learner; train the CUT $Y(\cdot; \widehat{\eta}_{k}^{\Lambda})$ on $\overline{\mathcal{V}}_{k}'$ and predict $\widehat{\mu}_{k}(\cdot, a, X; \widehat{\eta}^{\Lambda})$ on $\mathcal{V}_{k}'$.

\begin{step} \label{step7}
Transformed minimization. Estimate $E[Y^{*}(Y(\cdot, A; \widehat{\eta}^{\Lambda}))|X]$ via weighted regression with weights $w^{*}(\widehat{\pi})$ using data set $O''$. Obtain the estimates of $\widehat{\psi}^{*}(\cdot, x)$ by plugging in $x$ to the fitted regression, i.e. $\hat{E}[Y^{*}(Y(\cdot, A; \widehat{\eta}^{\Lambda}))|X=x]$.

% Third sample-splitting. For every $k = 1, \ldots, K_{3}$, add the outcome transformation $Y^{*}(Y(\cdot, a; \widehat{\eta}_{k})^{\Lambda}); (\widehat{\mu}_{k}, \widehat{\pi}_{k}))$ and weight $w^{*}(\widehat{\pi}_{k})$ to $\mathcal{V}_{k}$ and combine into one augmented dataset $O''$. Randomly split the data $O_1'', \ldots, O_n''$ into $K_{3}$ disjoint validation sets $\mathcal{V}_{1}'', \ldots, \mathcal{V}_{K_{3}}''$ with sizes $n_1, \ldots, n_{K_{3}}$, where $K_{3} \in\{2,3, \ldots,\lfloor n / 2\rfloor\}$. For each $k = 1, \ldots, K_{3}$, we define training set $\overline{\mathcal{V}}_{k}''=\{O_i: i \notin \mathcal{V}_{k}''\}$.
\end{step}

% Bind $Y^{*}(Y(\cdot; \widehat{\eta}^{\Lambda}); (\widehat{\mu}_{k}, \widehat{\pi}_{k}))$ and $w^{*}(\widehat{\pi}_{k})$ to $\mathcal{V}_{k}$ and combine $\mathcal{V}_{k}$ into one dataset. We 

% \begin{step} \label{step8}
% Cross-fitting transformed minimization. For every $k = 1, \ldots, K_{3}$, train a supervised learning algorithm with $Y^{*}(Y(\cdot, a; \widehat{\eta}_{k}^{\Lambda}); (\widehat{\mu}_{k}, \widehat{\pi}_{k}))$ as outcome, $X$ as covariates, and $w^{*}(\widehat{\pi}_{k})$ as weights using $\overline{\mathcal{V}}_{k}''$. Estimate $\widehat{\psi}_{k}^{*}(\cdot, x)$ and $\widehat{\psi}^{*}(\cdot, x)=$. % Estimate $\widehat{\psi}_{k}^{*}(\cdot, X)$ using $X$ from $\mathcal{V}_{k}''$.
% \end{step}

% For each learner from~\Cref{tab:weighted_minimization} except the IF-learner, train a supervised learning algorithm by taking $Y^{*}(Y(\cdot, a; \widehat{\eta}_{k}^{\Lambda}); (\widehat{\mu}_{k}, \widehat{\pi}_{k}))$ as the outcome, $X$ as covariates, and $w^{*}(\widehat{\pi})$ as sampling weights using $\overline{\mathcal{V}}_{k_{3}}$.

%\begin{remark}\label{remark:split}
%(Multiple sample splitting) To eschew Donsker conditions on learner-specific transformed outcome and weight, we borrow the spirit of double sample splitting from~\citet{newey2018cross,kennedy2020optimal} but modify the method by resplitting every time we train a new model, perhaps on censored outcomes, CUTs, or transformed outcomes.
%\end{remark}

\section{Uncertainty quantification and oracle properties}

% \begin{remark}\label{remark:normality}
% (Asymptotic normality) 
% \end{remark}

Valid confidence intervals for HTEs with time-to-event outcomes are only available in limited circumstances. Regarding confidence intervals, similar to~\citet[Theorem 3]{cui2023estimating}, under regularity conditions in ~\citet[Theorem 4.4]{oprescu2019orthogonal}, the estimated HTE is asymptotically normal for causal forest~\citep{wager2018estimation} applied to any CUT $Y(\cdot, A; \widehat{\eta}^{\Lambda})$ or for any of the transformed minimization HTE learners from ~\Cref{tab:weighted_minimization} applied to transformed CUT $Y^{*}(Y(\cdot, A; \widehat{\eta}^{\Lambda})); (\widehat{\mu}, \widehat{\pi}))$ and weights $w^{*}(\widehat{\pi})$ with generalized random forest~\citep{athey2019generalized} as the regression in ~\Cref{step7} of the algorithm in Section 3.

However, similar results have not been established for other learners except under additional assumptions. For example, one can obtain confidence bands by imposing linearity on the HTE~\citep{yadlowsky2021estimation,yang2023elastic,liang2022semiparametric} or by projecting the HTE on a subset (or transformation) of $X$~\citep{lee2017doubly,semenova2021debiased,cui2023estimating}. % ratkovic2023estimation; Though~\citet{kunzel2019metalearners} suggested the bootstrap to estimate the standard errors of HTE, the asymptotic normality of the X-learner has not been proved. 
~\citet{lei2021conformal} have developed conformal prediction for HTEs on uncensored outcomes. However, their approach cannot be directly applied to our CUT based HTE estimators because CUTs only preserve conditional mean (not conditional quantiles), and their method requires conditional quantiles as inputs. Bayesian credible intervals also leverage conditional quantiles, making them inapplicable to our setting~\citep{hahn2020bayesian}. For the majority of learners that lack valid confidence bands (absent parametric assumptions) or prediction sets, we derive bounds.% Recently,~\citet{cwiling2023comprehensive} proposed conformal prediction intervals for conditional RMST, but their approach has not been extended to counterfactual prediction. 

\subsection{Definitions}

In this work, we adopt ensemble learning with a convex combination of weak individual learners as the conditional distributions of CUTs are unknown and complicated. We assume
{\footnotesize
$$
\widehat{\mu}(\cdot, A, X; \rho(\widehat{\mu}, A))=\sum_{v=1}^{v^{*}} \rho_{v}(\widehat{\mu}, A) \widehat{\mu}_{v}(\cdot, A, X), \quad \rho_{v}(\widehat{\mu}, A) \geq 0, \quad \sum_{v=1}^{v^{*}} \rho_{v}(\widehat{\mu}, A)=1
$$
}
where $v=1, \ldots, v^{*}$, $v^{*}$ is the number of candidate learning algorithms, the contribution of the $v^{th}$ learner minimizes the cross-validated population prediction loss~\eqref{cmd_cv_loss} as
{\footnotesize
$$
\rho^{\star}(\widehat{\mu}, A)=\underset{\rho(\widehat{\mu}, A)}{\operatorname{arg min}} \; L_{2}(\widehat{\mu}; \widehat{\eta}^{\Lambda}, \rho(\widehat{\mu}, A)) \quad s.t. \quad \rho_{v}(\widehat{\mu}, A) \geq 0, \quad \sum_{v=1}^{v^{*}} \rho_{v}(\widehat{\mu}, A)=1
$$
}
and the corresponding cross-validated population prediction loss of $\widehat{\mu}(\cdot, A, X)$ is defined as
{\footnotesize
\begin{equation}\label{cmd_cv_loss}
L_{2}(\widehat{\mu}; \widehat{\eta}^{\Lambda}, \rho(\widehat{\mu}, A))=\frac{1}{n}\sum_{k=1}^{K} \sum_{i \in \mathcal{V}_k} \ell_{2}(Y(\cdot, A; \widehat{\eta}_{k}^{\Lambda}), \widehat{\mu}_{k}(\cdot, A, X; \rho(\widehat{\mu}, A))).
\end{equation}
}
% Denote subscript $\infty$ as the elementwise limiting distribution of nuisance functions as $n \rightarrow \infty$ that makes corresponding transformation consistent.
The oracle learner of a convex combination of weak individual learners is
{\footnotesize
$$
\widetilde{\mu}(\cdot, A, X; \rho(\widetilde{\mu}, A))=\sum_{v=1}^{v^{*}} \rho_{v}(\widetilde{\mu}, A) \widetilde{\mu}(\cdot, A, X; \rho_{v}(\widetilde{\mu}, A)), \quad \rho_{v}(\widetilde{\mu}, A) \geq 0, \quad \sum_{v=1}^{v^{*}} \rho_{v}(\widetilde{\mu}, A)=1
$$
}
and the oracle selector can then be defined as
{\footnotesize
$$
\rho^{\star}(\widetilde{\mu}, A)=\underset{\rho(\widetilde{\mu}, A)}{\operatorname{arg min}} \; L_{2}(\widetilde{\mu}; \eta_{\infty}^{\Lambda}, \rho(\widetilde{\mu}, A)) \quad s.t. \quad \rho_{v}(\widetilde{\mu}, A) \geq 0, \quad \sum_{v=1}^{v^{*}} \rho_{v}(\widetilde{\mu}, A)=1
$$
}
where the oracle loss is defined as
{\footnotesize
\begin{equation}\label{cmd_oracle_loss}
L_{2}(\widetilde{\mu}; \eta_{\infty}^{\Lambda}, \rho(\widetilde{\mu}, A))=\sum_{k=1}^{K} \frac{n_{k}}{n} \int \ell_{2}(Y(\cdot, A; \eta_{\infty}^{\Lambda}), \widetilde{\mu}_{k}(\cdot, A, X; \rho(\widetilde{\mu}, A))) \mathrm{d} P_{0}.
\end{equation}
}
Naturally, we define the optimal loss as
{\footnotesize
$$
L_{2}(\mu_{0}; \eta_{\infty}^{\Lambda})= \int \ell_{2}(Y(\cdot, A; \eta_{\infty}^{\Lambda}), \mu_{0}(\cdot, A, X)) \mathrm{d} P_{0}.
$$
}
In this work, we use the word ``risk'' to describe the asymptotic behavior of learners. The excess risk of the learner is defined as
{\footnotesize
$$
\mathcal{E}^{2}(\widehat{\psi}^{*})=E[P_{0}\{\widehat{\psi}^{*}(\cdot, X)-\psi_{0}(\cdot, X)\}^{2}] % =E\{L_{2}(\widehat{\psi}^{*}; \eta_{\infty}^{\Lambda})-L_{2}(\psi_{0}; \eta_{\infty}^{\Lambda})\},
$$
}
the excess risk of the conditional mean is
{\footnotesize
$$
\mathcal{E}^{2}(\widehat{\mu}; \rho(\widehat{\mu}, A))=E[P_{0}\{\widehat{\mu}(\cdot, A, X; \rho(\widehat{\mu}, A))-\mu_{0}(\cdot, A, X)\}^{2}], % =E\{L_{2}(\widehat{\psi}^{*}; \eta_{\infty}^{\Lambda})-L_{2}(\psi_{0}; \eta_{\infty}^{\Lambda})\}
$$
}
and the oracle risk of the conditional mean is
{\footnotesize
$$
\mathcal{E}^{2}(\widetilde{\mu}; \rho(\widetilde{\mu}, A))=E[P_{0}\{\widetilde{\mu}(\cdot, A, X; \rho(\widetilde{\mu}, A))-\mu_{0}(\cdot, A, X)\}^{2}] % =E\{L_{2}(\widetilde{\psi}^{*}; \eta_{\infty}^{\Lambda})-L_{2}(\psi_{0}; \eta_{\infty}^{\Lambda})\}
$$
}
where the expectation is taken over samples from $O$.

Regarding learners based on transformed optimization, the convex combination of weak individual learners is
{\footnotesize
$$
\widehat{\psi}^{*}(\cdot, X; \rho(\widehat{\psi}^{*}))=\sum_{v=1}^{v^{*}} 
\rho_{v}(\widehat{\psi}^{*}) \widehat{\psi}_{v}^{*}(\cdot, X), \quad \rho_{v}(\widehat{\psi}^{*}) \geq 0, \quad \sum_{v=1}^{v^{*}} \rho_{v}(\widehat{\psi}^{*})=1
$$
}
Note that this is differenet from the S and T-learners because the convex combination is defined directly over the estimator of interest $\widehat{\psi}^{*}(\cdot, X)$ instead of over the treatment-specific conditional outcome. The contribution of the $v^{th}$ learner on the last-step cross-validated population prediction loss is
{\footnotesize
$$
\rho^{\star}(\widehat{\psi}^{*})=\underset{\rho(\widehat{\psi}^{*})}{\operatorname{arg min}} \; L_{2}^{*}(\widehat{\psi}^{*}; \widehat{\eta}^{\Lambda}, (\widehat{\mu}, \widehat{\pi})) \quad s.t. \quad \rho_{v}(\widehat{\psi}^{*}) \geq 0, \quad \sum_{v=1}^{v^{*}} \rho_{v}(\widehat{\psi}^{*})=1,
$$
}
and the corresponding cross-validated population prediction loss of $\widehat{\psi}^{*}(\cdot, X)$ is defined as
{\footnotesize
\begin{equation}\label{trans_cv_loss}
L_{2}^{*}(\widehat{\psi}^{*}; Y(\cdot, A; \widehat{\eta}^{\Lambda}), (\widehat{\mu}, \widehat{\pi}), \rho(\widehat{\psi}^{*}))=\frac{1}{n}\sum_{k=1}^{K} \sum_{i \in \mathcal{V}_k} \ell_{2}^{*}(Y(\cdot, A; \widehat{\eta}_{k}^{\Lambda}), \widehat{\psi}_{k}^{*}(\cdot, X; \rho(\widehat{\psi}^{*})); (\widehat{\mu}_{k}, \widehat{\pi}_{k})).
\end{equation}
}
Transformed optimization procedures involve constructing transformed outcomes (and weights) twice--once to construct the CUT $Y(\cdot, A; \widehat{\eta}^{\Lambda}))$, and again to construct the transformed CUT $Y^{*}(Y(\cdot, A; \widehat{\eta}^{\Lambda})); (\widehat{\mu}, \widehat{\pi}))$. We use the term ``partial oracle'' to describe procedures based on $Y^{*}(Y(\cdot, A; \widehat{\eta}^{\Lambda})); (\mu_{\infty}, \pi_{\infty}))$ where the second stage transformation incorporates limiting nuisance functions $\mu_{\infty}$ and $\pi_{\infty}$. Assume a library of hypothetical partial-oracle learners $\widetilde{\psi}_{v}^{*}(\cdot, X; \widehat{\eta}^{\Lambda})$ is trained using $\widehat{\eta}^{\Lambda}$ and $(\mu_{\infty}, \pi_{\infty})$. Then the corresponding ensemble partial-oracle learner can be defined as
{\footnotesize
$$
\widetilde{\psi}^{*}(\cdot, X; \rho(\widetilde{\psi}^{*}; \widehat{\eta}^{\Lambda}))=\sum_{v=1}^{v^{*}} \rho_{v}(\widetilde{\psi}^{*}; \widehat{\eta}^{\Lambda}) \widetilde{\psi}_{v}^{*}(\cdot, X; \widehat{\eta}^{\Lambda}), \quad \rho_{v}(\widetilde{\psi}^{*}; \widehat{\eta}^{\Lambda}) \geq 0, \quad \sum_{v=1}^{v^{*}} \rho_{v}(\widetilde{\psi}^{*}; \widehat{\eta}^{\Lambda})=1
$$
}
with partial-oracle selector
{\footnotesize
$$
\rho^{\star}(\widetilde{\psi}^{*}; \widehat{\eta}^{\Lambda})=\underset{\rho(\widetilde{\psi}^{*}; \widehat{\eta}^{\Lambda})}{\operatorname{arg min}} \; L_{2}^{*}(\widetilde{\psi}^{*}; \widehat{\eta}^{\Lambda}, (\mu_{\infty}, \pi_{\infty}), \rho(\widetilde{\psi}^{*}; \widehat{\eta}^{\Lambda})) \quad s.t. \quad \rho_{v}(\widetilde{\psi}^{*}; \widehat{\eta}^{\Lambda}) \geq 0, \quad \sum_{v=1}^{v^{*}} \rho_{v}(\widetilde{\psi}^{*}; \widehat{\eta}^{\Lambda})=1
$$
}
minimizing a hypothetical population loss function
{\footnotesize
$$
L_{2}^{*}(\widetilde{\psi}^{*}; \widehat{\eta}^{\Lambda}, (\mu_{\infty}, \pi_{\infty}), \rho(\widetilde{\psi}^{*}; \widehat{\eta}^{\Lambda}))=\sum_{k=1}^{K} \frac{n_{k}}{n} \int \ell_{2}^{*}(Y^{\mathrm{AIPCW}}(\cdot, A; \widehat{\eta}_{k}^{\Lambda}), \widetilde{\psi}_{k}(\cdot, X; \rho(\widetilde{\psi}^{*}; \widehat{\eta}^{\Lambda})); (\mu_{\infty}, \pi_{\infty})) \mathrm{d} P_{0}.
$$
}
The partial-oracle loss function replaces $\widehat{\eta}^{\Lambda}$ in the loss function above by its probability limit in the CUT, i.e.
{\footnotesize
$$
L_{2}^{*}(\widetilde{\psi}^{*}(\widehat{\eta}^{\Lambda}); \eta_{\infty}^{\Lambda}, (\mu_{\infty}, \pi_{\infty}), \rho(\widetilde{\psi}^{*}; \widehat{\eta}^{\Lambda}))=\sum_{k=1}^{K} \frac{n_{k}}{n} \int \ell_{2}^{*}(Y^{\mathrm{AIPCW}}(\cdot, A; \eta_{\infty}^{\Lambda}), \widetilde{\psi}_{k}(\cdot, X; \rho(\widetilde{\psi}^{*}; \widehat{\eta}^{\Lambda})); (\mu_{\infty}, \pi_{\infty})) \mathrm{d} P_{0}
$$
}
The oracle learner can be constructed as
{\footnotesize
$$
\widetilde{\psi}^{*}(\cdot, X; \rho(\widetilde{\psi}^{*}))=\sum_{v=1}^{v^{*}} \rho_{v}(\widetilde{\psi}^{*}) \widetilde{\psi}_{v}^{*}(\cdot, X), \quad \rho_{v}(\widetilde{\psi}^{*}) \geq 0, \quad \sum_{v=1}^{v^{*}} \rho_{v}(\widetilde{\psi}^{*})=1
$$
}
with oracle selector
{\footnotesize
$$
\rho^{\star}(\widetilde{\psi}^{*})=\underset{\rho(\widetilde{\psi}^{*})}{\operatorname{arg min}} \; L_{2}^{*}(\widetilde{\psi}^{*}; \eta_{\infty}^{\Lambda}, (\mu_{\infty}, \pi_{\infty}), \rho(\widetilde{\psi}^{*})) \quad s.t. \quad \rho_{v}(\widetilde{\psi}^{*}) \geq 0, \quad \sum_{v=1}^{v^{*}} \rho_{v}(\widetilde{\psi}^{*})=1
$$
}
where the oracle loss is
{\footnotesize
$$
L_{2}^{*}(\widetilde{\psi}^{*}; \eta_{\infty}^{\Lambda}, (\mu_{\infty}, \pi_{\infty}), \rho(\widetilde{\psi}^{*}))=\sum_{k=1}^{K} \frac{n_{k}}{n} \int \ell_{2}^{*}(Y^{\mathrm{AIPCW}}(\cdot, A; \eta_{\infty}^{\Lambda}), \widetilde{\psi}_{k}^{*}(\cdot, X; \rho(\widetilde{\psi}^{*})); (\mu_{\infty}, \pi_{\infty})) \mathrm{d} P_{0}
$$
}
and the optimal loss is defined as
{\footnotesize
$$
L_{2}^{*}(\psi_{0}(\cdot, X); \eta_{\infty}^{\Lambda}, (\mu_{\infty}, \pi_{\infty}))= \int \ell_{2}^{*}(Y^{\mathrm{AIPCW}}(\cdot, A; \eta_{\infty}^{\Lambda}), \psi_{0}(\cdot, X); (\mu_{\infty}, \pi_{\infty})) \mathrm{d} P_{0}
$$
} % $(\mu_{\infty}, \pi_{\infty})=(\mu_{0}, \pi_{0})$, except for the AIPTW learner, where $(\mu_{\infty}, \pi_{\infty})$ represents either $\mu_{\infty}=\mu_{0}$ or $\pi_{\infty}=\pi_{0}$. Also suppose $Y^{\mathrm{AIPCW}}(\cdot; \eta_{\infty}^{\Lambda})=Y^{\mathrm{AIPCW}}(\cdot; \eta_{0}^{\Lambda})$ but $\eta_{\infty}^{\Lambda}$ does not necessarily equal to $\eta_{0}^{\Lambda}$.
Finally, we define the excess risk as % of the selector $\rho(\widehat{\psi}^{*})$ as
{\footnotesize
$$
\mathcal{E}^{2}(\widehat{\psi}^{*}; \rho(\widehat{\psi}^{*}))=E[P_{0}\{\widehat{\psi}^{*}(\cdot, X; \rho(\widehat{\psi}^{*}))-\psi_{0}(\cdot, X)\}^{2}],
$$
}
% the partial-oracle risk as
% {\footnotesize
% $$
% \mathcal{E}^{2}(\widetilde{\psi}^{*}(\widehat{\eta}^{\Lambda}); \rho(\widetilde{\psi}^{*}; \widehat{\eta}^{\Lambda}))=E[P_{0}\{\widetilde{\psi}^{*}(\cdot, X; \rho(\widetilde{\psi}^{*}; \widehat{\eta}^{\Lambda}))-\psi_{0}(\cdot, X)\}^{2}]
% $$
% }
and the oracle risk as
{\footnotesize
$$
\mathcal{E}^{2}(\widetilde{\psi}^{*}; \rho(\widetilde{\psi}^{*}))=E[P_{0}\{\widetilde{\psi}^{*}(\cdot, X; \rho(\widetilde{\psi}^{*}))-\psi_{0}(\cdot, X)\}^{2}]
$$
}
Asymptotic optimality is concerned with performance of the estimator with respect to the oracle loss, and oracle loss is the estimated loss on an infinite validation set when the outcome (and weight) is constructed consistently.

%\begin{remark}\label{remark:oracle}

%(Definition of oracle) In this work, ``oracle'' refers to the cross-validated selector among a group of candidate learners and its asymptotic optimality~\citep{luedtke2016super,diaz2018targeted,van2003unified}, whereas defining oracle as an estimator given access to the true or consistent learner-specific transformed outcome and weight~\citep[Remark 3]{kennedy2023towards,curth2020estimating,curth2021nonparametric,nie2021quasi,morzywolek2023general,foster2023orthogonal}, or the counterfactuals~\citep{kennedy2020optimal,kennedy2022minimax} can be considered as a special case. The asymptotic optimality refers to the cross-validation selector's performance is as good as the selector given access to the true data generating process~\citep{laan2006cross,dudoit2005asymptotics,mitchell2009general,lecue2012oracle,vaart2006oracle}.

% \end{remark}

% (question: shall I use censoring unbiased oracle loss, etc? when I disuss about def, lemma, and theorem that can be applied to continuous outcomes, shall I remove $\cdot$?)

% \begin{definition} \label{definition:oracle}

% \end{definition}

\subsection{Results}

We generalize Theorem 1 from~\citet{diaz2018targeted} to loss functions with non-unitary weights.

% (question: shall I use $(\mu, \pi)$ instead of $\eta$, $\pi$ in $w^{*}$? in lemma 1? $\eta$ is short and more general but can be confusing as $\eta$ has been defined as the full set of nuisance in the survival setting)

\begin{lemma} \label{lemma1}
% Assume the outcome is continuous and $\eta$ as the non-censored nuisance functions in this Lemma.
(Oracle inequality of the weighted minimization with selector) When $w^{*} \neq 1$, define
{\footnotesize
$$
\begin{aligned}
r_{1}^{*}((\widehat{\mu}, \widehat{\pi}), (\mu_{0}, \pi_{0}))=&\max_{k} \{E(P_{0}[\{w^{*}(\widehat{\mu}_{k}, \widehat{\pi}_{k})Y^{*}(\widehat{\mu}_{k}, \widehat{\pi}_{k})-w^{*}(\mu_{0}, \pi_{0})Y^{*}(\mu_{0}, \pi_{0})\}^{2}])\}^{1/2} \\
r_{2}^{*}((\widehat{\mu}, \widehat{\pi}), (\mu_{0}, \pi_{0}))=&\max_{k} \{E(P_{0}[\{w^{*}(\widehat{\mu}_{k}, \widehat{\pi}_{k})-w^{*}(\mu_{0}, \pi_{0})\}^{2}])\}^{1/2} \\
\mathcal{B}_{1}^{*}((\widehat{\mu}, \widehat{\pi}), (\mu_{0}, \pi_{0}))=&\max_{k} E(P_{0}[\{w^{*}(\widehat{\mu}_{k}, \widehat{\pi}_{k})Y^{*}(\widehat{\mu}_{k}, \widehat{\pi}_{k})-w^{*}(\mu_{\infty}, \pi_{\infty})Y^{*}(\mu_{\infty}, \pi_{\infty})\}^{2}]) \\
\mathcal{B}_{2}^{*}((\widehat{\mu}, \widehat{\pi}), (\mu_{0}, \pi_{0}))=&\max_{k} E(P_{0}[\{w^{*}(\widehat{\mu}_{k}, \widehat{\pi}_{k})Y^{*}(\widehat{\mu}_{k}, \widehat{\pi}_{k})-w^{*}(\mu_{\infty}, \pi_{\infty})Y^{*}(\mu_{\infty}, \pi_{\infty})\}\{w^{*}(\widehat{\mu}_{k}, \widehat{\pi}_{k})-w^{*}(\mu_{\infty}, \pi_{\infty})\}]) \\
\mathcal{B}_{3}^{*}((\widehat{\mu}, \widehat{\pi}), (\mu_{0}, \pi_{0}))=&\max_{k} E(P_{0}[\{w^{*}(\widehat{\mu}_{k}, \widehat{\pi}_{k})-w^{*}(\mu_{\infty}, \pi_{\infty})\}^{2}])
\end{aligned}
$$
}
For $\delta>0$,
{\footnotesize
$$
\begin{aligned}
&[E\{L_{2}^{*}(\widehat{\psi}^{*}; \mu_{\infty}, \pi_{\infty}, \rho^{\star}(\widehat{\psi}^{*}))-L_{2}^{*}(\psi_{0}; \mu_{\infty}, \pi_{\infty})\}]^{1/2} \\
\leq & \{(1+2 \delta) [E\{L_{2}^{*}(\widetilde{\psi}^{*}; \mu_{\infty}, \pi_{\infty}, \rho^{\star}(\widetilde{\psi}^{*}))-L_{2}^{*}(\psi_{0}; \mu_{\infty}, \pi_{\infty})\}]^{1/2}+\widetilde{c}_{1}^{*}(\delta)\{(1+\log n) / n\}^{1 / 2}+\{c_{r1}^{*}(\delta)r_{1}^{*}((\widehat{\mu}, \widehat{\pi}), (\mu_{0}, \pi_{0})) \\
&+c_{r2}^{*}(\delta)r_{2}^{*}((\widehat{\mu}, \widehat{\pi}), (\mu_{0}, \pi_{0}))\}+(\log n/n)^{1/4} \{c_{\mathcal{B}1}^{*, 2}(\delta) \mathcal{B}_{1}^{*}((\widehat{\mu}, \widehat{\pi}), (\mu_{0}, \pi_{0}))+c_{\mathcal{B}2}^{*, 2}(\delta)\mathcal{B}_{2}^{*}((\widehat{\mu}, \widehat{\pi}), (\mu_{0}, \pi_{0})) \\
&x+c_{\mathcal{B}3}^{*, 2}(\delta)\mathcal{B}_{3}^{*}((\widehat{\mu}, \widehat{\pi}), (\mu_{0}, \pi_{0}))\}^{1/2}
\end{aligned}
$$
}
where $\widetilde{c}_{1}^{*}(\delta), c_{r1}^{*}(\delta)$, $c_{r2}^{*}(\delta)$, $c_{\mathcal{B}1}^{*}(\delta)$, $c_{\mathcal{B}2}^{*}(\delta)$, and $c_{\mathcal{B}3}^{*}(\delta)$ are universal constants of $\delta$.
\end{lemma}

\begin{theorem} \label{theorem:oracle}

(Oracle inequality for censoring unbiased R-learner) The excess risk for a selector of a censoring unbiased R-learner is bounded relative to the excess risk for an oracle selector of a censoring unbiased R-learner:
{\footnotesize
$$
\begin{aligned}
\begin{split}
\mathcal{E}^{2}(\widehat{\psi}^{\mathrm{RL}}; \rho^{\star}(\widehat{\psi}^{\mathrm{RL}})) \leq \epsilon^{2} E\{L_{2}^{\mathrm{RL}}(\widehat{\psi}^{\mathrm{RL}}; \eta_{\infty}^{\Lambda}, (\mu_{\infty}, \pi_{\infty}), \rho^{\star}(\widehat{\psi}^{\mathrm{RL}}))-L_{2}^{\mathrm{RL}}(\psi_{0}; \eta_{\infty}^{\Lambda}, (\mu_{\infty}, \pi_{\infty}))\},\\
&\text{}
\end{split}\\
\begin{split}
&[E\{L_{2}^{\mathrm{RL}}(\widehat{\psi}^{\mathrm{RL}}; \eta_{\infty}^{\Lambda}, (\mu_{\infty}, \pi_{\infty}), \rho^{\star}(\widehat{\psi}^{\mathrm{RL}}))-L_{2}^{\mathrm{RL}}(\psi_{0}; \eta_{\infty}^{\Lambda}, (\mu_{\infty}, \pi_{\infty}))\}]^{1/2} \\
\leq& \{(1+2 \delta) [E\{L_{2}^{\mathrm{RL}}(\widetilde{\psi}^{\mathrm{RL}}(\widehat{\eta}^{\Lambda}); \eta_{\infty}^{\Lambda}, (\mu_{\infty}, \pi_{\infty}), \rho^{\star}(\widetilde{\psi}^{\mathrm{RL}}; \widehat{\eta}^{\Lambda}))-L_{2}^{\mathrm{RL}}(\psi_{0}; \eta_{\infty}^{\Lambda}, (\mu_{\infty}, \pi_{\infty}))\}]^{1/2} \\
&+\widetilde{c}_{1}^{\mathrm{RL}}(\delta)\{(1+\log n) / n\}^{1 / 2}+\{c_{r1}^{\mathrm{RL}}(\delta) r_{1}^{\mathrm{RL}}((\widehat{\mu}, \widehat{\pi}), (\mu_{0}, \pi_{0}))+c_{r2}^{\mathrm{RL}}(\delta)r_{2}^{\mathrm{RL}}((\widehat{\mu}, \widehat{\pi}), (\mu_{0}, \pi_{0}))\} \\
&+(\log n/n)^{1/4} \{c_{\mathcal{B}1}^{\mathrm{RL}, 2}(\delta) \mathcal{B}_{1}^{\mathrm{RL}}((\widehat{\mu}, \widehat{\pi}), (\mu_{0}, \pi_{0}))+c_{\mathcal{B}2}^{\mathrm{RL}, 2}(\delta)\mathcal{B}_{2}^{\mathrm{RL}}((\widehat{\mu}, \widehat{\pi}), (\mu_{0}, \pi_{0}))+c_{\mathcal{B}3}^{\mathrm{RL}, 2}(\delta)\mathcal{B}_{3}^{\mathrm{RL}}((\widehat{\mu}, \widehat{\pi}), (\mu_{0}, \pi_{0}))\}^{1/2},\\
&\text{}
\end{split}\\
\begin{split}
&[E\{L_{2}^{\mathrm{RL}}(\widetilde{\psi}^{\mathrm{RL}}(\widehat{\eta}^{\Lambda}); \eta_{\infty}^{\Lambda}, (\mu_{\infty}, \pi_{\infty}), \rho^{\star}(\widetilde{\psi}^{\mathrm{RL}}; \widehat{\eta}^{\Lambda}))-L_{2}^{\mathrm{RL}}(\psi_{0}; \eta_{\infty}^{\Lambda}, (\mu_{\infty}, \pi_{\infty}))\}]^{1/2}\\
&\leq (1+2 \delta)^{1 / 2} [E\{L_{2}^{\mathrm{RL}}(\widetilde{\psi}^{\mathrm{RL}}; \eta_{\infty}^{\Lambda}, (\mu_{\infty}, \pi_{\infty}), \rho^{\star}(\widetilde{\psi}^{\mathrm{RL}}))-L_{2}^{\mathrm{RL}}(\psi_{0}; \eta_{\infty}^{\Lambda}, (\mu_{\infty}, \pi_{\infty}))\}]^{1/2} +c_{1}^{\mathrm{RL}}(\delta)\{(1+\log n)/n\}^{1 / 2} \\
&+c_{2}^{\mathrm{RL}}(\delta) r(\widehat{\eta}^{\Lambda}, \eta_{0}^{\Lambda}) +c_{3}^{\mathrm{RL}}(\delta)(\log n / n)^{1 / 4}\mathcal{B}^{\mathrm{RMST}}(\widehat{\eta}^{\Lambda}, \eta_{0}^{\Lambda}),\\
&\text{and}
\end{split}\\
\begin{split}
&E\{L_{2}^{\mathrm{RL}}(\widetilde{\psi}^{\mathrm{RL}}; \eta_{\infty}^{\Lambda}, (\mu_{\infty}, \pi_{\infty}), \rho^{\star}(\widetilde{\psi}^{\mathrm{RL}}))-L_{2}^{\mathrm{RL}}(\psi_{0}; \eta_{\infty}^{\Lambda}, (\mu_{\infty}, \pi_{\infty}))\}\leq \left(1-\frac{1}{\epsilon}\right)^{2}\mathcal{E}^{2}(\widetilde{\psi}^{\mathrm{RL}}; \rho^{\star}(\widetilde{\psi}^{\mathrm{RL}}))
\end{split}
\end{aligned}
$$
}
\flushbottom
where
{\footnotesize
$$
\begin{aligned}
r_{1}^{\mathrm{RL}}((\widehat{\mu}, \widehat{\pi}), (\mu_{0}, \pi_{0}))=&\max_{k}\{ E(P_{0}[\{A-\widehat{\pi}_{k}(1 \mid X; \rho^{\star}(\widehat{\pi}))\}\{Y^{\mathrm{AIPCW}}(\cdot, A; \eta_{\infty}^{\Lambda})-\widehat{\mu}_{k}(\cdot, X; \rho^{\star}(\widehat{\mu}))\}-\{A-\pi_{0}(1 \mid X)\} \\
&\times \{Y^{\mathrm{AIPCW}}(\cdot, A; \eta_{\infty}^{\Lambda})-\mu_{0}(\cdot, X)\}]^2)\}^{1/2} \\
=&\max_{k} \{E(P_{0}[\{\widehat{\mu}_{k}(\cdot, X; \rho^{\star}(\widehat{\mu}))-\mu_{0}(\cdot, X)\}\{\widehat{\pi}_{k}(1 \mid X; \rho^{\star}(\widehat{\pi}))-\pi_{0}(1 \mid X)\}]^2)\}^{1/2}
r_{2}^{\mathrm{RL}}((\widehat{\mu}, \widehat{\pi}), (\mu_{0}, \pi_{0}))=&\max_{k} \{E(P_{0}[\{A-\widehat{\pi}_{k}(1 \mid X; \rho^{\star}(\widehat{\pi}))\}^{2}-\{A-\pi_{0}(1 \mid X)\}^{2}]^{2})\}^{1/2} \\
=&\max_{k} \{E(P_{0}[\{\widehat{\pi}_{k}(1 \mid X; \rho^{\star}(\widehat{\pi}))-\pi_{0}(1 \mid X)\}\{\widehat{\pi}_{k}(1 \mid X; \rho^{\star}(\widehat{\pi}))+\pi_{0}(1 \mid X)\} \\
&-2A\{\widehat{\pi}_{k}(1 \mid X; \rho^{\star}(\widehat{\pi}))-\pi_{0}(1 \mid X)\}]^2)\}^{1/2} \\
\leq& 4 \max_{k} (E[P_{0}\{\widehat{\pi}_{k}(1 \mid X; \rho^{\star}(\widehat{\pi}))-\pi_{0}(1 \mid X)\}^2])^{1/2} \\
\mathcal{B}_{1}^{\mathrm{RL}}((\widehat{\mu}, \widehat{\pi}), (\mu_{0}, \pi_{0}))=&\max_{k} E(P_{0}[\{\widehat{\mu}_{k}(\cdot, X; \rho^{\star}(\widehat{\mu}))-\mu_{\infty}(\cdot, X)\}\{\widehat{\pi}_{k}(1 \mid X; \rho^{\star}(\widehat{\pi}))-\pi_{\infty}(1 \mid X)\}]^2) \\
\mathcal{B}_{2}^{\mathrm{RL}}((\widehat{\mu}, \widehat{\pi}), (\mu_{0}, \pi_{0}))=&\max_{k} E(P_{0}[\{\widehat{\mu}_{k}(\cdot, X; \rho^{\star}(\widehat{\mu}))-\mu_{\infty}(\cdot, X)\}\{\widehat{\pi}_{k}(1 \mid X; \rho^{\star}(\widehat{\pi}))-\pi_{\infty}(1 \mid X)\}^{2} \\
&\times\{\widehat{\pi}_{k}(1 \mid X; \rho^{\star}(\widehat{\pi}))+\pi_{\infty}(1 \mid X)-2A\}] \\
\leq& 2\max_{k} E(P_{0}[|\widehat{\mu}_{k}(\cdot, X; \rho^{\star}(\widehat{\mu}))-\mu_{\infty}(\cdot, X)|\{\widehat{\pi}_{k}(1 \mid X; \rho^{\star}(\widehat{\pi}))-\pi_{\infty}(1 \mid X)\}^{2}]
\end{aligned}
$$
}   
{\footnotesize
$$
\begin{aligned}
\mathcal{B}_{3}^{\mathrm{RL}}((\widehat{\mu}, \widehat{\pi}), (\mu_{0}, \pi_{0}))=&16 \max_{k} E[P_{0}\{\widehat{\pi}_{k}(1 \mid X; \rho^{\star}(\widehat{\pi}))-\pi_{\infty}(1 \mid X)\}^2] \\
r(A; \widehat{\eta}_{k}^{\Lambda}, \eta_{0}^{\Lambda})=&\max_k E_{0}\left| \int_0^{\cdot} \left\{\frac{G_{0}(t- \mid A, X)}{\widehat{G}_{k}(t- \mid A, X)}-1\right\}\mathrm{d}\left\{\frac{S_{0}(t\mid A, X)}{\widehat{S}_{k}(t \mid A, X)}-1\right\}\right| \\
\mathcal{B}^{\mathrm{RMST}}(A; \widehat{\eta}_{k}^{\Lambda}, \eta_{0}^{\Lambda})=&\max_{k} E_{0} \left[I\{S_{\infty}(t \mid A, X)=S_{0}(t \mid A, X)\} \sup_{t \in [0, \cdot]} \left| \frac{\mathrm{R}\widehat{\mathrm{MS}}\mathrm{T}_{k}(\cdot \mid A, X)}{\widehat{S}_{k}(t \mid A, X)}-\frac{\mathrm{RMST}_{0}(\cdot \mid A, X)}{S_{0}(t \mid A, X)} \right| \right. \\
&\left.+I\{G_{\infty}(t \mid A, X)=G_{0}(t \mid A, X)\} \sup_{t \in [0, \cdot]} \left| \frac{1}{\widehat{G}_{k}(t- \mid A, X)}-\frac{1}{G_{0}(t- \mid A, X)} \right| \right]^2
\end{aligned}
$$
}
for universal constants $\widetilde{c}_{1}^{\mathrm{RL}}(\delta)$, $c_{r1}^{\mathrm{RL}}(\delta)$, $c_{r2}^{\mathrm{RL}}(\delta)$, $c_{\mathcal{B}1}^{\mathrm{RL}}(\delta)$, $c_{\mathcal{B}2}^{\mathrm{RL}}(\delta)$, $c_{\mathcal{B}3}^{\mathrm{RL}}(\delta)$, $c_{1}^{\mathrm{RL}}(\delta)$, $c_{2}^{\mathrm{RL}}(\delta)$, and $c_{3}^{\mathrm{RL}}(\delta)$ depending on $\delta>0$, and for any $\epsilon\in(1,\infty)$ satisfying $P_{0}(A=a \mid X) >\frac{1}{\epsilon}$ and $P_{0}(C \geq t \mid A=a, X)>\frac{1}{\epsilon}$ almost surely for all $a$ and $t$.

\end{theorem}

The bounds of the rest of the censoring unbiased orthogonal learners and their proofs are provided in the Supplementary Materials. Of particular interest, the bounds of the MC-learner and the MCEA-learner have not been previously derived even in a simple setting when the outcome is continuous.

% discuss the difference between censoring unbiased minimax rates and orthogonal loss minimax rates, hard to guess, especially when $w^{*} \neq=1$

% The minimax rates of survSuperLearner \citep{westling2023inference}. \\
% Theorem 4: minimax rates of optimal individualized treatment rules \\

\section{Simulation Studies}

We performed simulations to explore the finite sample behavior of CUT-based HTE learners. We considered settings with and without competing risks, estimating appropriate effect measures for each setting. We also explored settings with and without poor overlap, as poor overlap is a well known challenge for HTE learners. In the competing risks settings, we generated ground truth counterfactual outcomes corresponding to separable components of treatment underlying the definitions of separable direct and indirect effects. The data generating processes for the four simulation settings we considered are described in the Supplementary Materials. For good overlap settings, distributions of heterogeneous effect estimates were similar across all learners, and also similar to the ground truth. For poor overlap settings, there was considerable variability across learners with learners that incorporated inverse probability weights estimating a much wider distribution of effects, presumably due to instability of the weights. (See Figure ~\Cref{fig:sim_true}  and Supplemental Figure ~\Cref{fig:sim_hte}.) We also compared performance of the HTE learners as evaluated by the metric $\mathrm{PEHE}=P_{n} \{\psi_{0}(X)-\widehat{\psi}(X)\}^2$. Again, we saw similar performance across learners in the good overlap settings and considerable variation in the poor overlap settings with inverse probability weighted learners performing worst. See supplemental ~\Cref{fig:sim_pehe,fig:sim_grd}. Additional metrics and performance summaries are defined and presented in Supplemental ~\Cref{app:sim}. Also see~\Cref{app:sim} for details on the HTE learners used.

\begin{figure}[H]
\centering
\includegraphics[scale=0.55]{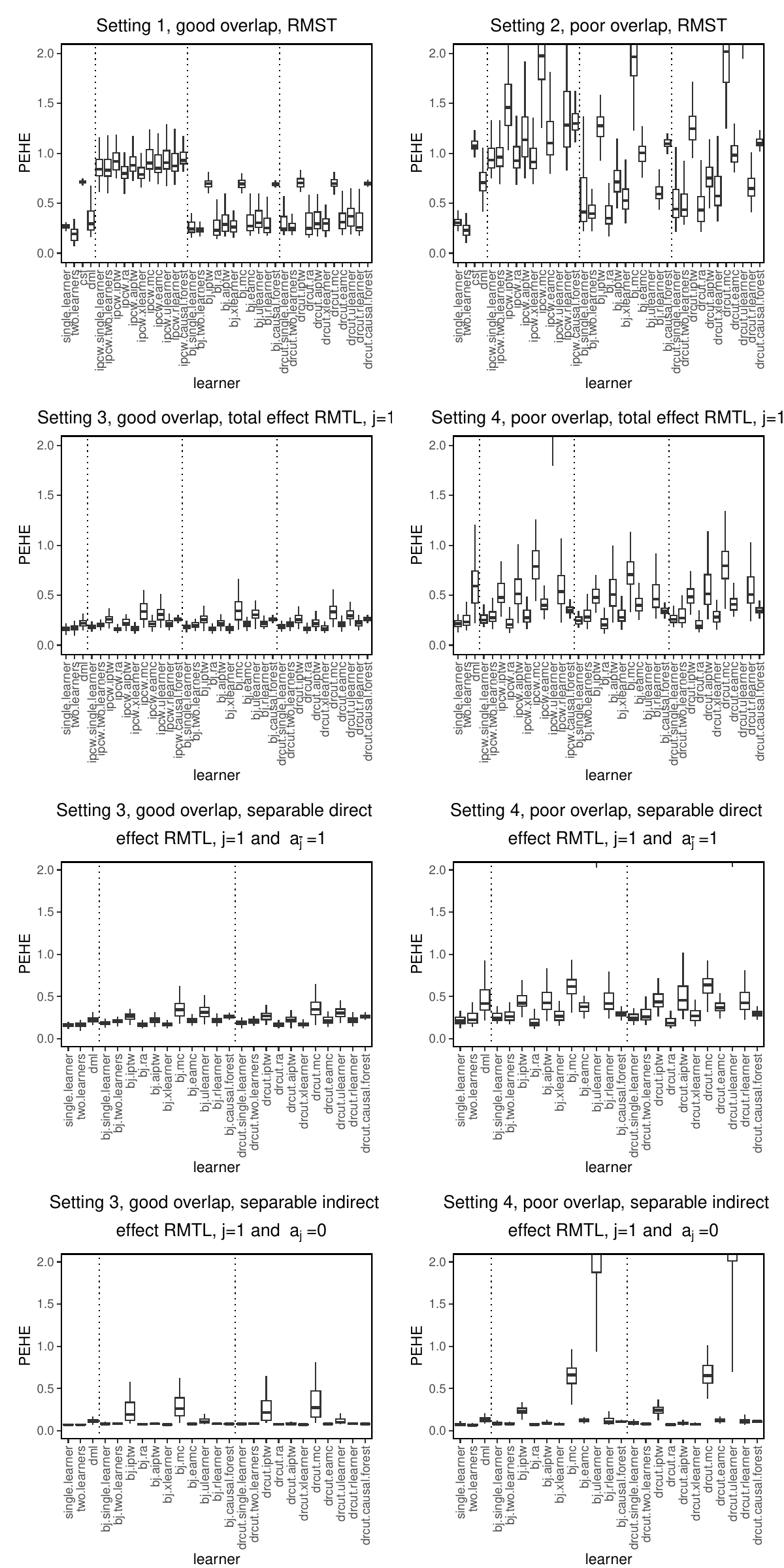}
\caption{PEHE of estimated HTE.}
\label{fig:sim_pehe}
\end{figure}

\clearpage

\clearpage

\section{Application}

Allogeneic hematopoietic cell transplantation (HCT) is a therapy for acute myeloid leukemia and (AML) and myelodysplastic syndromes (MDS). Before transplantation, the host receives a conditioning regimen comprising a mix of radiation or chemotherapy to eliminate their hematopoietic stem cells before they are replaced by HCT. Conventional high dose myeloablative conditioning (MAC) regimens can be toxic, so reduced intensity conditioning (RIC) regimens are sometimes used as an alternative. Compared to MAC, RIC might lead HCT to be less effective at preventing relapse but also decrease the likelihood of non-relapse mortality (NRM), which is a competing risk for relapse. From a decision making perspective, we are interested in effects of RIC on a composite outcome of relapse or mortality. From a scientific perspective, we are further interested in understanding the trade-off between direct harmful effects of RIC on relapse (from its  inferior elimination of host stem cells) and its beneficial effects on NRM (due to reduced toxicity elsewhere in the body). We could conceive of a modified version of RIC targeting stem cells like MAC but acting like RIC in the rest of the body, making the dismissible component assumptions reasonable. It therefore makes sense to consider separable direct and indirect effects of RIC on relapse. Applying CUT-based HTE learners to data from the Center for International Blood and Marrow Transplant
Research (CIBMTR) registry previously analyzed by ~\citep{bejanyan2021myeloablative}, we study the heterogeneous effects of RIC on the composite relapse/mortality outcome and the heterogeneous separable direct and indirect effects of RIC on relapse. 

\begin{figure}[H]
\centering
\includegraphics[scale=0.6]{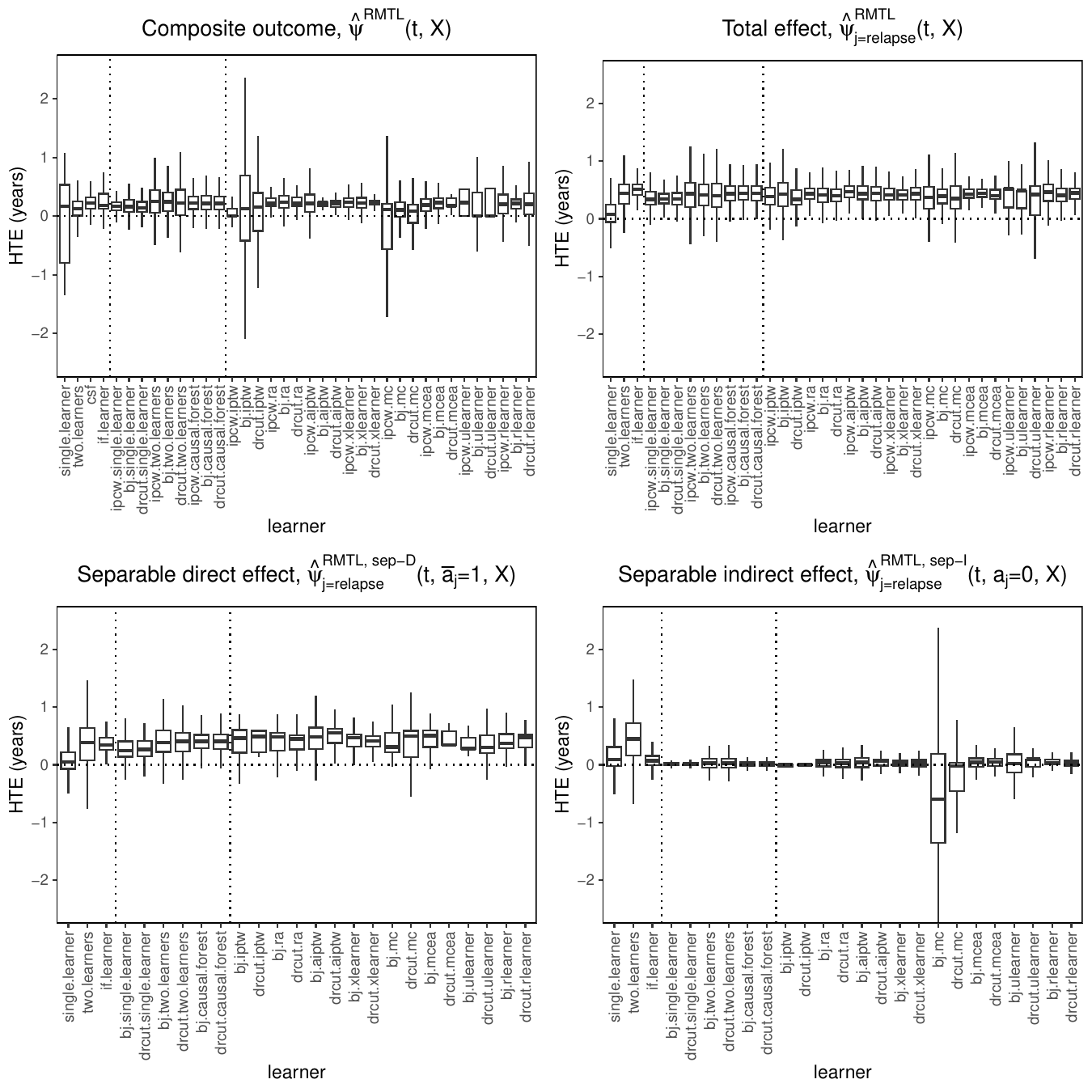}
\caption{Estimated HTE at $\tau=4$ years.}
\label{fig:app_box}
\end{figure}

We adjusted for and estimated effect heterogeneity as a function of the following covariates: donor/recipient sex match; treatment for AML or MDS; disease status prior to HCT; Karnofsky score; white blood count at diagnosis; (ATG/Alemtuzumab) for conditioning regimen or graft-versus-host disease (GVHD) prophylaxis; GVHD prophylaxis; donor/recipient CMV serostatus; donor type; age at HCT; graft source; and risk level - low/intermediate risk or high/very high risk. The top left panel of Figure \ref{fig:app_box} shows most learners estimate a moderately harmful conditional effect of RIC on the composite outcome across most of the covariate distribution. The bottom left panel shows most learners estimate the separable direct effect of RIC on relapse is harmful across most of the covariate distribution. The bottom right panel shows most learners estimate small separable indirect effects of RIC on relapse via pre-relapse mortality across most of the covariate distribution.

%\begin{figure}[H]
%\centering
%\includegraphics[scale=0.8]{figure/application/application_hist.pdf}
%\caption{Density of estimated propensity score; light gray: peripheral blood; dark gray: bone marrow: $n$ (MAC)$=2,570$ and $n$ (RIC)$=1658$.}
%\label{fig:application_ps}
%\end{figure}

\section{Discussion}

%% extension

To summarize, we have provided new tools and error bounds for a general CUT-based approach to estimating heterogeneous treatment effects in survival settings, both with and without competing events. For a large set of effect measures, counterfactual
CUTs (from Theorem 1) can be plugged into the HTE learners summarized in Table 1 according to the procedure described in Section 3.3. The resulting HTE estimates will be consistent, and the error bounds are as characterized in Section 4. This general approach opens up many additional options for estimating HTEs on time-to-event outcomes, particularly in competing risks settings.  

There are several dimensions along which this work might be extended or improved. One current shortcoming is that shape constraints are ignored. Some shape constraints include
{\footnotesize
$$
\begin{aligned}
&\psi_{0}^{S}(t, X)+\psi_{j, 0}^{F}(t, X)+\psi_{\overline{j}, 0}^{F}(t, X)=1 \\
&\psi_{0}^{\mathrm{RMST}}(\tau, X)+\psi_{j, 0}^{\mathrm{RMTL}}(\tau, X)+\psi_{\overline{j}, 0}^{\mathrm{RMTL}}(\tau, X)=\tau \\
&\psi_{j, 0}^{F}(t, a_{\overline{j}}=a^{*}, X)+\psi_{j, 0}^{F}(t, a_{\overline{j}}=1-a^{*}, X)=\psi_{j, 0}^{F}(t, X) \\
&\psi_{j, 0}^{\mathrm{RMTL}}(\tau, a_{\overline{j}}=a^{*}, X)+\psi_{j, 0}^{\mathrm{RMTL}}(\tau, a_{\overline{j}}=1-a^{*}, X)=\psi_{j, 0}^{\mathrm{RMTL}}(\tau, X),
\end{aligned}
$$
}
but none of these are enforced. A possible solution is to utilize a one-dimensional universal least favorable submodel~\citep{rytgaard2024one}. Another shortcoming is that all results assumed the censoring and competing risk processes are confounded only by baseline covariates. This is often not realistic, as loss to follow-up and competing events are frequently driven by post-treatment developments prognostic for the outcome. An extension of our results would modify Assumptions 4 and 7 to require sequential exchangeability of time-varying censoring and competing risks processes given treatment and time-varying covariates. Then corresponding machine learning estimators of these time-varying nuisance models ~\citep{lee2019dynamic,nagpal2021deep} would also need to be incorporated. % gupta2019cresa,kopper2022deeppamm, 

\section*{Acknowledgements}

This work was supported by IBM Research under Grant W1771646; and National Institutes of Health under Grant R01AG058063-04.

\section*{Conflict of Interest}

The authors report there are no competing interests to declare.

\bibliographystyle{apacite} % \bibliographystyle{Chicago}

\bibliography{main}

\newpage

\appendix

% \appendixone

% \renewcommand{\thesection}{\arabic{section}}

\begin{center}
{\Large \textbf{Supplementary material to ``Heterogeneous Treatment Effects on Survival Outcomes Using Counterfactual Censoring Unbiased Transformations''}}
\end{center}

\section{Simulations}
\textit{Setting 1} (No competing risks, good overlap) The exposure follows a binomial distribution $A \sim \operatorname{expit}(-(0.3+0.2 X_1+0.3 X_2+0.3 X_3-0.2 X_4-0.3 X_5-0.2 X_6))$. The counterfactual survival times and censoring times follow
{\footnotesize
$$
\begin{aligned}
T^{a=0}=& \exp [0.2\{\log U -log(1-U)\}+ 0.8-0.8X_1+X_2+0.8X_3+0.4X_4-0.4X_5+0.8X_6] \\
T^{a=1}=& \exp \{\Phi^{-1}(U)+0.4+0.6X_1-0.8X_2+1.2X_3+0.6X_4-0.3X_5+0.5X_6\} \\
C=&I(A=0) \exp \{0.8\Phi^{-1}(U)+1.8+0.6X_1-0.8X_2+0.5X_3+0.7X_4-0.4X_5-0.2X_6\} \\
&+I(A=1) \exp [0.8\{\log U -log(1-U)\}+ 2.2+0.6X_1-0.8X_2+0.5X_3+0.7X_4+0.8X_5+1.2X_6]
\end{aligned}
$$
}
where $U$ follows a uniform distribution over [0, 1] and $\Phi$ is the standard normal cumulative distribution.

\textit{Setting 2} (No competing risks, poor overlap) The exposure follows a binomial distribution $A \sim \operatorname{expit}(-(-1+ X_1+1.5 X_2+1.5 X_3- X_4-1.5 X_5- X_6))$, where the coefficients are five times that of \textit{Setting 1} and the intercept is adjusted to balance the number of exposure groups to 1:1. The counterfactual survival times and censoring times follow \textit{Setting 1}.

% \textit{Setting} 3 $A \sim \operatorname{expit}(-(4.3+ X_1+1.5 X_2+1.5 X_3- X_4-1.5 X_5- X_6))$

\textit{Setting 3} (Competing risks, good overlap) The exposure follows \textit{Setting 1} while the counterfactual competing failure times and censoring times follow
{\footnotesize
$$
\begin{aligned}
&T_{j=1}^{a_{j}=0, a_{\overline{j}}=0}, T_{j=1}^{a_{j}=0, a_{\overline{j}}=1} \sim \lambda_{j=1}^{a=0}(t \mid X)= 0.12\exp \{0.1+ 0.1X_1-0.2 X_2+0.2 X_3+0.1 X_4+0.8 X_5-0.2 X_6\} \\
&T_{j=1}^{a_{j}=1, a_{\overline{j}}=0}, T_{j=1}^{a_{j}=1, a_{\overline{j}}=1} \sim \lambda_{j=1}^{a=1}(t \mid X)= 0.15\exp \{0.17+ 0.2X_1-0.1 X_2+0.4 X_3+0.2 X_4+0.3 X_5+0.4 X_6\} \\
&T_{j=2}^{a_{j}=0, a_{\overline{j}}=0}, T_{j=2}^{a_{j}=1, a_{\overline{j}}=0} \sim \lambda_{j=2}^{a=0}(t \mid X)= 0.1\exp \{0.12-0.1 X_1+0.3 X_2+0.1 X_3+0.2 X_4-0.4 X_5+0.5 X_6\} \\
&T_{j=2}^{a_{j}=0, a_{\overline{j}}=1}, T_{j=2}^{a_{j}=1, a_{\overline{j}}=1} \sim \lambda_{j=2}^{a=1}(t \mid X)= 0.08\exp \{0.1-0.2 X_1-0.1 X_2+0.2 X_3+0.3 X_4+0.3 X_5-0.3 X_6\} \\
&C=I(A=0) \exp \{0.8\Phi^{-1}(U)+2.5+0.6X_1-0.4X_2+0.7X_3+1.5X_4+1.2X_5+1.6X_6\} \\
&\quad \quad +I(A=1) \exp [0.8\{\log U -log(1-U)\}+ 2+0.6X_1+0.8X_2+0.5X_3+1.2  X_4+1.6X_5+1.2X_6]
\end{aligned}
$$
}
We let $A_{j}=A_{\overline{j}}=A$.

\textit{Setting 4} (Competing risks, poor overlap) The exposure follows \textit{Setting 2} and the counterfactual competing failure times and censoring times follow \textit{Setting 3}. 
% \textit{Setting} 6 $A \sim \operatorname{expit}(-(4.3+ X_1+1.5 X_2+1.5 X_3- X_4-1.5 X_5- X_6))$

In \textit{Setting 3} and \textit{4}, we computed the ground truth separable direct and indirect effects as follows:
{\footnotesize
$$
\begin{aligned}
&T^{a_{j}=0, a_{\overline{j}}=0}=\min\{T_{j=1}^{a_{j}=0, a_{\overline{j}}=0}, T_{j=2}^{a_{j}=0, a_{\overline{j}}=0}\}, \quad J^{a_{j}=0, a_{\overline{j}}=0}=\arg \min_{j}\{T_{j=1}^{a_{j}=0, a_{\overline{j}}=0}, T_{j=2}^{a_{j}=0, a_{\overline{j}}=0}\} \\
&T^{a_{j}=0, a_{\overline{j}}=1}=\min\{T_{j=1}^{a_{j}=0, a_{\overline{j}}=1}, T_{j=2}^{a_{j}=0, a_{\overline{j}}=1}\}, \quad J^{a_{j}=0, a_{\overline{j}}=1}=\arg \min_{j}\{T_{j=1}^{a_{j}=0, a_{\overline{j}}=1}, T_{j=2}^{a_{j}=0, a_{\overline{j}}=1}\} \\
&T^{a_{j}=1, a_{\overline{j}}=0}=\min\{T_{j=1}^{a_{j}=1, a_{\overline{j}}=0}, T_{j=2}^{a_{j}=1, a_{\overline{j}}=0}\}, \quad J^{a_{j}=1, a_{\overline{j}}=0}=\arg \min_{j}\{T_{j=1}^{a_{j}=1, a_{\overline{j}}=0}, T_{j=2}^{a_{j}=1, a_{\overline{j}}=0}\} \\
&T^{a_{j}=1, a_{\overline{j}}=1}=\min\{T_{j=1}^{a_{j}=1, a_{\overline{j}}=1}, T_{j=2}^{a_{j}=1, a_{\overline{j}}=1}\}, \quad J^{a_{j}=1, a_{\overline{j}}=1}=\arg \min_{j}\{T_{j=1}^{a_{j}=0, a_{\overline{j}}=0}, T_{j=2}^{a_{j}=1, a_{\overline{j}}=1}\}
\end{aligned}
$$
}
Assume we are interested in cause $j=1$ and $a_{\overline{j}}=1$. The true separable direct effect can be computed as
{\footnotesize
$$
\begin{aligned}
&F_{j, 0}(t, A_{j}=1, A_{\overline{j}}=1)-F_{j, 0}(t, A_{j}=0, A_{\overline{j}}=1) \\
=&P_{0}(T^{a_{j}=1, a_{\overline{j}}=1} \leq t, J^{a_{j}=1, a_{\overline{j}}=1}=j)-P_{0}(T^{a_{j}=0, a_{\overline{j}}=1} \leq t, J^{a_{j}=0, a_{\overline{j}}=1}=j)
\end{aligned}
$$
}
and the true separable indirect effect as
{\footnotesize
$$
\begin{aligned}
&F_{j, 0}(t, A_{j}=0, A_{\overline{j}}=1)-F_{j, 0}(t, A_{j}=0, A_{\overline{j}}=0) \\
=&P_{0}(T^{a_{j}=0, a_{\overline{j}}=1} \leq t, J^{a_{j}=0, a_{\overline{j}}=1}=j)-P_{0}(T^{a_{j}=0, a_{\overline{j}}=0} \leq t, J^{a_{j}=0, a_{\overline{j}}=0}=j)
\end{aligned}
$$
}

\begin{figure}[H]
\centering
\includegraphics[scale=0.6]{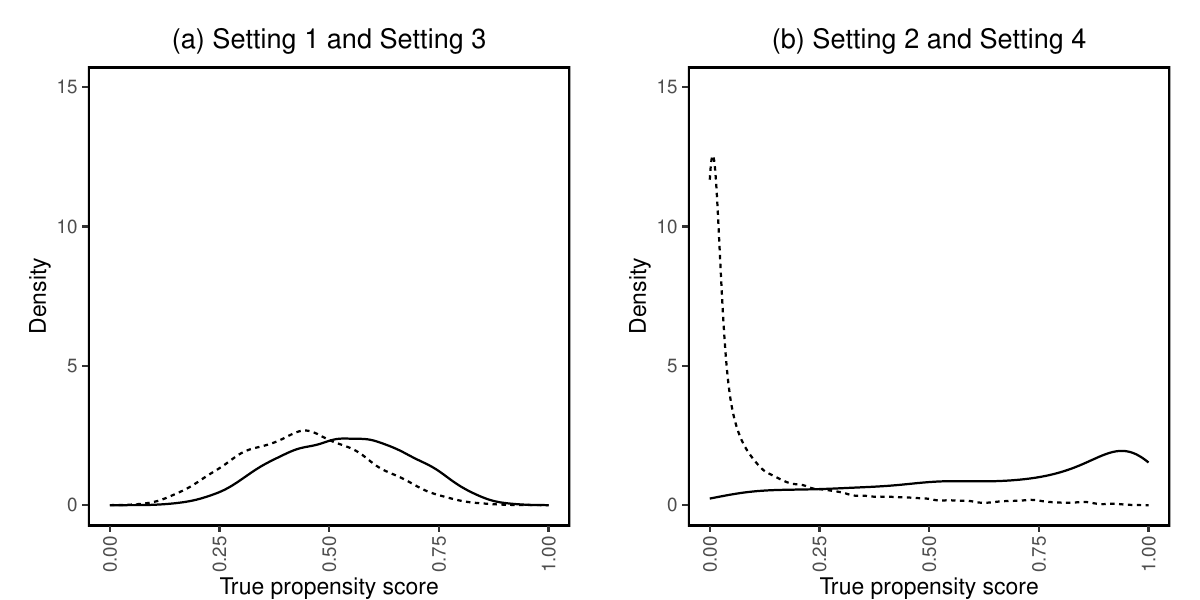}
\caption{Density of true propensity score where \textit{Setting} 1 and \textit{Setting} 4, \textit{Setting} 2 and \textit{Setting} 5, \textit{Setting} 3 and \textit{Setting} 6, are identical; dotted: treatment; solid: control.}
\label{fig:sim_ps}
\end{figure}

\begin{figure}[H]
\centering
\includegraphics[scale=0.6]{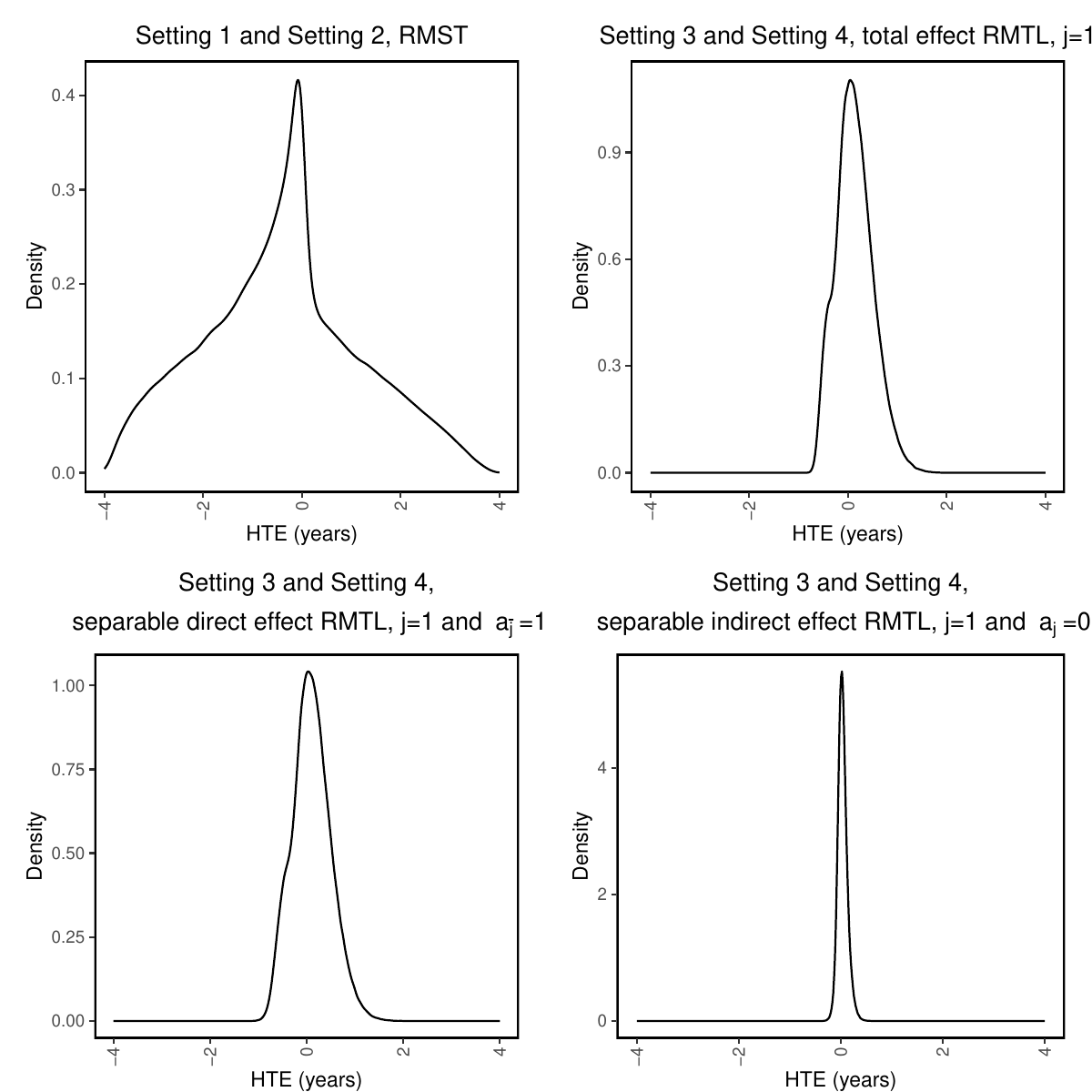}
\caption{Density of true HTE.}
\label{fig:sim_true}
\end{figure}

\begin{figure}[H]
\centering
\includegraphics[scale=0.55]{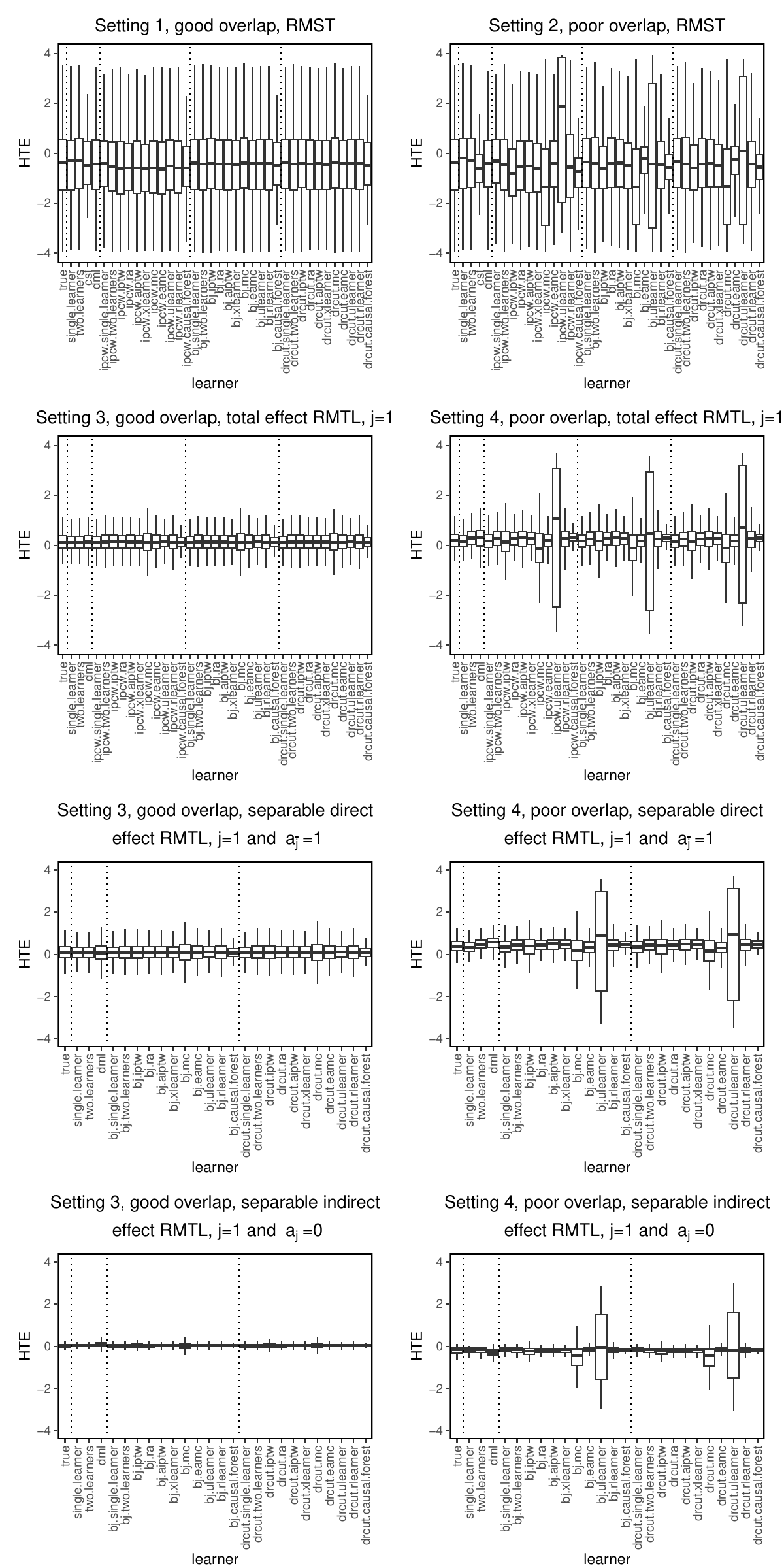}
\caption{Estimated HTE.}
\label{fig:sim_hte}
\end{figure}

\begin{figure}[H]
\centering
\includegraphics[scale=0.55]{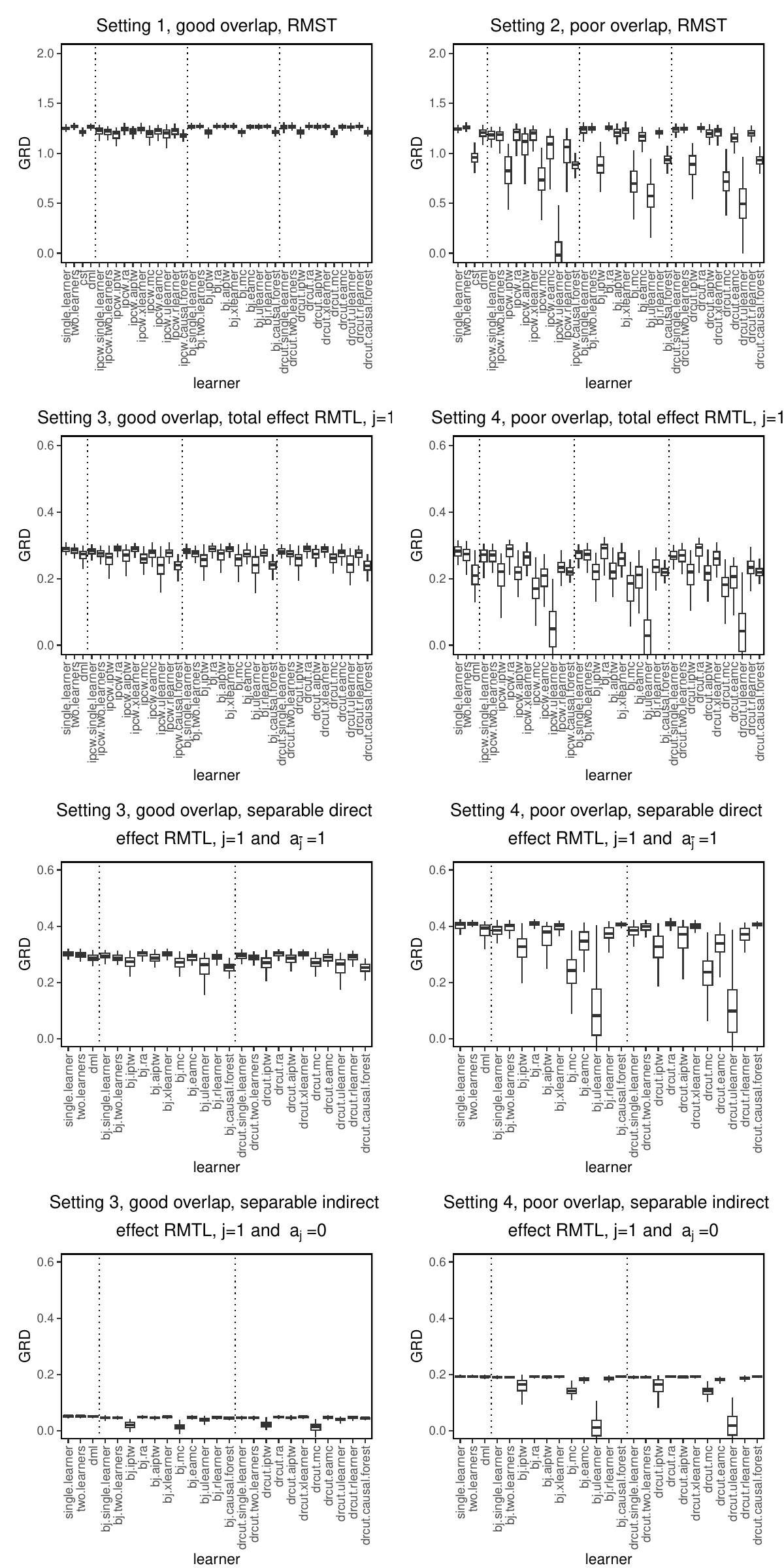}
\caption{GRD of estimated HTE.}
\label{fig:sim_grd}
\end{figure}

\section{Learning algorithm for performance evaluation}

\setcounter{step}{0}
\begin{step} % \label{step1}
First sample-splitting. We randomly split the data $O_1, \ldots, O_n$ into $K_{1}$ disjoint validation sets $\mathcal{V}_{1}, \ldots, \mathcal{V}_{K_{1}}$ with sizes $n_1, \ldots, n_{K_{1}}$, where $K_{1} \in\{2,3, \ldots,\lfloor n / 2\rfloor\}$. For each $k_{1} = 1, \ldots, K_{1}$, we define training set $\overline{\mathcal{V}}_{k_{1}}=\{O_i: i \notin \mathcal{V}_{k_{1}}\}$ and treatment-specific training set $\overline{\mathcal{V}}_{a, k_{1}}=\{O_i: i \in \overline{\mathcal{V}}_{k_{1}}, A_i=a\}$.
\end{step}

\begin{step} % \label{step2}
Survival conditional mean difference HTE learners and CUTs construction. For every $k_{1} = 1, \ldots, K_{1}$,
\begin{enumerate}
  \item when $j^{*}=1$, train time-to-event learners on $\overline{\mathcal{V}}_{k_{1}}$ and evaluate $\widehat{\Lambda}_{k_{1}}(t \mid A, X)$ using $\mathcal{V}_{k_{1}}$ at unique observed survival times $\widetilde{T}$ in $\mathcal{V}_{k_{1}}$ to capture jump information from observed counting processes $\mathrm{d} N(t)$ and $\mathrm{d} N^C(t)$; when $j^{*}>1$, train a time-to-event model with $(A, X, AX)$ or a MT-learner with $X$ on $\overline{\mathcal{V}}_{k_{1}}$ for cause $j$ and all causes, and evaluate $\widehat{\Lambda}_{j, k_{1}}(t \mid A, X)$ and $\widehat{\Lambda}_{k_{1}}(t \mid A, X)$ using $\mathcal{V}_{k_{1}}$ at unique observed survival times $\widetilde{T}$ in $\mathcal{V}_{k_{1}}$ to capture jump information from observed counting processes $\mathrm{d} N(t)$, $\mathrm{d} N^C(t)$, and $\mathrm{d} N_{j}(t)$;

  \item train the propensity model on $\overline{\mathcal{V}}_{k_{1}}$ and predict $\widehat{\pi}_{k_{1}}(a \mid X)$ on $\mathcal{V}_{k_{1}}$;

  \item construct transformations by plugging in $\widehat{\eta}_{k_{1}}^{\Lambda}$ into $Y(\cdot, A; \widehat{\eta}_{k_{1}}^{\Lambda})$ and $\widehat{\eta}_{k_{1}}$ into $\phi(\cdot; \widehat{\eta}_{k_{1}})$;

\end{enumerate}
Bind $\widehat{\eta}_{k_{1}}$, $Y(\cdot, A; \widehat{\eta}_{k_{1}}^{\Lambda})$, and $\phi(\cdot; \widehat{\eta}_{k_{1}})$ to $\mathcal{V}_{k_{1}}$ and combine $\mathcal{V}_{k_{1}}$ into one dataset.
\end{step}

\begin{step} % \label{step3}
Second sample-splitting. We randomly split the data $O_1, \ldots, O_n$ into $K_{2}$ disjoint validation sets $\mathcal{V}_{1}, \ldots, \mathcal{V}_{K_{2}}$ with sizes $n_1, \ldots, n_{K_{2}}$, where $K_{2} \in\{2,3, \ldots,\lfloor n / 2\rfloor\}$. For each $k_{2} = 1, \ldots, K_{2}$, we define training set $\overline{\mathcal{V}}_{k_{2}}=\{O_i: i \notin \mathcal{V}_{k_{2}}\}$ and treatment-specific training set $\overline{\mathcal{V}}_{a, k_{2}}=\{O_i: i \in \overline{\mathcal{V}}_{k_{2}}, A_i=a\}$.
\end{step}

\begin{step} % \label{step4}
CUTs conditional mean difference HTE learners and transformed minimization construction. For every $k_{2} = 1, \ldots, K_{2}$,

\begin{enumerate}
  \item train the CUT outcome model on $\overline{\mathcal{V}}_{k_{2}}$ and predict $\widehat{\mu}_{k_{2}}(\cdot, a, X; \widehat{\eta}^{\Lambda})$ on $\mathcal{V}_{k_{2}}$;

  \item train the propensity model on $\overline{\mathcal{V}}_{k_{2}}$ and predict $\widehat{\pi}_{k_{2}}(a \mid X)$ on $\mathcal{V}_{k_{2}}$;

  \item construct outcome transformations by plugging in $\widehat{\mu}_{k_{2}}(\cdot, a, X; \widehat{\eta}^{\Lambda})$ and $\widehat{\pi}_{k_{2}}(a \mid X)$ into $Y^{*}(Y(\cdot, A; \widehat{\eta}^{\Lambda}); (\widehat{\mu}_{k_{2}}, \widehat{\pi}_{k_{2}}))$ and weights by $\widehat{\pi}_{k_{2}}(a \mid X)$ into $w^{*}(\widehat{\pi}_{k_{2}})$. \\

  we take $\widehat{\mu}_{k_{2}}(\cdot, X; \widehat{\eta}^{\Lambda})=\widehat{\pi}_{k_{2}}(0 \mid X)\widehat{\mu}_{k_{2}}(\cdot, 0, X; \widehat{\eta}^{\Lambda})+\widehat{\pi}_{k_{2}}(1 \mid X)\widehat{\mu}_{k_{2}}(\cdot, 1, X; \widehat{\eta}^{\Lambda})$ to tackle potential limited overlap in observational studies. Alternatively, one can consider regressing $X$ on $Y(\cdot, A; \widehat{\eta}^{\Lambda})$. We can also take corresponding quantity of $\widehat{\eta}^{\Lambda}$ as $\widehat{\mu}_{k_{2}}(\cdot, a, X; \widehat{\eta}^{\Lambda})$ or $\widehat{\mu}_{k_{2}}(\cdot, X; \widehat{\eta}^{\Lambda})$ without regression on the CUTs;

\end{enumerate}
Bind $Y^{*}(Y(\cdot, A; \widehat{\eta}^{\Lambda}); (\widehat{\mu}_{k_{2}}, \widehat{\pi}_{k_{2}}))$ and $w^{*}(\widehat{\pi}_{k_{2}})$ to $\mathcal{V}_{k_{2}}$ and combine $\mathcal{V}_{k_{2}}$ into one dataset.
\end{step}

\begin{step} % \label{step5}
Third sample-splitting. We randomly split the data $O_1, \ldots, O_n$ into $K_{3}$ disjoint validation sets $\mathcal{V}_{1}, \ldots, \mathcal{V}_{K_{3}}$ with sizes $n_1, \ldots, n_{K_{3}}$, where $K_{3} \in\{2,3, \ldots,\lfloor n / 2\rfloor\}$. For each $k_{3} = 1, \ldots, K_{3}$, we define training set $\overline{\mathcal{V}}_{k_{3}}=\{O_i: i \notin \mathcal{V}_{k_{3}}\}$.
\end{step}

\begin{step} % \label{step6}
Cross-fitting transformed minimization. For every $k_{3} = 1, \ldots, K_{3}$,

\begin{enumerate}
  \item For each learner from~\Cref{tab:weighted_minimization} except the IF-learner, train a supervised learning algorithm by taking $Y^{*}(Y(\cdot, A; \widehat{\eta}^{\Lambda}); (\widehat{\mu}_{k_{2}}, \widehat{\pi}_{k_{2}}))$ as the outcome, $X$ as covariates, and $w^{*}(\widehat{\pi})$ as sampling weights using $\overline{\mathcal{V}}_{k_{3}}$.

  \item predict $\widehat{\psi}_{k_{3}}^{*}(\cdot, X)$ using $X$ from $\mathcal{V}_{k_{3}}$;

\end{enumerate}
Bind all $\widehat{\psi}_{k_{3}}^{*}(\cdot, X)$ to $\mathcal{V}_{k_{3}}$ and combine $\mathcal{V}_{k_{3}}$ into one dataset.
\end{step}

\section{Additional CUTs of Theorem~\ref{theorem:cut}} \label{additional:cut}

(v) (Counterfactual CUTs to facilitate estimation of $\psi_0^{RMST}(\tau,X)$) Define:
{\footnotesize
$$
\begin{aligned} \label{rmst_cut}
Y^{\mathrm{RMST}, \mathrm{BJ}}(\tau, a; S)=&\min(\widetilde{T}, \tau)+\frac{\{1-\Delta(\tau)\}\{\mathrm{RMST}(\tau \mid a, X)-\mathrm{RMST}(\tau \wedge \widetilde{T}\mid a, X)\}}{S(\tau \wedge \widetilde{T} \mid a, X)} \\
Y^{\mathrm{RMST}, \mathrm{IPCW}}(\tau, a; G)=&\frac{\Delta(\tau)\min(\widetilde{T}, \tau)}{G(\tau \wedge \widetilde{T}- \mid a, X)} \\
Y^{\mathrm{RMST}, \mathrm{AIPCW}}(\tau, a; \eta^{\Lambda})=&\frac{\Delta(\tau)\min(\widetilde{T}, \tau)}{G(\tau \wedge \widetilde{T}- \mid a, X)}+\int_{0}^{\tau \wedge \widetilde{T}} \frac{t \mathrm{d} M^{C}(t \mid a, X)}{G(t- \mid a, X)} \\
&+\int^{\tau \wedge \widetilde{T}} \frac{\{\mathrm{RMST}(\tau \mid a, X)-\mathrm{RMST}(t \mid a, X)\} \mathrm{d} M^{C}(t \mid a, X)}{S(t \mid a, X)G(t- \mid a, X)} \\
=& \mathrm{RMST}(\tau \mid a, X)-\int_{0}^{\tau \wedge \widetilde{T}}\frac{\{\mathrm{RMST}(\tau \mid a, X)-\mathrm{RMST}(t \mid a, X)\} \mathrm{d} M(t \mid A, X)}{S(t \mid a, X)G(t- \mid a, X)}.
\end{aligned}
$$
}
Under ~\Cref{assumption:positivity}-\Cref{assumption:depcens_survival}: if $S_{\infty}(t \mid a, X)=S_{0}(t \mid a, X)$, then $E_{0}\{Y^{\mathrm{RMST}, \mathrm{BJ}}(\tau, A; S_{\infty}) \mid A=a, X\}=E_{0}\{\min(T^a, \tau) \mid X\}$; if $G_{\infty}(t \mid a, X)=G_{0}(t \mid a, X)$, then $E_{0}\{Y^{\mathrm{RMST}, \mathrm{IPCW}}(\tau, A; G_{\infty}) \mid A=a, X\}=E_{0}\{\min(T^a, \tau) \mid X\}$; if  either $S_{\infty}(t \mid a, X)=S_{0}(t \mid a, X)$ or $G_{\infty}(t \mid a, X)=G_{0}(t \mid a, X)$, then $E_{0}\{Y^{\mathrm{RMST}, \mathrm{AIPCW}}(\tau, A; \eta_{\infty}^{\Lambda}) \mid A=a, X\}=E_{0}\{\min(T^a, \tau) \mid X\}$, where $E_{0}\{\min(T^a, \tau) \mid X\}$ is the true conditional counterfactual RMST. \\

(vi) (Counterfactual CUTs to facilitate estimation of $\psi_{j,0}^{RMTL}(\tau,X)$) Define: % The CUTs for cause-specifc RMTL are
{\footnotesize
$$
\begin{aligned} \label{rmtlj_cut}
Y_{j}^{\mathrm{RMTL}, \mathrm{BJ}}(\tau,a; (S, \Lambda_{j}))=&\{\tau-\min(\widetilde{T}, \tau)\}I(\widetilde{J}=j)+\frac{\{1-\Delta(\tau)\}}{S(\tau \wedge \widetilde{T} \mid a, X)}\times[\mathrm{RMTL}_{j}(\tau \mid a, X) \\
&-\mathrm{RMTL}_{j}(\tau \wedge \widetilde{T} \mid a, X)-F_{j}(\tau \wedge \widetilde{T} \mid a, X)\{\tau-\min(\widetilde{T}, \tau)\}] \\
Y_{j}^{\mathrm{RMTL}, \mathrm{IPCW}}(\tau,a; G)=&\frac{\{\tau-\min(\widetilde{T}, \tau)\}I(\widetilde{J}=j)}{G(\widetilde{T}- \mid a, X)} \\
Y_{j}^{\mathrm{RMTL}, \mathrm{AIPCW}}(\tau,a; \eta_{j}^{\Lambda})=&\frac{\{\tau-\min(\widetilde{T}, \tau)\}I(\widetilde{J}=j)}{G(\widetilde{T}- \mid a, X)}-\int_{0}^{\tau \wedge \widetilde{T}} \frac{(\tau-t) F_{j}(t \mid a, X) \mathrm{d} M^{C}(t \mid a, X)}{S(t \mid a, X)G(t- \mid a, X)} \\
&+\int_{0}^{\tau \wedge \widetilde{T}} \frac{\{\mathrm{RMTL}_{j}(\tau \mid a, X)-\mathrm{RMTL}_{j}(t \mid a, X)\} \mathrm{d} M^{C}(t \mid a, X)}{S(t \mid a, X)G(t- \mid a, X)} \\
=&\mathrm{RMTL}_{j}(\tau \mid a, X)+\int_0^{\tau \wedge \widetilde{T}} \frac{(\tau-t)\mathrm{d}M_{j}(t \mid a, X)}{G(t- \mid a, X)} \\
&-\int_0^{\tau \wedge \widetilde{T}}\frac{\{\mathrm{RMTL}_{j}(\tau \mid a, X)-\mathrm{RMTL}_{j}(t \mid a, X)\} \mathrm{d}M(t \mid a, X)}{S(t \mid a, X)G(t- \mid a, X)} \\
&+\int_0^{\tau \wedge \widetilde{T}}\frac{(\tau-t)F_{j}(t \mid a, X)\mathrm{d}M(t \mid a, X)}{S(t \mid a, X)G(t- \mid a, X)}.
\end{aligned}
$$
}
Under ~\Cref{assumption:positivity} and~\Cref{assumption:SUTVA_competing}-\Cref{assumption:depcens_competing}: if $S_{\infty}(t \mid a, X)=S_{0}(t \mid a, X)$ and $\Lambda_{j, \infty}(t \mid a, X)=\Lambda_{j, 0}(t \mid a, X)$, then $E_{0}\{Y_{j}^{\mathrm{RMTL}, \mathrm{BJ}}(\tau, A; (S_{\infty}, \Lambda_{j, \infty})) \mid A=a, X\}=E_{0}[\{\tau-\min (T^a, \tau)\} I(J^a=j) \mid X]$; if $G_{\infty}(t \mid a, X)=G_{0}(t \mid a, X)$, then $E_{0}\{Y_{j}^{\mathrm{RMTL}, \mathrm{IPCW}}(\tau, A; G_{\infty}) \mid A=a, X\}=E_{0}[\{\tau-\min (T^a, \tau)\} I(J^a=j) \mid X]$; if either $S_{\infty}(t \mid a, X)=S_{0}(t \mid a, X)$ and $\Lambda_{j, \infty}(t \mid a, X)=\Lambda_{j, 0}(t \mid a, X)$ or $G_{\infty}(t \mid a, X)=G_{0}(t \mid a, X)$, then $E_{0}\{Y_{j}^{\mathrm{RMTL}, \mathrm{AIPCW}}(\tau, A; \eta_{j, \infty}^{\Lambda}) \mid A=a, X\}=E_{0}[\{\tau-\min (T^a, \tau)\} I(J^a=j) \mid X]$, where $E_{0}[\{\tau-\min (T^a, \tau)\} I(J^a=j) \mid X]$ is the true conditional counterfactual cause-specific RMTL. \\

(vii) Under ~\Cref{assumption:positivity} and~\Cref{assumption:SUTVA_competing}-\Cref{assumption:dismissible}, the following is a counterfactual CUT for $\{\tau-\min (T^{a, a^{*}}, \tau)\} I(J^{a, a^{*}}=j)$:
% counterfactual cause-specific separable direct CIF (vi) the following are counterfactual CUTs for cause-specific separable direct RMTL 
% $\mathrm{RMTL}(\tau \mid A, a_{\overline{j}}=a^{*}, X)=E_{0}[\{\tau-\min(T,\tau)\}I(J=j) \mid A_{j}=A, A_{\overline{j}}=a^{*}, X]$
% and the CUTs for cumulative cause-specific separable direct effect are
{\footnotesize
$$
\begin{aligned} \label{rmtlj_sep_direct_cut}
&Y_{j}^{\mathrm{RMTL}, \mathrm{AIPCW}}(\tau, A_{j}=a, A_{\overline{j}}=a^{*}; \eta_{j}^{\Lambda}) \\
=&\mathrm{RMTL}_{j}(\tau \mid a, a^{*}, X)+\int_0^{\tau \wedge \widetilde{T}} \frac{(\tau-t) S_{\overline{j}}(t \mid a^{*}, X)\mathrm{d} M_{j}(t \mid a, X)}{S_{\overline{j}}(t \mid a, X)G(t- \mid a, X)}-\int_0^{\tau \wedge \widetilde{T}} \frac{S_{\overline{j}}(t \mid a^{*}, X)\mathrm{d} M(t \mid a, X)}{S_{\overline{j}}(t \mid a, X)S(t \mid a, X)G(t- \mid a, X)} \\
&\times\{\mathrm{RMTL}_{j}(\tau \mid a, X)-\mathrm{RMTL}_{j}(t \mid a, X)\}+ \int_0^{\tau \wedge \widetilde{T}} \frac{(\tau-t)S_{\overline{j}}(t \mid a^{*}, X)F_{j}(t \mid a, X)\mathrm{d} M(t \mid a, X)}{S_{\overline{j}}(t \mid a, X)S(t \mid a, X)G(t- \mid a, X)} \\
&+\left\{1-\frac{I(A=a^{*}) \pi(a \mid X)}{\pi(a^{*} \mid X) I(A=a)} \right\} \left[\{\mathrm{RMTL}_{j}(\tau \mid 1-a^{*}, a^{*}, X)-\mathrm{RMTL}_{j}(t \mid A_{j}=1-a^{*}, a^{*}, X)\} \right. \\
&\left. \times \int_0^{\tau \wedge \widetilde{T}}\frac{\mathrm{d} M_{\overline{j}}(t \mid a, X)}{S(t \mid a, X)G(t- \mid a, X)} -\int_0^{\tau \wedge \widetilde{T}}\frac{(\tau-t) F_{j}(t \mid 1-a^{*}, a^{*}, X)\mathrm{d} M_{\overline{j}}(t \mid a, X)}{S(t \mid a, X)G(t- \mid a, X)} \right].
\end{aligned}
$$
}
Under ~\Cref{assumption:positivity} and~\Cref{assumption:SUTVA_competing}-\Cref{assumption:dismissible}: if $\Lambda_{j, \infty}(t \mid a, X)=\Lambda_{j, 0}(t \mid a, X)$ and $\Lambda_{\overline{j}, \infty}(t \mid a, X)=\Lambda_{\overline{j}, 0}(t \mid a, X)$, then $E_{0}\{Y_{j}^{\mathrm{RMTL}, \mathrm{sep-D}, \mathrm{AIPCW}}(\tau,A,a^{*}; \eta_{j, \infty}^{\Lambda}) \mid A=a, X\}=E_{0}[\{\tau-\min (T^{a, a^{*}}, \tau)\} I(J^{a, a^{*}}=j) \mid X]$, which is the true conditional counterfactual separable cause-specific RMTL.

% That is when $a = 1-a^{*}$, if  $\Lambda_{j, \infty}(t \mid a, X)=\Lambda_{j, 0}(t \mid a, X)$ and $\Lambda_{\overline{j}, \infty}(t \mid a, X)=\Lambda_{\overline{j}, 0}(t \mid a, X)$, then $E_{0}\{Y_{j}^{\mathrm{RMTL}, \mathrm{AIPCW}}(\tau, A_{j}=A, A_{\overline{j}}=a^{*}; \eta_{j, \infty}^{\Lambda}) \mid A=a, X\}=E_{0}[\{\tau-\min (T^{a, a^{*}}, \tau)\} I(J^{a, a^{*}}=j) \mid X]$, which is the true conditional counterfactual separable direct cause-specific RMTL. \\

% That is, \sloppy $E\{Y_{j}^{\mathrm{RMTL}, \mathrm{AIPCW}}(\tau, A_{j}=A, A_{\overline{j}}=a^{*}; \eta_{j}^{\Lambda}) \mid A=a, X\}=E[\{\tau-\min (T^{a, a^{*}}, \tau)\} I(J^{a, a^{*}}=j) \mid X]$, which is the separable direct counterfactual RMTL.

(viii) Under ~\Cref{assumption:positivity} and~\Cref{assumption:SUTVA_competing}-\Cref{assumption:dismissible}, the following is a counterfactual CUT for $\{\tau-\min (T^{1-a^{*}, a}, \tau)\} I(J^{1-a^{*}, a}=j)$:
% the following are counterfactual CUTs for cause-specific separable indirect RMTL $\mathrm{RMTL}(\tau \mid a_{j}=1-a^{*}, A, X)=E_{0}[\{\tau-\min(T,\tau)\}I(J=j) \mid A_{j}=1-a^{*}, A_{\overline{j}}=A, X]$ % and the CUTs for cumulative cause-specific separable indirect effect are
{\footnotesize
$$
\begin{aligned} \label{rmtlj_sep_indirect_cut}
&Y_{j}^{\mathrm{RMTL}, \mathrm{AIPCW}}(\tau, A_{j}=1-a^{*}, A_{\overline{j}}=a; \eta_{j}^{\Lambda}) \\
=&\mathrm{RMTL}_{j}(\tau \mid a_{j}=1-a^{*}, a, X)+\int_0^{\tau \wedge \widetilde{T}} \frac{(\tau-t)\mathrm{d} M_{j}(t \mid a, X)}{G(t- \mid a, X)}\left\{1-\frac{S_{\overline{j}}(t \mid a^{*}, X)}{S_{\overline{j}}(t \mid a, X)}\right\} \\
&-\int_0^{\tau \wedge \widetilde{T}} \frac{\mathrm{d} M(t \mid a, X)}{S(t \mid a, X)G(t- \mid a, X)}\left\{1-\frac{S_{\overline{j}}(t \mid a^{*}, X)}{S_{\overline{j}}(t \mid a, X)}\right\}\{\mathrm{RMTL}_{j}(\tau \mid a, X)-\{\mathrm{RMTL}_{j}(t \mid a, X)\} \\
&+\int_0^{\tau \wedge \widetilde{T}} \frac{(\tau-t)F_{j}(t \mid a, X)\mathrm{d} M(t \mid a, X)}{S(t \mid a, X)G(t- \mid a, X)}\left\{1-\frac{S_{\overline{j}}(t \mid a^{*}, X)}{S_{\overline{j}}(t \mid a, X)}\right\} \\
&-\left\{1-\frac{I(A=a^{*}) \pi(a \mid X)}{\pi(a^{*} \mid X) I(A=a)} \right\} \int_0^{\tau \wedge \widetilde{T}}\frac{\mathrm{d} M_{\overline{j}}(t \mid a, X)}{S(t \mid a, X)G(t- \mid a, X)}\{\mathrm{RMTL}_{j}(\tau \mid 1-a^{*}, a^{*}, X)-\mathrm{RMTL}_{j}(t \mid 1-a^{*}, a^{*}, X)\} \\
&+\left\{1-\frac{I(A=a^{*}) \pi(a \mid X)}{\pi(a^{*} \mid X) I(A=a)} \right\} \int_0^{\tau \wedge \widetilde{T}}\frac{(\tau-t) F_{j}(t \mid 1-a^{*}, a^{*}, X)\mathrm{d} M_{\overline{j}}(t \mid a, X)}{S(t \mid a, X)G(t- \mid a, X)}.
\end{aligned}
$$
}
Under ~\Cref{assumption:positivity} and~\Cref{assumption:SUTVA_competing}-\Cref{assumption:dismissible}: if  $\Lambda_{j, \infty}(t \mid a, X)=\Lambda_{j, 0}(t \mid a, X)$ and $\Lambda_{\overline{j}, \infty}(t \mid a, X)=\Lambda_{\overline{j}, 0}(t \mid a, X)$, then $E_{0}\{Y_{j}^{\mathrm{RMTL}, \mathrm{sep-I}, \mathrm{AIPCW}}(t,a^{*},A; \eta_{j}^{\Lambda, \infty}) \mid A=a, X\}=E_0[\{\tau-\min (T^{a^{*}, a}, \tau)\} I(J^{a^{*}, a}=j) \mid X]$, which is the true conditional counterfactual separable cause-specific RMTL.

% That is when $a = 1-a^{*}$, if  $\Lambda_{j, \infty}(t \mid a, X)=\Lambda_{j, 0}(t \mid a, X)$ and $\Lambda_{\overline{j}, \infty}(t \mid a, X)=\Lambda_{\overline{j}, 0}(t \mid a, X)$, then $E_{0}\{Y_{j}^{\mathrm{RMTL}, \mathrm{AIPCW}}(\tau, A_{j}=1-a^{*}, A_{\overline{j}}=A; \eta_{j}^{\Lambda}) \mid A=a, X\}=E_0[\{\tau-\min (T^{1-a^{*}, a}, \tau)\} I(J^{1-a^{*}, a}=j) \mid X]$, which is the true conditional counterfactual separable indirect cause-specific RMTL.

% That is, \sloppy $E\{Y_{j}^{\mathrm{RMTL}, \mathrm{AIPCW}}(\tau, a_{j}=1-a^{*}, a_{\overline{j}}=a; \eta_{j}^{\Lambda}) \mid X\}=E[\{\tau-\min (T^{1-a^{*}, a}, \tau)\} I(J^{1-a^{*}, a}=j) \mid X]$, which is the separable indirect counterfactual RMTL.

\section{Additional influence function transformations of Theorem~\ref{theorem:cut}}

The IF transformation for each estimand is
{\footnotesize
$$
\begin{aligned} \label{eq:ift}
\phi^{S}(t, \eta)=&\phi^{S}(t, 1; \eta)-\phi^{S}(t, 0; \eta) \\
\phi^{S}(t, a; \eta)=&\frac{I(A=a)I(\widetilde{T}>t)}{\pi(a \mid X) G(t \mid a, X)}+\left\{1-\frac{I(A=a)}{\pi(a \mid X)}\right\}S(t \mid a, X) \\
&+S(t \mid a, X) \int_{0}^{t \wedge \widetilde{T}} \frac{\mathrm{d} M^{C}(u \mid a, X)}{S(u \mid a, X) G(u- \mid a, X)} \\
=&S(t \mid a, X)-\frac{I(A=a)S(t \mid a, X)}{\pi(a \mid X)}\int_0^{t \wedge \widetilde{T}} \frac{\mathrm{d} M(u \mid a, X)}{S(u \mid a, X) G(u-\mid a, X)} \\
\phi^{\mathrm{RMST}}(\tau; \eta)=&\phi^{\mathrm{RMST}}(\tau, 1; \eta)-\phi^{\mathrm{RMST}}(\tau, 0; \eta) \\
\phi^{\mathrm{RMST}}(\tau, a; \eta)=&\frac{I(A=a)\Delta(\tau)\min(\widetilde{T}, \tau)}{\pi(a \mid X) G(\tau \wedge \widetilde{T}- \mid a, X)}+\left\{1-\frac{I(A=a)}{\pi(a \mid X)}\right\}\mathrm{RMST}(\tau \mid a, X) \\
&+\int^{\tau \wedge \widetilde{T}} \frac{\{\mathrm{RMST}(\tau \mid a, X)-\mathrm{RMST}(t \mid a, X)\} \mathrm{d} M^{C}(t \mid a, X)}{S(t \mid a, X)G(t- \mid a, X)}+\int_{0}^{\tau \wedge \widetilde{T}} \frac{t \mathrm{d} M^{C}(t \mid a, X)}{G(t- \mid a, X)} \\
=&\mathrm{RMST}(\tau \mid a, X)-\frac{I(A=a)}{\pi(a \mid X)}\int_0^{\tau \wedge \widetilde{T}} \frac{\{\mathrm{RMST}(\tau \mid a, X)-\mathrm{RMST}(t \mid a, X)\} \mathrm{d} M(u \mid a, X)}{S(u \mid a, X) G(u-\mid a, X)}
\end{aligned}
$$
}

{\footnotesize
$$
\begin{aligned}
\phi^{F}_{j}(t; \eta_{j})=&\phi^{F}_{j}(t, 1; \eta_{j})-\phi^{F}_{j}(t, 0; \eta_{j}) \\
\phi^{F}_{j}(t, a; \eta_{j})=&\frac{I(A=a)I(\widetilde{T} \leq t, \widetilde{J}=j)}{\pi(a \mid X)G(\widetilde{T}-\mid a, X)}+\left\{1-\frac{I(A=a)}{\pi(a \mid X)}\right\}F_{j}(t \mid a, X) \\
&+\int_0^{t \wedge \widetilde{T}} \frac{\{F_{j}(t \mid a, X)-F_{j}(u \mid a, X)\} \mathrm{d} M^C(u \mid a, X)}{S(u \mid a, X) G(u-\mid a, X)} \\
=&F_{j}(t \mid a, X)+\frac{I(A=a)}{\pi(a \mid X)} \int_0^{t \wedge \widetilde{T}} \frac{\mathrm{d}M_{j}(u \mid a, X)}{G(u \mid a, X)} \\
&-\frac{I(A=a)}{\pi(a \mid X)} \int_0^{t \wedge \widetilde{T}} \frac{\{F_{j}(t \mid a, X)-F_{j}(u \mid a, X)\}\mathrm{d} M(u \mid a, X)}{S(u \mid a, X) G(u-\mid a, X)}
\end{aligned}
$$
}

{\footnotesize
$$
\begin{aligned}
\phi^{\mathrm{RMTL}}_{j}(\tau; \eta_{j})=&\phi^{\mathrm{RMTL}}_{j}(\tau, 1; \eta_{j})-\phi^{\mathrm{RMTL}}_{j}(\tau, 0; \eta_{j}) \\
\phi^{\mathrm{RMTL}}_{j}(\tau, a; \eta_{j})=&\frac{I(A=a)\{\tau-\min(\widetilde{T}, \tau)\}I(\widetilde{J}=j)}{\pi(a \mid X) G(\widetilde{T}- \mid a, X)}+\left\{1-\frac{I(A=a)}{\pi(a \mid X)}\right\}\mathrm{RMTL}_{j}(\tau \mid a, X) \\
&-\int_{0}^{\tau \wedge \widetilde{T}} \frac{(\tau-t) F_{j}(t \mid a, X) \mathrm{d} M^{C}(t \mid a, X)}{S(t \mid a, X)G(t- \mid a, X)} \\
&+\int_{0}^{\tau \wedge \widetilde{T}} \frac{\{\mathrm{RMTL}_{j}(\tau \mid a, X)-\mathrm{RMTL}_{j}(t \mid a, X)\} \mathrm{d} M^{C}(t \mid a, X)}{S(t \mid a, X)G(t- \mid a, X)} \\
=&\mathrm{RMTL}_{j}(\tau \mid a, X)+\frac{I(A=a)}{\pi(a \mid X)}\int_0^{\tau \wedge \widetilde{T}} \frac{(\tau-t)\mathrm{d}M_{j}(t \mid a, X)}{G(t- \mid a, X)} \\
&-\frac{I(A=a)}{\pi(a \mid X)}\int_0^{\tau \wedge \widetilde{T}}\frac{\{\mathrm{RMTL}_{j}(\tau \mid a, X)-\mathrm{RMTL}_{j}(t \mid a, X)\} \mathrm{d}M(t \mid a, X)}{S(t \mid a, X)G(t- \mid a, X)} \\
&+\frac{I(A=a)}{\pi(a \mid X)}\int_0^{\tau \wedge \widetilde{T}}\frac{(\tau-t)F_{j}(t \mid a, X)\mathrm{d}M(t \mid a, X)}{S(t \mid a, X)G(t- \mid a, X)}
\end{aligned}
$$
}

{\footnotesize
$$
\begin{aligned}
&\phi^{F}_{j}(t, A_{\overline{j}}=a^{*}; \eta_{j})=\phi^{F}_{j}(t, A_{j}=1, A_{\overline{j}}=a^{*}; \eta_{j})-\phi^{F}_{j}(t, A_{j}=0, A_{\overline{j}}=a^{*}; \eta_{j}) \\
&\phi^{F}_{j}(t, A_{j}=a, A_{\overline{j}}=a^{*}; \eta_{j}) \\
% =&F_{j, 0}(t \mid a, a^{*}, X)+\frac{I(A=a)}{\pi_{0}(a \mid X)}\int_0^{t \wedge \widetilde{T}} \frac{S_{\overline{j}, 0}(u \mid a^{*}, X)\mathrm{d} M_{j, 0}(u \mid a, X)}{S_{\overline{j}, 0}(u \mid a, X)G_{0}(u- \mid a, X)}\left[1-\frac{\{F_{j, 0}(t \mid a, X)-F_{j, 0}(u \mid a, X)\}}{S_{0}(u \mid a, X)}\right] \\
% &+\frac{I(A=a)}{\pi_{0}(a \mid X)}\int_0^{t \wedge \widetilde{T}}\frac{\mathrm{d} M_{\overline{j}, 0}(u \mid a, X)}{S_{0}(u \mid a, X)G_{0}(u- \mid a, X)}\left[F_{j, 0}(t \mid 1-a^{*}, a^{*}, X)-F_{j, 0}(u \mid 1-a^{*}, a^{*}, X) \right. \\
% &\left.-\frac{S_{\overline{j}, 0}(u \mid 1, X)}{S_{\overline{j}, 0}(u \mid a, X)}\{F_{j, 0}(t \mid a, X)-F_{j, 0}(u \mid a, X)\}\right] \\
=&F_{j}(t \mid a, a^{*}, X)+\frac{I(A=a)}{\pi(a \mid X)}\int_0^{t \wedge \widetilde{T}} \frac{S_{\overline{j}}(u \mid a^{*}, X)\mathrm{d} M_{j}(u \mid a, X)}{S_{\overline{j}}(u \mid a, X)G(u- \mid a, X)} \\
&-\frac{I(A=a)}{\pi(a \mid X)}\int_0^{t \wedge \widetilde{T}} \frac{S_{\overline{j}}(u \mid a^{*}, X)\mathrm{d} M(u \mid a, X)}{S_{\overline{j}}(u \mid a, X)S(u \mid a, X)G(u- \mid a, X)}\{F_{j}(t \mid a, X)-F_{j}(u \mid a, X)\} \\
&+\left\{\frac{I(A=a)}{\pi(a \mid X)} -\frac{I(A=a^{*})}{\pi(a^{*} \mid X)} \right\} \int_0^{t \wedge \widetilde{T}}\frac{\mathrm{d} M_{\overline{j}}(u \mid a, X)}{S(u \mid a, X)G(u- \mid a, X)}\{F_{j}(t \mid 1-a^{*}, a^{*}, X)-F_{j}(u \mid 1-a^{*}, a^{*}, X)\}
\end{aligned}
$$
}

{\footnotesize
$$
\begin{aligned}
&\phi^{\mathrm{RMTL}}_{j}(t, A_{\overline{j}}=a^{*}; \eta_{j})=\phi^{\mathrm{RMTL}}_{j}(t, A_{j}=1, A_{\overline{j}}=a^{*}; \eta_{j})-\phi^{\mathrm{RMTL}}_{j}(t, A_{j}=0, A_{\overline{j}}=a^{*}; \eta_{j}) \\
&\phi^{\mathrm{RMTL}}_{j}(t, A_{j}=a, A_{\overline{j}}=a^{*}; \eta_{j}) \\
=&\mathrm{RMTL}_{j}(t \mid a, a^{*}, X)+\frac{I(A=a)}{\pi(a \mid X)}\int_0^{\tau \wedge \widetilde{T}} \frac{(\tau-t) S_{\overline{j}}(t \mid a^{*}, X)\mathrm{d} M_{j}(t \mid a, X)}{S_{\overline{j}}(t \mid a, X)G(t- \mid a, X)} \\
&-\frac{I(A=a)}{\pi(a \mid X)}\int_0^{\tau \wedge \widetilde{T}} \frac{S_{\overline{j}}(t \mid a^{*}, X)\mathrm{d} M(t \mid a, X)}{S_{\overline{j}}(t \mid a, X)S(t \mid a, X)G(t- \mid a, X)}\{\mathrm{RMTL}_{j}(\tau \mid a, X)-\mathrm{RMTL}_{j}(t \mid a, X)\} \\
&+\frac{I(A=a)}{\pi(a \mid X)}\int_0^{\tau \wedge \widetilde{T}} \frac{(\tau-t)S_{\overline{j}}(t \mid a^{*}, X)F_{j}(t \mid a, X)\mathrm{d} M(t \mid a, X)}{S_{\overline{j}}(t \mid a, X)S(t \mid a, X)G(t- \mid a, X)}+\left\{\frac{I(A=a)}{\pi(a \mid X)} -\frac{I(A=a^{*})}{\pi(a^{*} \mid X)} \right\} \\
&\times \int_0^{\tau \wedge \widetilde{T}}\frac{\mathrm{d} M_{\overline{j}}(t \mid a, X)}{S(t \mid a, X)G(t- \mid a, X)}\{\mathrm{RMTL}_{j}(\tau \mid 1-a^{*}, a^{*}, X)-\mathrm{RMTL}_{j}(t \mid A_{j}=1-a^{*}, a^{*}, X)\} \\
&-\left\{\frac{I(A=a)}{\pi(a \mid X)} -\frac{I(A=a^{*})}{\pi(a^{*} \mid X)} \right\}\int_0^{\tau \wedge \widetilde{T}}\frac{(\tau-t) F_{j}(t \mid 1-a^{*}, a^{*}, X)\mathrm{d} M_{\overline{j}}(t \mid a, X)}{S(t \mid a, X)G(t- \mid a, X)}
\end{aligned}
$$
}

{\footnotesize
$$
\begin{aligned}
&\phi^{F}_{j}(t, A_{j}=1-a^{*}; \eta_{j})=\phi^{F}_{j}(t, A_{j}=1-a^{*}, A_{\overline{j}}=1; \eta_{j})-\phi^{F}_{j}(t, A_{j}=1-a^{*}, A_{\overline{j}}=0; \eta_{j}) \\
&\phi^{F}_{j}(t, A_{j}=1-a^{*}, A_{\overline{j}}=a; \eta_{j}) \\
% =&F_{j, 0}(t \mid 1-a^{*}, A, X)+\frac{I(A=a)}{\pi_{0}(a \mid X)}\int_0^{t \wedge \widetilde{T}} \frac{\mathrm{d} M_{j, 0}(u \mid A, X)}{G_{0}(u- \mid A, X)} \left\{1-\frac{S_{\overline{j}, 0}(u \mid a^{*}, X)}{S_{\overline{j}, 0}(u \mid A, X)}\right\}\left[1-\frac{\{F_{j, 0}(t \mid A, X)-F_{j, 0}(u \mid A, X)\}}{S_{0}(u \mid A, X)}\right] \\
% &+\frac{I(A=a)}{\pi_{0}(a \mid X)}\int_0^{t \wedge \widetilde{T}}\frac{\mathrm{d} M_{\overline{j}, 0}(u \mid A, X)}{S_{0}(u \mid A, X)G_{0}(u- \mid A, X)}\left[-F_{j, 0}(t \mid 1-a^{*}, a^{*}, X)+F_{j, 0}(u \mid 1-a^{*}, a^{*}, X) \right. \\
% &\left.-\left\{1-\frac{S_{\overline{j}, 0}(u \mid a^{*}, X)}{S_{\overline{j}, 0}(u \mid A, X)}\right\}\{F_{j, 0}(t \mid A, X)-F_{j, 0}(u \mid A, X)\}\right] \\
=&F_{j}(t \mid 1-a^{*}, a, X)+\frac{I(A=a)}{\pi(a \mid X)}\int_0^{t \wedge \widetilde{T}} \frac{\mathrm{d} M_{j}(u \mid a, X)}{G(u- \mid a, X)}\left\{1-\frac{S_{\overline{j}}(u \mid a^{*}, X)}{S_{\overline{j}}(u \mid a, X)}\right\} \\
&-\frac{I(A=a)}{\pi(a \mid X)}\int_0^{t \wedge \widetilde{T}}\frac{\mathrm{d} M(u \mid a, X)}{S(u \mid a, X)G(u- \mid a, X)}\left\{1-\frac{S_{\overline{j}}(u \mid a^{*}, X)}{S_{\overline{j}}(u \mid a, X)}\right\}\{F_{j}(t \mid a, X)-F_{j}(u \mid a, X)\} \\
&-\left\{\frac{I(A=a)}{\pi(a \mid X)} -\frac{I(A=a^{*})}{\pi(a^{*} \mid X)} \right\} \int_0^{t \wedge \widetilde{T}}\frac{\mathrm{d} M_{\overline{j}}(u \mid a, X)}{S(u \mid a, X)G(u- \mid a, X)}\{F_{j}(t \mid 1-a^{*}, a^{*}, X)-F_{j}(u \mid 1-a^{*}, a^{*}, X)\}
\end{aligned}
$$
}

{\footnotesize
$$
\begin{aligned}
&\phi^{\mathrm{RMTL}}_{j}(t, A_{j}=1-a^{*}; \eta_{j})=\phi^{\mathrm{RMTL}}_{j}(t, A_{j}=1-a^{*}, A_{\overline{j}}=1; \eta_{j})-\phi^{\mathrm{RMTL}}_{j}(t, A_{j}=1-a^{*}, A_{\overline{j}}=0; \eta_{j}) \\
&\phi^{\mathrm{RMTL}}_{j}(t, A_{j}=1-a^{*}, A_{\overline{j}}=a; \eta_{j}) \\
=&\mathrm{RMTL}_{j}(\tau \mid 1-a^{*}, a, X)+\frac{I(A=a)}{\pi(a \mid X)}\int_0^{\tau \wedge \widetilde{T}} \frac{(\tau-t)\mathrm{d} M_{j}(t \mid a, X)}{G(t- \mid a, X)}\left\{1-\frac{S_{\overline{j}}(t \mid a^{*}, X)}{S_{\overline{j}}(t \mid a, X)}\right\} \\
&-\frac{I(A=a)}{\pi(a \mid X)}\int_0^{\tau \wedge \widetilde{T}} \frac{\mathrm{d} M(t \mid a, X)}{S(t \mid a, X)G(t- \mid a, X)}\left\{1-\frac{S_{\overline{j}}(t \mid a^{*}, X)}{S_{\overline{j}}(t \mid a, X)}\right\}\{\mathrm{RMTL}_{j}(\tau \mid a, X)-\{\mathrm{RMTL}_{j}(t \mid a, X)\} \\
&+\frac{I(A=a)}{\pi(a \mid X)}\int_0^{\tau \wedge \widetilde{T}} \frac{(\tau-t)F_{j}(t \mid a, X)\mathrm{d} M(t \mid a, X)}{S(t \mid a, X)G(t- \mid a, X)}\left\{1-\frac{S_{\overline{j}}(t \mid a^{*}, X)}{S_{\overline{j}}(t \mid a, X)}\right\}-\left\{\frac{I(A=a)}{\pi(a \mid X)} -\frac{I(A=a^{*})}{\pi(a^{*} \mid X)} \right\} \\
&\times \int_0^{\tau \wedge \widetilde{T}}\frac{\mathrm{d} M_{\overline{j}}(t \mid a, X)}{S(t \mid a, X)G(t- \mid a, X)}\{\mathrm{RMTL}_{j}(\tau \mid 1-a^{*}, a^{*}, X)-\mathrm{RMTL}_{j}(t \mid 1-a^{*}, a^{*}, X)\} \\
&+\left\{\frac{I(A=a)}{\pi(a \mid X)} -\frac{I(A=a^{*})}{\pi(a^{*} \mid X)} \right\} \int_0^{\tau \wedge \widetilde{T}}\frac{(\tau-t) F_{j}(t \mid 1-a^{*}, a^{*}, X)\mathrm{d} M_{\overline{j}}(t \mid a, X)}{S(t \mid a, X)G(t- \mid a, X)}
\end{aligned}
$$
}

\section{Proof to Theorem~\ref{theorem:cut}} \label{theorem:cut:proof}

Proof: The first property of counterfactual CUTs is the treatment-specific conditional expectation of CUTs reduce to the treatment-specific quantity-of-interest. For (i), under $S_{\infty}(t \mid a, X)=S_0(t \mid a, X)$
{\footnotesize
$$
\begin{aligned} \label{S_cut:demo}
&E_{0}\{Y^{S, \mathrm{BJ}}(t, A; S_{\infty}) \mid A=a, X\}= E_{0}\{Y^{S, \mathrm{BJ}}(t, A; S_{0}) \mid A=a, X\} \\
=&E_{0}\{\frac{S_{0}(t \mid A, X)-\Delta I(\widetilde{T} \leq t) S_{0}(t \mid A, X)}{S_{0}(t \wedge \widetilde{T} \mid A, X)} \mid A=a, X\} \\
=&E_{0}\{\frac{I(\widetilde{T} > t)S_{0}(t \wedge \widetilde{T} \mid A, X)+I(\widetilde{T} \leq t)S_{0}(t \mid A, X)-\Delta I(\widetilde{T} \leq t) S_{0}(t \mid A, X)}{S_{0}(t \wedge \widetilde{T} \mid A, X)} \mid A=a, X\} \\
=&E_{0}\{I(\widetilde{T} > t)+\frac{(1-\Delta) I(\widetilde{T} \leq t) S_{0}(t \mid A, X)}{S_{0}(t \wedge \widetilde{T} \mid A, X)} \mid A=a, X\} \\
=&E_{0}\{I(\widetilde{T} > t) \mid A=a, X\}+E_{0}\{\frac{I(C<T) I(C \leq t) S_{0}(t \mid A, X)}{P_{0}(T>C \mid A=a, X)} \mid A=a, X\} \\
=&S_{0}(t \mid a, X)P_{0}(C > t \mid A=a, X)+S_{0}(t \mid a, X)P_{0}(C \leq t \mid A=a, X)=S_{0}(t \mid a, X)
\end{aligned}
$$
}

% =&I(\widetilde{T}>t)+S_{0}(t \mid A, X) \int_{0}^{t \wedge \widetilde{T}} \frac{\mathrm{d} N_{0}^{C}(u \mid A, X)}{S_{0}(u \mid A, X)} \\
% =&I(\widetilde{T}>t)+\frac{(1-\Delta)I(\widetilde{T} \leq t)S_{0}(t \mid A, X)}{S_{0}(t \wedge \widetilde{T} \mid A, X)}
Under $G_{\infty}(t \mid a, X)=G_0(t \mid a, X)$,
{\footnotesize
$$
\begin{aligned}
&E_{0}\{Y^{S, \mathrm{IPCW1}}(t, A; G_{\infty}) \mid A=a, X\}=E_{0}\{Y^{S, \mathrm{IPCW1}}(t, A; G_{0}) \mid A=a, X\}=E_{0}\{\frac{I(\widetilde{T}>t)}{G_{0}(t \mid A, X)}\mid A=a, X\} \\
=&\frac{E_{0}\{I(T>t)\mid A=a, X\} E_{0}\{I(C>t)\mid A=a, X\}}{G_{0}(t \mid a, X)}=S_{0}(t \mid a, X) \\
&E_{0}\{Y^{S, \mathrm{IPCW2}}(t, A; G_{\infty}) \mid A=a, X\}=E_{0}\{Y^{S, \mathrm{IPCW2}}(t, A; G_{0}) \mid A=a, X\}=E_{0}\{\frac{\Delta I(\widetilde{T}>t)}{G_{0}(\widetilde{T}- \mid A, X)}\mid A=a, X\} \\
=&\frac{E_{0}\{I(T \leq C) \mid A=a, X\} E_{0}\{I(T > t) \mid A=a, X\}}{G_{0}(\widetilde{T}- \mid a, X)}=S_{0}(t \mid a, X)
\end{aligned}
$$
}

Under $S_{\infty}(t \mid a, X)=S_0(t \mid a, X)$,
{\footnotesize
$$
\begin{aligned}
&E_{0}\{Y^{S, \mathrm{AIPCW}}(t, A; (S_{\infty}, G_{\infty})) \mid A=a, X\}=E_{0}\{Y^{S, \mathrm{AIPCW}}(t, A; (S_{0}, G_{\infty})) \mid A=a, X\} \\
=&E_{0}\{S_{0}(t \mid A, X)-S_{0}(t \mid A, X)\int_0^{t \wedge \widetilde{T}} \frac{\mathrm{d} M_{0}(u \mid A, X)}{S_{0}(u \mid A, X) G_{\infty}(u- \mid A, X)}\mid A=a, X\}=S_{0}(t \mid a, X)
\end{aligned}
$$
}

Under $G_{\infty}(t \mid a, X)=G_{0}(t \mid a, X)$,
{\footnotesize
$$
\begin{aligned}
&E_{0}\{Y^{S, \mathrm{AIPCW}}(t, A; (S_{\infty}, G_{\infty})) \mid A=a, X\}=E_{0}\{Y^{S, \mathrm{AIPCW}}(t, A; (S_{\infty}, G_{0})) \mid A=a, X\} \\
=&E_{0}\{\frac{I(\widetilde{T}>t)}{G_{0}(t \mid A, X)}+S_{\infty}(t \mid A, X) \int_{0}^{t \wedge \widetilde{T}} \frac{\mathrm{d} M_{0}^{C}(u \mid A, X)}{S_{\infty}(u \mid A, X) G_{0}(u- \mid A, X)} \mid A=a, X\} \\
=&E_{0}\{\frac{I(\widetilde{T}>t)}{G_{0}(t \mid A, X)} \mid A=a, X\}=S_{0}(t \mid a, X)
\end{aligned}
$$
}
By~\Cref{assumption:positivity}-\Cref{assumption:depcens_survival}, $S_{0}(t \mid a, X)=P_{0}(T^{a}>t \mid X)$.

For (iii), under $S_{\infty}(t \mid a, X)=S_{0}(t \mid a, X)$ and $\Lambda_{j, \infty}(t \mid a, X)=\Lambda_{j, 0}(t \mid a, X)$,
{\footnotesize
$$
\begin{aligned} % \label{S_cut:demo}
&E_{0}\{Y_{j}^{F, \mathrm{BJ}}(t, A; (S_{\infty}, \Lambda_{j, \infty})) \mid A=a, X\}= E_{0}\{Y_{j}^{F, \mathrm{BJ}}(t, A; (S_{0}, \Lambda_{j, 0})) \mid A=a, X\} \\
=& E_{0}\{I(\widetilde{T} \leq t, \widetilde{J}=j)+\frac{(1-\Delta)I(\widetilde{T} \leq t) \{F_{j, 0}(t \mid A, X)-F_{j, 0}(t \wedge \widetilde{T} \mid A, X)\}}{S_{0}(t \wedge \widetilde{T} \mid A, X)} \mid A=a, X\} \\
=&F_{j, 0}(t \mid a, X)P_{0}(C \geq T \mid A=a, X)+E_{0}\{\frac{I(C<T) I(C \leq t) \{F_{j}(t \mid a, X)-P_{0}(T \leq C, J=j \mid A=a, X)\}}{P_{0}(T>C \mid A=a, X)} \mid A=a, X\} \\
=&F_{j, 0}(t \mid a, X)P_{0}(C \geq T \mid A=a, X)+P_{0}(C \leq t \mid A=a, X) \{F_{j, 0}(t \mid a, X)-P_{0}(T \leq C, J=j \mid A=a, X)\} \\
=&F_{j, 0}(t \mid a, X)P_{0}(C \geq T \mid A=a, X)+P_{0}(C \leq t \mid A=a, X) F_{j, 0}(t \mid a, X)-P_{0}(T \leq C \leq t, J=j \mid A=a, X) \\
=&F_{j, 0}(t \mid a, X)P_{0}(C \geq T \mid A=a, X)+P_{0}(C \leq t \mid A=a, X) F_{j, 0}(t \mid a, X)-P_{0}(T \leq t, J=j \mid A=a, X) \\
&\times \{P_{0}(C \leq t \mid A=a, X)-P_{0}(C < T \mid A=a, X)\}=F_{j, 0}(t \mid a, X)
\end{aligned}
$$
}

Under $G_{\infty}(t \mid a, X)=G_{0}(t \mid a, X)$,
{\footnotesize
$$
\begin{aligned}
&E_{0}\{Y^{F, \mathrm{IPCW}}(t, A; G_{\infty}) \mid A=a, X\}=E_{0}\{Y^{F, \mathrm{IPCW}}(t, A; G_{0}) \mid A=a, X\}=E_{0}\{\frac{I(\widetilde{T} \leq t, \widetilde{J}=j)}{G_{0}(\widetilde{T}- \mid A, X)}\mid A=a, X\} \\
=&\frac{E_{0}\{\Delta I(T \leq t, J=j)\mid A=a, X\}}{G_{0}(\widetilde{T}- \mid A, X)}=F_{j, 0}(t \mid a, X)
\end{aligned}
$$
}

Under $S_{\infty}(t \mid a, X)=S_{0}(t \mid a, X)$ and $\Lambda_{j, \infty}(t \mid a, X)=\Lambda_{j, 0}(t \mid a, X)$,
{\footnotesize
$$
\begin{aligned}
&E_{0}\{Y_{j}^{F, \mathrm{AIPCW}}(t, A; (S_{\infty}, \Lambda_{j, \infty}, G_{\infty})) \mid A=a, X\}=E_{0}\{Y_{j}^{F, \mathrm{AIPCW}}(t, A; (S_{0}, \Lambda_{j, 0}, G_{\infty})) \mid A=a, X\} \\
=&E_{0}\{ F_{j, 0}(t \mid A, X)-\int_0^{t \wedge \widetilde{T}} \frac{\{F_{j, 0}(t \mid A, X)-F_{j, 0}(u \mid A, X)\}\mathrm{d} M_{0}(u \mid A, X)}{S_{0}(u \mid A, X) G_{\infty}(u- \mid A, X)}+\int_0^{t \wedge \widetilde{T}} \frac{\mathrm{d} M_{j, 0}(u \mid A, X)}{G_{\infty}(u-\mid A, X)}\mid A=a, X\} \\
=& F_{j, 0}(t \mid a, X)
\end{aligned}
$$
}

Under $G_{\infty}(t \mid a, X)=G_{0}(t \mid a, X)$,
{\footnotesize
$$
\begin{aligned}
&E_{0}\{Y_{j}^{F, \mathrm{AIPCW}}(t, A; (S_{\infty}, \Lambda_{j, \infty}, G_{\infty})) \mid A=a, X\}=E_{0}\{Y_{j}^{F, \mathrm{AIPCW}}(t, A; (S_{\infty}, \Lambda_{j, \infty}, G_{0})) \mid A=a, X\} \\
=&E_{0}\{\frac{I(\widetilde{T} \leq t, \widetilde{J}=j)}{G_{0}(\widetilde{T}-\mid A, X)}+\int_0^{t \wedge \widetilde{T}} \frac{\{F_{j, \infty}(t \mid A, X)-F_{j, \infty}(u \mid A, X)\} \mathrm{d} M^C_{0}(u \mid A, X)}{S_{\infty}(u \mid A, X) G_{0}(u-\mid A, X)} \mid A=a, X\} \\
=&E_{0}\{\frac{I(\widetilde{T} \leq t, \widetilde{J}=j)}{G_{0}(\widetilde{T}-\mid A, X)} \mid A=a, X\}=F_{j, 0}(t \mid a, X)
\end{aligned}
$$
}

By~\Cref{assumption:positivity} and~\Cref{assumption:SUTVA_competing}-\Cref{assumption:depcens_competing}, $F_{j, 0}(t \mid a, X)=P_{0}(T^{a} \leq t, J^{a}=j \mid X)$.

For (v), when $A_{j}=a^{*}$, the CUT reduces to (iii). Here, we discuss the case $A_{j} = 1-a^{*}$. Under $\Lambda_{j, \infty}(t \mid a, X)=\Lambda_{j, 0}(t \mid a, X)$ and $\Lambda_{\overline{j}, \infty}(t \mid a, X)=\Lambda_{\overline{j}, 0}(t \mid a, X)$,
{\footnotesize
$$
\begin{aligned}
&E_{0}\{Y_{j}^{F, \mathrm{AIPCW}}(t, A_{j}=A, A_{\overline{j}}=a^{*}; (\Lambda_{j, \infty}, \Lambda_{\overline{j}, \infty}, G_{\infty})) \mid A=a, X\} \\
=&E_{0}\{Y_{j}^{F, \mathrm{AIPCW}}(t, A_{j}=A, A_{\overline{j}}=a^{*}; (\Lambda_{j, 0}, \Lambda_{\overline{j}, 0}, G_{\infty})) \mid A=a, X\} \\
=&E_{0} \left\{F_{j, 0}(t \mid a, a^{*}, X)+\int_0^{t \wedge \widetilde{T}} \frac{S_{\overline{j}, 0}(u \mid a^{*}, X)\mathrm{d} M_{j, 0}(u \mid a, X)}{S_{\overline{j}, 0}(u \mid a, X)G_{\infty}(u- \mid a, X)} \right. \\
&-\int_0^{t \wedge \widetilde{T}} \frac{S_{\overline{j}, 0}(u \mid a^{*}, X)\mathrm{d} M_{0}(u \mid a, X)}{S_{\overline{j}, 0}(u \mid a, X)S_{0}(u \mid a, X)G_{\infty}(u- \mid a, X)}\{F_{j, 0}(t \mid a, X)-F_{j, 0}(u \mid a, X)\} +I(A=1-a^{*}) \\
&\left. \times \int_0^{t \wedge \widetilde{T}}\frac{\mathrm{d} M_{\overline{j}, 0}(u \mid a, X)}{S_{0}(u \mid a, X)G_{\infty}(u- \mid a, X)}\{F_{j, 0}(t \mid 1-a^{*}, a^{*}, X)-F_{j, 0}(u \mid 1-a^{*}, a^{*}, X)\} \mid A=a, X\right\} \\
=&F_{j, 0}(t \mid a, a^{*}, X)
% =&E_{0}\left\{F_{j, 0}(t \mid a, a^{*}, X)+\int_0^{t \wedge \widetilde{T}} \frac{S_{\overline{j}, 0}(u \mid a^{*}, X)\mathrm{d} M_{j, 0}(u \mid a, X)}{S_{\overline{j}, 0}(u \mid a, X)G(u- \mid a, X)}\left[1-\frac{\{F_{j, 0}(t \mid a, X)-F_{j, 0}(u \mid a, X)\}}{S_{0}(u \mid a, X)}\right]\right. \\
% &+\int_0^{t \wedge \widetilde{T}}\frac{\mathrm{d} M_{\overline{j}, 0}(u \mid a, X)}{S_{0}(u \mid a, X)G(u- \mid a, X)}\left[F_{j, 0}(t \mid 1-a^{*}, a^{*}, X)-F_{j, 0}(u \mid 1-a^{*}, a^{*}, X) \right. \\
% &\left.-\frac{S_{\overline{j}, 0}(u \mid a^{*}, X)}{S_{\overline{j}, 0}(u \mid a, X)}\{F_{j, 0}(t \mid a, X)-F_{j, 0}(u \mid a, X)\} ] \mid A=a, X\right\}=F_{j, 0}(t \mid a, a^{*}, X)
\end{aligned}
$$
}
By~\Cref{assumption:positivity} and~\Cref{assumption:SUTVA_competing}-\Cref{assumption:dismissible}, $F_{j, 0}(t \mid a, a^{*}, X)=P_{0}(T^{a, a^{*}} \leq t, J^{a, a^{*}}=j \mid X)$. \\

Note that the consistency conditions for $Y_{j}^{F, \mathrm{AIPCW}}(t, A_{j}=a, A_{\overline{j}}=a^{*}; \eta_{j})$ is the same as the consistency conditions~\citep{martinussen2023estimation}, but such discussion is pointless since this counterfactual CUT will not even be defined. It's straightforward to see that when $\pi(\cdot \mid X)$ appear in $\frac{I(A=a^{*}) \pi(a \mid X)}{\pi(a^{*} \mid X) I(A=a)}$, we need $a=1-a^{*}$ and  $A=a^{*}$, but this implies $A=1-a$, leading to a zero in the denominator. \\

(vi), (vii), and (viii) follow similar arguments.

\section{Details on ~\Cref{remark:consistency}}

Here, we enumerate the consistency conditions for counterfactual CUTs and IF transformations for each estimand from~\Cref{theorem:cut} as discussed in~\Cref{remark:consistency}

For (i) and (ii) in~\Cref{theorem:cut}, $E_{0}\{Y^{S, \mathrm{AIPCW}}(t, A; \eta_{\infty}^{\Lambda}) \mid A=a, X\}=P_{0}(T^a>t \mid X)$ and $E_{0}\{Y^{\mathrm{RMST}, \mathrm{AIPCW}}(\tau, A; \eta_{\infty}^{\Lambda}) \mid A=a, X\}=E_{0}\{\min(T^a, \tau) \mid X\}$ if either $S_{\infty}=S_{0}$ or $G_{\infty}=G_{0}$, while $E_{0}\{\phi^{S}(t; \eta_{\infty}) \mid X\}=\psi_{0}^{S}(t, X)$ and $E_{0}\{\phi^{\mathrm{RMST}}(\tau; \eta_{\infty}) \mid X\}=\psi_{0}^{\mathrm{RMST}}(\tau, X)$ if either $S_{\infty}=S_{0}$ or both $G_{\infty}=G_{0}$ and $\pi_{\infty}=\pi_{0}$; \\

for (iii) and (iv) in~\Cref{theorem:cut}, $E_{0}\{Y_{j}^{F, \mathrm{AIPCW}}(t, A; \eta_{j, \infty}^{\Lambda}) \mid A=a, X\}=P_{0}(T^{a} \leq t, J^{a}=j \mid X)$ and $E_{0}\{Y_{j}^{\mathrm{RMTL}, \mathrm{AIPCW}}(\tau, A; \eta_{j, \infty}^{\Lambda}) \mid A=a, X\}=E_{0}[\{\tau-\min (T^a, \tau)\} I(J^a=j) \mid X]$ if either $S_{\infty}=S_{0}$ and $\Lambda_{j, \infty}=\Lambda_{j, 0}$ or $G_{\infty}=G_{0}$, while $E_{0}\{\phi^{F}_{j}(t; \eta_{\infty}) \mid X\}=\psi_{0, j}^{F}(t, X)$ and $E_{0}\{\phi_{j}^{\mathrm{RMTL}}(\tau; \eta_{\infty}) \mid X\}=\psi_{0, j}^{\mathrm{RMTL}}(\tau, X)$ if  either $S_{\infty}=S_{0}$ and $F_{j, \infty}=F_{j, 0}$ or $G_{\infty}=G_{0}$ and $\pi_{\infty}=\pi_{0}$ or $G_{\infty}=G_{0}$ and $F_{j, \infty}=F_{j, 0}$~\citep{ozenne2020estimation};

\section{\Cref{proposition:augmented} extension} \label{proposition:augmented:extension}

Plugging in the AIPCW CUT for RMST into the AIPTW transformation yields
{\footnotesize
$$
\begin{aligned}
&\frac{I(A=a)Y^{\mathrm{RMST}, \mathrm{AIPCW}}(\tau, A; \eta_{0}^{\Lambda})}{\pi_{0}(a \mid X)}+\left\{1-\frac{I(A=a)}{\pi_{0}(a \mid X)}\right\}E_{0}\{Y_{0}^{\mathrm{RMST}, \mathrm{AIPCW}}(\tau, A; \eta_{0}^{\Lambda}) \mid A=a, X\} \\
=&\frac{I(A=a)\Delta(\tau)\min(\widetilde{T}, \tau)}{\pi_{0}(a \mid X)G_{0}(\tau \wedge \widetilde{T}- \mid a, X)}+\left\{1-\frac{I(A=a)}{\pi_{0}(a \mid X)}\right\} \mathrm{RMST}_{0}(\tau \mid a, X)+\frac{I(A=a)}{\pi_{0}(a \mid X)} \int_{0}^{\tau \wedge \widetilde{T}} \frac{t \mathrm{d} M_{0}^{C}(t \mid a, X)}{G_{0}(t- \mid a, X)} \\
&+\frac{I(A=a)}{\pi(a \mid X)}\int_{0}^{\tau \wedge \widetilde{T}} \frac{\{\mathrm{RMST}_{0}(\tau \mid a, X)-\mathrm{RMST}_{0}(t \mid a, X)\} \mathrm{d} M_{0}^{C}(t \mid a, X)}{S_{0}(t \mid a, X)G_{0}(t- \mid a, X)}=\phi^{\mathrm{RMST}}(\tau, a; \eta_{0})
\end{aligned}
$$
}
where we used
{\footnotesize
$$
\begin{aligned}
&E_{0}\{Y^{\mathrm{RMST}, \mathrm{AIPCW}}(\tau, A; \eta_{0}^{\Lambda}) \mid A=a, X\} \\
=&E_{0} \left\{\frac{\Delta(\tau)\min(\widetilde{T}, \tau)}{G_{0}(\tau \wedge \widetilde{T}- \mid A, X)} \mid A=a, X\right\}+E_{0} \left\{\int_{0}^{\tau \wedge \widetilde{T}} \frac{t \mathrm{d} M_{0}^{C}(t \mid A, X)}{G_{0}(t- \mid A, X)} \right. \\
&\left.+\int_{0}^{\tau \wedge \widetilde{T}} \frac{\{\mathrm{RMST}_{0}(\tau \mid A, X)-\mathrm{RMST}_{0}(t \mid A, X)\} \mathrm{d} M_{0}^{C}(t \mid A, X)}{S_{0}(t \mid A, X)G_{0}(t- \mid A, X)} \mid A=a, X \right\} \\
=&E_{0}\left\{\frac{I(C > \tau \wedge \widetilde{T}-)\min(T, \tau)}{G_{0}(\tau \wedge \widetilde{T}- \mid A, X)} \mid A=a, X\right\}=\mathrm{RMST}_{0}(\tau \mid a, X)
\end{aligned}
$$
}
Similar steps apply to all estimands.

\section{Proof to Theorem~\ref{theorem:equivalence}}

Proof: Without loss of generality, we consider the quantity $\psi_{0}^{\mathrm{RMST}}(\tau, X)$ under AIPCW since IPCW and BJ are its special cases. Its AIPCW Brier squared orthogonal loss~\citep{ausset2022empirical,kvamme2023brier} takes the form
{\footnotesize
\begin{equation}\label{theorem:equivalence:step1}
\begin{aligned}
\ell_{2}^{\mathrm{AIPCW}, *}(\min(T, \tau), \psi_{0}^{\mathrm{RMST}}(\tau, X); \eta_{0})=&\frac{\Delta w^{*}(\pi_{0}) \{Y^{*}(\min(T, \tau); \eta_{0})-\psi_{0}^{\mathrm{RMST}}(\tau, X)\}^2}{G_{0}(\widetilde{T}- \mid A, X)}+\int_{0}^{\infty} \frac{\mathrm{d}M_{0}^C(t \mid A, X)}{G_{0}(t- \mid A, X)} \\
&\times E_{0}[w^{*}(\pi_{0}) \{Y^{*}(\min(T, \tau); \eta_{0})-\psi_{0}^{\mathrm{RMST}}(\tau, X)\}^2 \mid A, X, T>t]
\end{aligned}
\end{equation}
}
However, this AIPCW Brier loss function is not optimizable directly as the second term integrates to infinity. We first simplify it to an observed loss function. Note that
{\footnotesize
\begin{equation}\label{theorem:equivalence:step2}
\begin{aligned}
&\int_{0}^{\infty} \frac{\mathrm{d}M_{0}^C(t \mid A, X)}{G_{0}(t- \mid A, X)} E_{0}[w^{*}(\pi_{0}) \{Y^{*}(\min(T, \tau); \eta_{0})-\psi_{0}^{\mathrm{RMST}}(\tau, X)\}^2 \mid A, X, T>t] \\
=&\int_{0}^{\tau} (t \leq \tau) \frac{\mathrm{d}M_{0}^C(t \mid A, X)}{G_{0}(t- \mid A, X)} E_{0}[w^{*}(\pi_{0}) \{Y^{*}(\min(T, \tau); \eta_{0})-\psi_{0}^{\mathrm{RMST}}(\tau, X)\}^2 \mid A, X, T>t] \\
&+\int_{\tau+}^{\infty} (t >\tau) \frac{\mathrm{d}M_{0}^C(t \mid A, X)}{G_{0}(t- \mid A, X)} E_{0}[w^{*}(\pi_{0}) \{Y^{*}(\min(T, \tau); \eta_{0})-\psi_{0}^{\mathrm{RMST}}(\tau, X)\}^2 \mid A, X, T>t]
\end{aligned}
\end{equation}
}

and
{\footnotesize
\begin{equation}\label{theorem:equivalence:step3}
\begin{aligned}
&\int_{\tau+}^{\infty} (t >\tau) \frac{\mathrm{d}M_{0}^C(t \mid A, X)}{G_{0}(t- \mid A, X)} E_{0}[w^{*}(\pi_{0}) \{Y^{*}(\min(T, \tau); \eta_{0})-\psi_{0}^{\mathrm{RMST}}(\tau, X)\}^2 \mid A, X, T>t] \\
=&w^{*}(\pi_{0})\{Y^{*}(\tau, \eta_{0})-\psi_{0}^{\mathrm{RMST}}(\tau, X)\}^2 \int_{\tau+}^{\infty} \frac{\mathrm{d}M_{0}^C(t \mid A, X)}{G_{0}(t- \mid A, X)} \\
=&w^{*}(\pi_{0})\{Y^{*}(\min(\widetilde{T}, \tau), \eta_{0})-\psi_{0}^{\mathrm{RMST}}(\tau, X)\}^2 \left\{\frac{I(\widetilde{T}>\tau)}{G_{0}(\tau \mid A, X)}-\frac{\Delta I(\widetilde{T}>\tau)}{G_{0}(\widetilde{T}-\mid A, X)}\right\} \\
\end{aligned}
\end{equation}
}

since
{\footnotesize
$$
\begin{aligned}
&\int_{\tau+}^{\infty} \frac{\mathrm{d}M_{0}^C(t \mid A, X)}{G_{0}(t- \mid A, X)}=\int_{\tau+}^{\infty} \frac{\mathrm{d} N^C(t)}{G_{0}(t- \mid A, X)}-\int_{\tau+}^{\infty} \frac{R(t) \mathrm{d} \Lambda_{0}^C(t \mid A, X)}{G_{0}(t- \mid A, X)} \\
=&\frac{(1-\Delta)I(\widetilde{T}>\tau)}{G_{0}(\widetilde{T}-\mid A, X)}-\left.\frac{I(\widetilde{T}>\tau)}{G_{0}(t- \mid A, X)}\right|_{\tau+}^{\widetilde{T}}=\frac{I(\widetilde{T}>\tau)}{G_{0}(\tau \mid A, X)}-\frac{\Delta I(\widetilde{T}>\tau)}{G_{0}(\widetilde{T}-\mid A, X)}
\end{aligned}
$$
}

Plug~\eqref{theorem:equivalence:step2} and~\eqref{theorem:equivalence:step3} into~\eqref{theorem:equivalence:step1} yields
{\footnotesize
\begin{equation}\label{theorem:equivalence:step4}
\begin{aligned}
&\ell_{2}^{\mathrm{AIPCW}, *}(\psi_{0}^{\mathrm{RMST}}(\tau, X); \eta_{0}) \\
=&\frac{\Delta w^{*}(\pi_{0}) \{Y^{*}(\min(\widetilde{T}, \tau), \eta_{0})-\psi_{0}(\tau, X)\}^2}{G_{0}(\widetilde{T}- \mid A, X)}+w^{*}(\pi_{0})\{Y^{*}(\min(\widetilde{T}, \tau), \eta_{0})-\psi_{0}^{\mathrm{RMST}}(\tau, X)\}^2 \\
&\times \left\{\frac{I(\widetilde{T}>\tau)}{G_{0}(\tau \mid A, X)}-\frac{\Delta I(\widetilde{T}>\tau)}{G_{0}(\widetilde{T}-\mid A, X)}\right\}+\int_{0}^{\tau} \frac{\mathrm{d}M_{0}^C(t \mid A, X)}{G_{0}(t- \mid A, X)} E_{0}[w^{*}(\pi_{0}) \{Y^{*}(\min(T, \tau), \eta_{0})-\psi_{0}^{\mathrm{RMST}}(\tau, X)\}^2 \mid A, X, T>t] \\
=&\frac{\Delta(\tau) w^{*}(\pi_{0}) \{Y^{*}(\min(\widetilde{T}, \tau), \eta_{0})-\psi_{0}(\tau, X)\}^2}{G_{0}(\tau \wedge\widetilde{T}- \mid A, X)} \\
&+\int_{0}^{\tau} \frac{\mathrm{d}M_{0}^C(t \mid A, X)}{G_{0}(t- \mid A, X)} E_{0}[w^{*}(\pi_{0}) \{Y^{*}(\min(T, \tau), \eta_{0})-\psi_{0}^{\mathrm{RMST}}(\tau, X)\}^2 \mid A, X, T>t]
\end{aligned}
\end{equation}
}

where $\Delta(\tau)=I(C>\tau \wedge T-)$ and
{\footnotesize
$$
\begin{aligned}
& \frac{\Delta}{G_{0}(\widetilde{T}-\mid A, X)}+\frac{I(\widetilde{T}>\tau)}{G_{0}(\tau \mid A, X)}-\frac{\Delta I(\widetilde{T}>\tau)}{G_{0}(\widetilde{T}-\mid A, X)}=\frac{\Delta I(\widetilde{T} \leq \tau)}{G_{0}(\tau \wedge \widetilde{T}-\mid A, X)}+\frac{I(\widetilde{T}>\tau)}{G_{0}(\tau \wedge \widetilde{T}-\mid A, X)} \\
= & \frac{I(C>\tau \wedge T-)}{G_{0}(\tau \wedge \widetilde{T}-\mid A, X)}=\frac{\Delta(\tau)}{G_{0}(\tau \wedge \widetilde{T}-\mid A, X)}
\end{aligned}
$$
}
Now, $\ell_{2}^{\mathrm{AIPCW}, *}(\psi_{0}^{\mathrm{RMST}}(\tau, X); \eta_{0})$ has been reduced to observed data orthogonal loss function. Expanding the squared bracket of~\eqref{theorem:equivalence:step4}, we have
{\footnotesize
$$
\begin{aligned}
&\ell_{2}^{\mathrm{AIPCW}, *}(\psi_{0}^{\mathrm{RMST}}(\tau, X); \eta_{0}) \\
=&w^{*}(\pi_{0}) \left(\frac{\Delta(\tau) Y^{*, 2}(\min(\widetilde{T}, \tau), \eta_{0})}{G_{0}(\tau \wedge\widetilde{T}- \mid A, X)}+\int_{0}^{\tau} \frac{\mathrm{d}M_{0}^C(t \mid A, X)}{G_{0}(t- \mid A, X)} E_{0}\{Y^{*, 2}(\min(T, \tau), \eta_{0}) \mid A, X, T>t\} \right. \\
&-2\psi_{0}^{\mathrm{RMST}}(\tau, X) \left[\frac{\Delta(\tau) Y^{*}(\min(\widetilde{T}, \tau), \eta_{0})}{G_{0}(\tau \wedge\widetilde{T}- \mid A, X)}+\int_{0}^{\tau} \frac{\mathrm{d}M_{0}^C(t \mid A, X)}{G_{0}(t- \mid A, X)} E_{0}\{Y^{*}(\min(T, \tau), \eta_{0}) \mid A, X, T>t\} \right] \\
&+\left. \psi_{0}^{\mathrm{RMST}, 2}(\tau, X) \left\{ \frac{\Delta(\tau)}{G_{0}(\tau \wedge\widetilde{T}- \mid A, X)}+\int_{0}^{\tau} \frac{\mathrm{d}M_{0}^C(t \mid A, X)}{G_{0}(t- \mid A, X)} \right\} \right)
\end{aligned}
$$
}
Note that
{\footnotesize
$$
\frac{\Delta(\tau)}{G_{0}(\tau \wedge\widetilde{T}- \mid A, X)}+\int_{0}^{\tau} \frac{\mathrm{d}M_{0}^C(t \mid A, X)}{G_{0}(t- \mid A, X)}=1
$$
}
Now, we consider the term
{\footnotesize
$$
\begin{aligned}
&\int_{0}^{\tau} \frac{\mathrm{d}M_{0}^C(t \mid A, X)}{G_{0}(t- \mid A, X)} E_{0}\{Y^{*}(\min(T, \tau), \eta_{0}) \mid A, X, T>t\} \\
=& \int_{t}^\infty Y^{*}(\min(u, \tau), \eta_{0}) \mathrm{d} P_{0}(T \leq u \mid A, X, T>t) \\
=&\int_t^\infty Y^{*}(\min(u, \tau), \eta_{0}) \mathrm{d} \left\{ \frac{P_{0}(t < T \leq u \mid A, X)}{P_{0}(T > t \mid A, X)} \right\} \\
=& \int_t^\infty Y^{*}(\min(u, \tau), \eta_{0}) \mathrm{d} \left\{ \frac{S_{0}(t \mid A, X)-S_{0}(u \mid A, X)}{S_{0}(t \mid A, X)} \right\}= \frac{\int_t^\infty Y^{*}(\min(u, \tau), \eta_{0}) \{-\mathrm{d} S_{0}(u \mid A, X)\}}{S_{0}(t \mid A, X)} \\
=& \frac{\int_{t}^{\tau} Y^{*}(u, \eta_{0})\{-\mathrm{d} S_{0}(u \mid A, X)\}+\int_{\tau+}^{\infty} Y^{*}(\tau, \eta_{0}) \{-\mathrm{d} S_{0}(u \mid A, X)\}}{S_{0} (t \mid A, X)} \\
=&\frac{-Y^{*}(\tau, \eta_{0})S_{0}(\tau \mid A, X)+Y^{*}(t, \eta_{0})S_{0}(t \mid A, X)+\int_{t}^{\tau}S_{0}(u \mid A, X) \mathrm{d}Y^{*}(u, \eta_{0}) +Y^{*}(\tau, \eta_{0}) S_{0}(\tau \mid A, X)}{S_{0}(t \mid A, X)} \\
=&Y^{*}(t, \eta_{0})+\frac{\mathrm{d}Y^{*}(u, \eta_{0})}{\mathrm{d} u} \frac{\int_{t}^{\tau}S_{0}(u \mid A, X) \mathrm{d}u}{S_{0}(t \mid A, X)} \\
=&Y^{*}(t, \eta_{0})+\frac{\mathrm{d}Y^{*}(u, \eta_{0})}{\mathrm{d} u} \frac{\mathrm{RMST}_{0}(\tau \mid A, X)-\mathrm{RMST}_{0}(t \mid A, X)}{S_{0}(t \mid A, X)}
\end{aligned}
$$
}
such that
{\footnotesize
\begin{equation}\label{theorem:equivalence:step5}
\begin{aligned}
&\frac{\Delta(\tau) Y^{*}(\min(\widetilde{T}, \tau), \eta_{0})}{G_{0}(\tau \wedge\widetilde{T}- \mid A, X)}+\int_{0}^{\tau} \frac{\mathrm{d}M_{0}^C(t \mid A, X)}{G_{0}(t- \mid A, X)} E_{0}\{Y^{*}(\min(T, \tau), \eta_{0}) \mid A, X, T>t\} \\
=&\frac{\Delta(\tau) Y^{*}(\min(\widetilde{T}, \tau), \eta_{0})}{G_{0}(\tau \wedge\widetilde{T}- \mid A, X)}+\int_{0}^{\tau} \frac{Y^{*}(t, \eta_{0}) \mathrm{d}M_{0}^C(t \mid A, X)}{G_{0}(t- \mid A, X)}+\frac{\mathrm{d}Y^{*}(u, \eta_{0})}{\mathrm{d} u}\int_{0}^{\tau} \frac{\mathrm{d}M_{0}^C(t \mid A, X)}{S_{0}(t \mid A, X) G_{0}(t- \mid A, X)} \\
& \times \{\mathrm{RMST}_{0}(\tau \mid A, X)-\mathrm{RMST}_{0}(t \mid A, X)\}
\end{aligned}
\end{equation}
}
Then we consider the squared orthogonal loss
$\ell_{2}^{*}(Y^{\mathrm{RMST, AIPCW}}(\tau, A; \eta_{0}^{\Lambda}), \psi_{0}^{\mathrm{RMST}}(\tau, X); (\mu_{0}, \pi_{0}))$ with plugged in AIPCW CUT
{\footnotesize
$$
\begin{aligned}
&\ell_{2}^{*}(Y^{\mathrm{RMST, AIPCW}}(\tau, A; \eta_{0}^{\Lambda}), \psi_{0}^{\mathrm{RMST}}(\tau, X); (\mu_{0}, \pi_{0})) \\
=& w^{*}(\pi_{0}) \{Y^{*}(Y^{\mathrm{RMST}, \mathrm{AIPCW}}(\tau, A; \eta_{0}^{\Lambda}); (\mu_{0}, \pi_{0}))-\psi_{0}^{\mathrm{RMST}}(\tau, X)\}^2 \\
=& w^{*}(\pi_{0}) \{Y^{*, 2}(Y^{\mathrm{RMST}, \mathrm{AIPCW}}(\tau, A; \eta_{0}^{\Lambda}); (\mu_{0}, \pi_{0}))-2\psi_{0}^{\mathrm{RMST}}(\tau, X) Y^{*}(Y^{\mathrm{RMST}, \mathrm{AIPCW}}(\tau, A; \eta_{0}^{\Lambda}); (\mu_{0}, \pi_{0}))+\psi_{0}^{\mathrm{RMST}, 2}(\tau, X)\}
\end{aligned}
$$
}

where
{\footnotesize
$$
\begin{aligned}
&Y^{*}\{Y^{\mathrm{RMST}, \mathrm{AIPCW}}(\tau, A; \eta_{0}^{\Lambda}); (\mu_{0}, \pi_{0})\}=Y^{*}\left\{\frac{\Delta(\tau)\min(\widetilde{T}, \tau)}{G_{0}(\tau \wedge \widetilde{T}- \mid A, X)}+\int_{0}^{\tau \wedge \widetilde{T}} \frac{t \mathrm{d} M_{0}^{C}(t \mid A, X)}{G_{0}(t- \mid A, X)} \right. \\
&+\left. \int_{0}^{\tau \wedge \widetilde{T}} \frac{\{\mathrm{RMST}_{0}(\tau \mid A, X)-\mathrm{RMST}_{0}(t \mid A, X)\} \mathrm{d} M_{0}^{C}(t \mid A, X)}{S_{0}(t \mid A, X)G_{0}(t- \mid A, X)}; (\mu_{0}, \pi_{0})\right\} \\
=&Y^{*}\left\{\frac{\Delta(\tau)\min(\widetilde{T}, \tau)}{G_{0}(\tau \wedge \widetilde{T}- \mid A, X)}; \eta_{0}\right\}+Y^{*}\left\{\int_{0}^{\tau \wedge \widetilde{T}} \frac{t \mathrm{d} M_{0}^{C}(t \mid A, X)}{G_{0}(t- \mid A, X)}; \eta_{0}\right\} \\
&+Y^{*}\left[\int_{0}^{\tau \wedge \widetilde{T}} \frac{\{\mathrm{RMST}_{0}(\tau \mid A, X)-\mathrm{RMST}_{0}(t \mid A, X)\} \mathrm{d} M_{0}^{C}(t \mid A, X)}{S_{0}(t \mid A, X)G_{0}(t- \mid A, X)}; \eta_{0}\right]
\end{aligned}
$$
}
With
{\footnotesize
$$
E_{0}\left\{\frac{\Delta(\tau)}{G_{0}(\tau \wedge \widetilde{T}- \mid A, X)}\right\}=1
$$
}
it is easy to verify that
{\footnotesize
\begin{equation}\label{theorem:equivalence:step6}
E_{0}\left\{Y^{*}\left(\frac{\Delta(\tau)\min(\widetilde{T}, \tau)}{G_{0}(\tau \wedge \widetilde{T}- \mid A, X)}; \eta_{0}\right)\right\}=E_{0}\left\{\frac{\Delta(\tau) Y^{*}(\min(\widetilde{T}; \tau), \eta_{0})}{G_{0}(\tau \wedge\widetilde{T}- \mid A, X)}\right\}
\end{equation}
}
Since
{\footnotesize
$$
\begin{aligned}
\int_{0}^{\tau} \frac{E_{0}(t \mid X) \mathrm{d} M_{0}^{C}(t \mid A, X)}{G_{0}(t- \mid A, X)}=&\pi_{0}(0 \mid X)E_{0} \left\{\int_{0}^{\tau} \frac{t \mathrm{d} M_{0}^{C}(t \mid A, X)}{G_{0}(t- \mid A, X)} \mid A=0, X\right\} \\
&+\pi_{0}(1 \mid X)E_{0} \left\{\int_{0}^{\tau} \frac{t \mathrm{d} M_{0}^{C}(t \mid A, X)}{G_{0}(t- \mid A, X)} \mid A=1, X\right\}=0
\end{aligned}
$$
}
then
{\footnotesize
\begin{equation}\label{theorem:equivalence:step7}
Y^{*}\left(\int_{0}^{\tau \wedge \widetilde{T}} \frac{t \mathrm{d} M_{0}^{C}(t \mid A, X)}{G_{0}(t- \mid A, X)}; \eta_{0}\right)=\int_{0}^{\tau} \frac{Y^{*}(t; \eta_{0}) \mathrm{d}M_{0}^C(t \mid A, X)}{G_{0}(t- \mid A, X)}
\end{equation}
}
Similarly, because of
{\footnotesize
$$
E_{0}\left[\int_{0}^{\tau \wedge \widetilde{T}} \frac{\{\mathrm{RMST}_{0}(\tau \mid A, X)-\mathrm{RMST}_{0}(t \mid A, X)\} \mathrm{d} M_{0}^{C}(t \mid A, X)}{S_{0}(t \mid A, X)G_{0}(t- \mid A, X)} \mid X \right]=0
$$
}
and the outcome transformation satisfies
{\footnotesize
$$
\frac{\mathrm{d}Y^{*}(Y; \eta_{0})}{\mathrm{d} Y} Y = Y^{*}(Y; \eta_{0})
$$
}
we have
{\footnotesize
\begin{equation}\label{theorem:equivalence:step8}
\begin{aligned}
&Y^{*}\left[\int_{0}^{\tau \wedge \widetilde{T}} \frac{\{\mathrm{RMST}_{0}(\tau \mid A, X)-\mathrm{RMST}_{0}(t \mid A, X)\} \mathrm{d} M_{0}^{C}(t \mid A, X)}{S_{0}(t \mid A, X)G_{0}(t- \mid A, X)}; \eta_{0}\right] \\
=& \frac{\mathrm{d}Y^{*}(u; \eta_{0})}{\mathrm{d} u}\int_{0}^{\tau} \frac{\{\mathrm{RMST}_{0}(\tau \mid A, X)-\mathrm{RMST}_{0}(t \mid A, X)\}\mathrm{d}M_{0}^C(t \mid A, X)}{S_{0}(t \mid A, X) G_{0}(t- \mid A, X)}
\end{aligned}
\end{equation}
}
By summing up \eqref{theorem:equivalence:step6}+\eqref{theorem:equivalence:step7}+\eqref{theorem:equivalence:step8}, we have
{\footnotesize
$$
\begin{aligned}
&\frac{\Delta(\tau) Y^{*}(\min(\widetilde{T}, \tau); \eta_{0})}{G_{0}(\tau \wedge\widetilde{T}- \mid A, X)}+\int_{0}^{\tau} \frac{Y^{*}(t; \eta_{0}) \mathrm{d}M_{0}^C(t \mid A, X)}{G_{0}(t- \mid A, X)}+\frac{\mathrm{d}Y^{*}(u; \eta_{0})}{\mathrm{d} u}\int_{0}^{\tau} \frac{\mathrm{d}M_{0}^C(t \mid A, X)}{S_{0}(t \mid A, X) G_{0}(t- \mid A, X)} \\
& \times \{\mathrm{RMST}_{0}(\tau \mid A, X)-\mathrm{RMST}_{0}(t \mid A, X)\}=Y^{*}\{Y^{\mathrm{RMST}, \mathrm{AIPCW}}(\tau, A; \eta_{0}); (\mu_{0}, \pi_{0})\}
\end{aligned}
$$
}

This implies that
{\footnotesize
$$
\begin{aligned}
&\underset{\psi_{0}^{\mathrm{RMST}}(\tau, X)}{\operatorname{arg min}} L_{2}^{\mathrm{AIPCW}, *}(\min(T, \tau), \psi_{0}^{\mathrm{RMST}}(\tau, X); \eta_{0}) \\
=&\underset{\psi_{0}^{\mathrm{RMST}}(\tau, X)}{\operatorname{arg min}} L_{2}(Y^{\mathrm{RMST, AIPCW}}(\tau, A; \eta_{0}^{\Lambda}), \psi_{0}^{\mathrm{RMST}}(\tau, X); (\mu_{0}, \pi_{0}))
\end{aligned}
$$
}

% \section{Proof to Theorem~\ref{theorem:augmented}}

\section{Proof to Lemma~\ref{lemma1}}

Proof: Following notations from~\citet{diaz2018targeted}, we have

{\footnotesize
\begin{equation}\label{theorem:oracle:step1}
\begin{aligned}
&E\{\mathcal{Q}_{1}(\rho^{\star}(\widehat{\psi}^{*}))\} \\
=&(1+\delta)E[\sum_{k=1}^{K} \frac{n_{k}}{n}P_{0}\{\ell_{2}^{*}(Y, \widehat{\psi}_{k}^{*}(X; \rho^{\star}(\widehat{\psi}^{*})); \widehat{\eta}_{k})-\ell_{2}^{*}(Y, \psi_{0}(X); \widehat{\eta}_{k})-\ell_{2}^{*}(Y, \widehat{\psi}_{k}^{*}(X; \rho^{\star}(\widehat{\psi}^{*})); \eta_{\infty})+\ell_{2}^{*}(Y, \psi_{0}(X); \eta_{\infty})\}] \\
=&(1+\delta)E(\sum_{k=1}^{K} \frac{n_{k}}{n}P_{0}[w^{*}(\widehat{\eta}_{k}) \{Y^{*}(\widehat{\eta}_{k})-\widehat{\psi}_{k}^{*}(X; \rho^{\star}(\widehat{\psi}^{*}))\}^{2}-w^{*}(\widehat{\eta}_{k}) \{Y^{*}(\widehat{\eta}_{k})-\psi_{0}(X)\}^{2} \\
&-w^{*}(\eta_{\infty}) \{Y^{*}(\eta_{\infty})-\widehat{\psi}_{k}^{*}(X; \rho^{\star}(\widehat{\psi}^{*}))\}^{2}+ w^{*}(\eta_{\infty}) \{Y^{*}(\eta_{\infty})-\psi_{0}(X)\}^{2}]) \\
=&(1+\delta)E\{\sum_{k=1}^{K} \frac{n_{k}}{n}P_{0}(w^{*}(\widehat{\eta}_{k})\{\psi_{0}(X)-\widehat{\psi}_{k}^{*}(X; \rho^{\star}(\widehat{\psi}^{*}))\}[2Y^{*}(\widehat{\eta}_{k})-\{\psi_{0}(X)+\widehat{\psi}_{k}^{*}(X; \rho^{\star}(\widehat{\psi}^{*}))\}] \\
&-w^{*}(\eta_{\infty})\{\psi_{0}(X)-\widehat{\psi}_{k}^{*}(X; \rho^{\star}(\widehat{\psi}^{*}))\}[2Y^{*}(\eta_{\infty})-\{\psi_{0}(X)+\widehat{\psi}_{k}^{*}(X; \rho^{\star}(\widehat{\psi}^{*}))\}])\} \\
=&2(1+\delta)E(\sum_{k=1}^{K} \frac{n_{k}}{n}P_{0}[\{\psi_{0}(X)-\widehat{\psi}_{k}^{*}(X; \rho^{\star}(\widehat{\psi}^{*}))\}\{w^{*}(\widehat{\eta}_{k})Y^{*}(\widehat{\eta}_{k})-w^{*}(\eta_{\infty})Y^{*}(\eta_{\infty})\}]) \\
&+(1+\delta)E(\sum_{k=1}^{K} \frac{n_{k}}{n}P_{0}[\{\psi_{0}(X)-\widehat{\psi}_{k}^{*}(X; \rho^{\star}(\widehat{\psi}^{*}))\}\{w^{*}(\eta_{\infty})-w^{*}(\widehat{\eta}_{k})\}\{\psi_{0}(X)+\widehat{\psi}_{k}^{*}(X; \rho^{\star}(\widehat{\psi}^{*}))\}]) \\
=&2(1+\delta)E(\sum_{k=1}^{K} \frac{n_{k}}{n}P_{0}[\{\psi_{0}(X)-\widehat{\psi}_{k}^{*}(X; \rho^{\star}(\widehat{\psi}^{*}))\}\{w^{*}(\widehat{\eta}_{k})Y^{*}(\widehat{\eta}_{k})-w^{*}(\eta_{0})Y^{*}(\eta_{0})\}]) \\
&+2(1+\delta)E(\sum_{k=1}^{K} \frac{n_{k}}{n}P_{0}[\{\psi_{0}(X)-\widehat{\psi}_{k}^{*}(X; \rho^{\star}(\widehat{\psi}^{*}))\}\{w^{*}(\eta_{0})Y^{*}(\eta_{0})-w^{*}(\eta_{\infty})Y^{*}(\eta_{\infty})\}]) \\
&+(1+\delta)E(\sum_{k=1}^{K} \frac{n_{k}}{n}P_{0}[\{\psi_{0}(X)-\widehat{\psi}_{k}^{*}(X; \rho^{\star}(\widehat{\psi}^{*}))\}\{w^{*}(\eta_{0})-w^{*}(\widehat{\eta}_{k})\}\{\psi_{0}(X)+\widehat{\psi}_{k}^{*}(X; \rho^{\star}(\widehat{\psi}^{*}))\}]) \\
&+(1+\delta)E(\sum_{k=1}^{K} \frac{n_{k}}{n}P_{0}[\{\psi_{0}(X)-\widehat{\psi}_{k}^{*}(X; \rho^{\star}(\widehat{\psi}^{*}))\}\{w^{*}(\eta_{\infty})-w^{*}(\eta_{0})\}\{\psi_{0}(X)+\widehat{\psi}_{k}^{*}(X; \rho^{\star}(\widehat{\psi}^{*}))\}]) \\
=&2(1+\delta)E(\sum_{k=1}^{K} \frac{n_{k}}{n}P_{0}[\{\psi_{0}(X)-\widehat{\psi}_{k}^{*}(X; \rho^{\star}(\widehat{\psi}^{*}))\}\{w^{*}(\widehat{\eta}_{k})Y^{*}(\widehat{\eta}_{k})-w^{*}(\eta_{0})Y^{*}(\eta_{0})\}]) \\
&+(1+\delta)E(\sum_{k=1}^{K} \frac{n_{k}}{n}P_{0}[\{\psi_{0}(X)-\widehat{\psi}_{k}^{*}(X; \rho^{\star}(\widehat{\psi}^{*}))\}\{w^{*}(\eta_{0})-w^{*}(\widehat{\eta}_{k})\}\{\psi_{0}(X)+\widehat{\psi}_{k}^{*}(X; \rho^{\star}(\widehat{\psi}^{*}))\}])
\end{aligned}
\end{equation}
}
Since $E(\sum_{k=1}^{K} \frac{n_{k}}{n}P_{0}[\{\psi_{0}(X)-\widehat{\psi}_{k}^{*}(X; \rho^{\star}(\widehat{\psi}^{*}))\}\{w^{*}(\eta_{0})Y^{*}(\eta_{0})-w^{*}(\eta_{\infty})Y^{*}(\eta_{\infty})\}])=0$ and $E(\sum_{k=1}^{K} \frac{n_{k}}{n}P_{0}[\{\psi_{0}(X)-\widehat{\psi}_{k}^{*}(X; \rho^{\star}(\widehat{\psi}^{*}))\}\{w^{*}(\eta_{\infty})-w^{*}(\eta_{0})\}\{\psi_{0}(X)+\widehat{\psi}_{k}^{*}(X; \rho^{\star}(\widehat{\psi}^{*}))\}])=0$ as the subscript $\infty$ ensures corresponding transformation consistent by its definition after conditioning on $X$. Then by the Cauchy–Schwarz inequality,
{\footnotesize
\begin{equation}
\begin{aligned}
E\{\mathcal{Q}_{1}(\rho^{\star}(\widehat{\psi}^{*}))\} \leq& 2(1+\delta)E(\sum_{k=1}^{K} \frac{n_{k}}{n}P_{0}[ \{\widehat{\psi}_{k}^{*}(X; \rho^{\star}(\widehat{\psi}^{*}))-\psi_{0}(X)\}^{2}])^{1/2} r_{1}^{*}(\widehat{\eta}, \eta_{0}) \\
&+(1+\delta) E(\sum_{k=1}^{K} \frac{n_{k}}{n}P_{0}[ \{\widehat{\psi}_{k}^{*}(X; \rho^{\star}(\widehat{\psi}^{*}))-\psi_{0}(X)\}^{2}])^{1/2} r_{2}^{*}(\widehat{\eta}, \eta_{0}) \times 2c(\psi) \\
\leq& \{c_{r1}^{*}(\delta)r_{1}^{*}(\widehat{\eta}, \eta_{0})+c_{r2}^{*}(\delta)r_{2}^{*}(\widehat{\eta}, \eta_{0})\} [E\{L_{2}^{*}(\widehat{\psi}^{*}; \eta_{\infty})-L_{2}^{*}(\psi_{0}; \eta_{\infty})\}]^{1/2}
\end{aligned}
\end{equation}
}
where in the last inequality we assumed
{\footnotesize
$$
E[P_{0}\{\widehat{\psi}^{*}(X)-\psi_{0}(X)\}^{2}] \leq c^{*}(\mathcal{E}) E\{L_{2}^{*}(\widehat{\psi}^{*}; \eta_{\infty})-L_{2}^{*}(\psi_{0}; \eta_{\infty})\}
$$
}
$c(\psi)=\max_{k}\{|\widehat{\psi}_{k}^{*}(X; \rho^{\star}(\widehat{\psi}^{*}))|, |\psi_{0}(X)|\}$ and
{\footnotesize
$$
\begin{aligned}
r_{1}^{*}(\widehat{\eta}, \eta_{0})=&\max_{k} E\{(P_{0}[\{w^{*}(\widehat{\eta}_{k})Y^{*}(\widehat{\eta}_{k})-w^{*}(\eta_{0})Y^{*}(\eta_{0})\}^{2}])\}^{1/2} \\
r_{2}^{*}(\widehat{\eta}, \eta_{0})=&\max_{k} E\{(P_{0}[\{w^{*}(\widehat{\eta}_{k})-w^{*}(\eta_{0})\}^{2}])\}^{1/2}
\end{aligned}
$$
}
Then we consider
{\footnotesize
$$
\begin{aligned}
&P_{0}\{\ell_{2}^{*}(Y, \widehat{\psi}_{k}^{*}(X; \rho^{\star}(\widehat{\psi}^{*})); \widehat{\eta}_{k})-\ell_{2}^{*}(Y, \psi_{0}(X); \widehat{\eta}_{k})-\ell_{2}^{*}(Y, \widehat{\psi}_{k}^{*}(X; \rho^{\star}(\widehat{\psi}^{*})); \eta_{\infty})+\ell_{2}^{*}(Y, \psi_{0}(X); \eta_{\infty})\}^{2} \\
=&P_{0}[\{\psi_{0}(X)-\widehat{\psi}_{k}^{*}(X; \rho^{\star}(\widehat{\psi}^{*}))\}\{w^{*}(\widehat{\eta}_{k})Y^{*}(\widehat{\eta}_{k})-w^{*}(\eta_{\infty})Y^{*}(\eta_{\infty})\} \\
&+\{\psi_{0}(X)-\widehat{\psi}_{k}^{*}(X; \rho^{\star}(\widehat{\psi}^{*}))\}\{w^{*}(\eta_{\infty})-w^{*}(\widehat{\eta}_{k})\}\{\psi_{0}(X)+\widehat{\psi}_{k}^{*}(X; \rho^{\star}(\widehat{\psi}^{*}))\}]^{2} \\
=&P_{0}[\{\psi_{0}(X)-\widehat{\psi}_{k}^{*}(X; \rho^{\star}(\widehat{\psi}^{*}))\}^{2}\{w^{*}(\widehat{\eta}_{k})Y^{*}(\widehat{\eta}_{k})-w^{*}(\eta_{\infty})Y^{*}(\eta_{\infty})\}^{2}] \\
&+2 P_{0}[\{\psi_{0}(X)-\widehat{\psi}_{k}^{*}(X; \rho^{\star}(\widehat{\psi}^{*}))\}^{2}\{w^{*}(\widehat{\eta}_{k})Y^{*}(\widehat{\eta}_{k})-w^{*}(\eta_{\infty})Y^{*}(\eta_{\infty})\}\{w^{*}(\eta_{\infty})-w^{*}(\widehat{\eta}_{k})\}\{\psi_{0}(X)+\widehat{\psi}_{k}^{*}(X; \rho^{\star}(\widehat{\psi}^{*}))\}] \\
&+P_{0}[\{\psi_{0}(X)-\widehat{\psi}_{k}^{*}(X; \rho^{\star}(\widehat{\psi}^{*}))\}^{2}\{w^{*}(\eta_{\infty})-w^{*}(\widehat{\eta}_{k})\}^{2}\{\psi_{0}(X)+\widehat{\psi}_{k}^{*}(X; \rho^{\star}(\widehat{\psi}^{*}))\}^{2}] \\
\leq & \widetilde{c}_{2}^{*, 2} P_{0}[\{w^{*}(\widehat{\eta}_{k})Y^{*}(\widehat{\eta}_{k})-w^{*}(\eta_{\infty})Y^{*}(\eta_{\infty})\}^{2}]+4\widetilde{c}_{2}^{*, 2}c(\psi) P_{0}[\{w^{*}(\widehat{\eta}_{k})Y^{*}(\widehat{\eta}_{k})-w^{*}(\eta_{\infty})Y^{*}(\eta_{\infty})\} \\
&\times \{w^{*}(\eta_{\infty})-w^{*}(\widehat{\eta}_{k})\}]+4\widetilde{c}_{2}^{*, 2} c^{2}(\psi) P_{0}[\{w^{*}(\eta_{\infty})-w^{*}(\widehat{\eta}_{k})\}^{2}] \\
\leq & \widetilde{c}_{2}^{*, 2}\mathcal{B}_{1}^{*}(\widehat{\eta}, \eta_{0}) + 4\widetilde{c}_{2}^{*, 2}c(\psi)\mathcal{B}_{2}^{*}(\widehat{\eta}, \eta_{0}) + 4\widetilde{c}_{2}^{*, 2} c^{2}(\psi)\mathcal{B}_{3}^{*}(\widehat{\eta}, \eta_{0})
\end{aligned}
$$
}
where we assume $\sup_{O} (P_{0}[\{\psi_{0}(X)-\widehat{\psi}_{k}^{*}(X; \rho^{\star}(\widehat{\psi}^{*}))\}^{2}]) \leq \widetilde{c}_{2}^{*, 2}$ for some constant $\widetilde{c}_{2}^{*}$, and
{\footnotesize
$$
\begin{aligned}
\mathcal{B}_{1}^{*}(\widehat{\eta}, \eta_{0})=&\max_{k} E(P_{0}[\{w^{*}(\widehat{\eta}_{k})Y^{*}(\widehat{\eta}_{k})-w^{*}(\eta_{\infty})Y^{*}(\eta_{\infty})\}^{2}]) \\
\mathcal{B}_{2}^{*}(\widehat{\eta}, \eta_{0})=&\max_{k} E(P_{0}[\{w^{*}(\widehat{\eta}_{k})Y^{*}(\widehat{\eta}_{k})-w^{*}(\eta_{\infty})Y^{*}(\eta_{\infty})\}\{w^{*}(\widehat{\eta}_{k})-w^{*}(\eta_{\infty})\}]) \\
\mathcal{B}_{3}^{*}(\widehat{\eta}, \eta_{0})=&\max_{k} E(P_{0}[\{w^{*}(\widehat{\eta}_{k})-w^{*}(\eta_{\infty})\}^{2}])
\end{aligned}
$$
}
such that
{\footnotesize
$$
E\{\mathcal{Q}_{2}(\rho^{\star}(\widehat{\psi}^{*}))\} \leq \{(\log n)/n\}^{1/2} \{c_{\mathcal{B}1}^{2}(\delta) \mathcal{B}_{1}^{*}(\widehat{\eta}, \eta_{0})+c_{\mathcal{B}2}^{2}(\delta) \mathcal{B}_{2}^{*}(\widehat{\eta}, \eta_{0})+c_{\mathcal{B}3}^{2}(\delta) \mathcal{B}_{3}^{*}(\widehat{\eta}, \eta_{0})\}
$$
}
By solving a quadratic equation on $E\{L_{2}^{*}(\widehat{\psi}^{*}; \eta_{\infty})-L_{2}^{*}(\psi_{0}; \eta_{\infty})\}$, we have that
{\footnotesize
$$
\begin{aligned}
&[E\{L_{2}^{*}(\widehat{\psi}^{*}; \eta_{\infty})-L_{2}^{*}(\psi_{0}; \eta_{\infty})\}]^{1/2} \\
\leq & \{(1+2 \delta) [E\{L_{2}^{*}(\widetilde{\psi}^{*}; \eta_{\infty})-L_{2}^{*}(\psi_{0}; \eta_{\infty})\}]^{1/2}+\widetilde{c}_{1}^{*}(\delta)\{(1+\log n) / n\}^{1 / 2}+\{c_{r1}^{*}(\delta)r_{1}^{*}(\widehat{\eta}, \eta_{0})+c_{r2}^{*}(\delta)r_{2}^{*}(\widehat{\eta}, \eta_{0})\} \\
&+(\log n/n)^{1/4} \{c_{\mathcal{B}1}^{*, 2}(\delta) \mathcal{B}_{1}^{*}(\widehat{\eta}, \eta_{0})+c_{\mathcal{B}2}^{*, 2}(\delta)\mathcal{B}_{2}^{*}(\widehat{\eta}, \eta_{0})+c_{\mathcal{B}3}^{*, 2}(\delta)\mathcal{B}_{3}^{*}(\widehat{\eta}, \eta_{0})\}^{1/2}
\end{aligned}
$$
}

\section{Proof to Theorem~\ref{theorem:oracle}}

Proof: for S-learner, we can apply Theorem 1 from \citet{diaz2018targeted} by modifying regularity conditions.

Define
{\footnotesize
$$
\begin{aligned}
r(A; \widehat{\eta}_{k}^{\Lambda}, \eta_{0}^{\Lambda})=&\max_k E_{0}\left| \int_0^{\tau} \left\{\frac{G_{0}(t- \mid A, X)}{\widehat{G}_{k}(t- \mid A, X)}-1\right\}\mathrm{d}\left\{\frac{S_{0}(t\mid A, X)}{\widehat{S}_{k}(t \mid A, X)}-1\right\}\right| \\
\mathcal{B}^{\mathrm{RMST}}(A; \widehat{\eta}_{k}^{\Lambda}, \eta_{0}^{\Lambda})=&\max_{k} E_{0} \left[I\{S_{\infty}(t \mid A, X)=S_{0}(t \mid A, X)\} \sup_{t \in [0, \tau]} \left| \frac{\mathrm{R}\widehat{\mathrm{MS}}\mathrm{T}_{k}(\tau \mid A, X)}{\widehat{S}_{k}(t \mid A, X)}-\frac{\mathrm{RMST}_{0}(\tau \mid A, X)}{S_{0}(t \mid A, X)} \right| \right. \\
&\left.+I\{G_{\infty}(t \mid A, X)=G_{0}(t \mid A, X)\} \sup_{t \in [0, \tau]} \left| \frac{1}{\widehat{G}_{k}(t- \mid A, X)}-\frac{1}{G_{0}(t- \mid A, X)} \right| \right]^2
\end{aligned}
$$
}
For S-learner, the selector operates solely on conditional mean by taking treatment $A$ as a covariate,
{\footnotesize
$$
\begin{aligned}
\mathcal{E}^{2}(\widehat{\psi}^{\mathrm{SL}}; \rho^{\star}(\widehat{\mu}, A))=&E(P_{0}[\{\widehat{\mu}^{\mathrm{SL}}(\tau, 1, X; \rho^{\star}(\widehat{\mu}, A))-\widehat{\mu}^{\mathrm{SL}}(\tau, 0, X; \rho^{\star}(\widehat{\mu}, A))\}-\{\mu_{0}(\tau, 1, X)-\mu_{0}(\tau, 0, X)\}]^2) \\
=&E(P_{0}[\{\widehat{\mu}^{\mathrm{SL}}(\tau, 1, X; \rho^{\star}(\widehat{\mu}, A))-\mu_{0}(\tau, 1, X)\}-\{\widehat{\mu}^{\mathrm{SL}}(\tau, 0, X; \rho^{\star}(\widehat{\mu}, A))-\mu_{0}(\tau, 0, X)\}]^2) \\
\leq&2 E(P_{0}[\{\widehat{\mu}^{\mathrm{SL}}(\tau, 1, X; \rho^{\star}(\widehat{\mu}, A))-\mu_{0}(\tau, 1, X)\}]^2)+2 E(P_{0}[\{\widehat{\mu}^{\mathrm{SL}}(\tau, 0, X; \rho^{\star}(\widehat{\mu}, A))-\mu_{0}(\tau, 0, X)\}]^2) \\
=&2\mathcal{E}^{2}(\widehat{\mu}^{\mathrm{SL}}; \rho^{\star}(\widehat{\mu}, A))+2\mathcal{E}^{2}(\widehat{\mu}^{\mathrm{SL}}; \rho^{\star}(\widehat{\mu}, A))=4\mathcal{E}^{2}(\widehat{\mu}^{\mathrm{SL}}; \rho^{\star}(\widehat{\mu}, A))
\end{aligned}
$$
}
the inequality above adopts $(a+b)^2\leq 2 (a^2+b^2)$ and
{\footnotesize
$$
\begin{aligned}
\mathcal{E}(\widehat{\mu}^{\mathrm{SL}}; \rho^{\star}(\widehat{\mu}, A)) \leq & (1+2 \delta)^{1 / 2} \mathcal{E}(\widetilde{\mu}^{\mathrm{SL}}; \rho^{\star}(\widetilde{\mu}, A))+c_{1}^{\mathrm{SL}}(\delta)\{(1+\log n)/n\}^{1 / 2}+c_{2}^{\mathrm{SL}}(\delta) r(A; \widehat{\eta}^{\Lambda}, \eta_{0}^{\Lambda}) \\
&+c_{3}^{\mathrm{SL}}(\delta)(\log n / n)^{1 / 4}\{\mathcal{B}^{\mathrm{RMST}}(A; \widehat{\eta}^{\Lambda}, \eta_{0}^{\Lambda})\}^{1/2}
\end{aligned}
$$
}
with $\delta>0$.

For T-learner, the selector operates on treatment-specific conditional outcomes separately as
{\footnotesize
$$
\begin{aligned}
&\mathcal{E}^{2}(\widehat{\psi}^{\mathrm{TL}}; \rho^{\star}(\widehat{\mu}, 0), \rho^{\star}(\widehat{\mu}, 1)) \\
=&E(P_{0}[\{\widehat{\mu}^{\mathrm{TL}}(\tau, 1, X; \rho^{\star}(\widehat{\mu}, 1))-\widehat{\mu}^{\mathrm{TL}}(\tau, 0, X; \rho^{\star}(\widehat{\mu}, 0))\}-\{\mu_{0}(\tau, 1, X)-\mu_{0}(\tau, 0, X)\}]^2) \\
=&E(P_{0}[\{\widehat{\mu}^{\mathrm{TL}}(\tau, 1, X; \rho^{\star}(\widehat{\mu}, 1))-\mu_{0}(\tau, 1, X)\}-\{\widehat{\mu}^{\mathrm{TL}}(\tau, 0, X; \rho^{\star}(\widehat{\mu}, 0))-\mu_{0}(\tau, 0, X)\}]^2) \\
\leq&2 E(P_{0}[\{\widehat{\mu}^{\mathrm{TL}}(\tau, 1, X; \rho^{\star}(\widehat{\mu}, 1))-\mu_{0}(\tau, 1, X)\}]^2)+2 E(P_{0}[\{\widehat{\mu}^{\mathrm{TL}}(\tau, 0, X; \rho^{\star}(\widehat{\mu}, 0))-\mu_{0}(\tau, 0, X)\}]^2) \\
=&2\mathcal{E}^{2}(\widehat{\mu}^{\mathrm{TL}}; \rho^{\star}(\widehat{\mu}, 1))+2\mathcal{E}^{2}(\widehat{\mu}^{\mathrm{TL}}; \rho^{\star}(\widehat{\mu}, 0))
\end{aligned}
$$
}
and
{\footnotesize
\begin{equation} \label{eq:TL}
\begin{aligned}
\mathcal{E}(\widehat{\mu}^{\mathrm{TL}}; \rho^{\star}(\widehat{\mu}, a)) \leq & (1+2 \delta)^{1 / 2} \mathcal{E}(\widetilde{\mu}^{\mathrm{TL}}; \rho^{\star}(\widetilde{\mu}, a))+c_{1}^{\mathrm{TL}}(a)\{(1+\log n)/n\}^{1 / 2} \\
& +c_{2}^{\mathrm{TL}}(a) r(a; \widehat{\eta}^{\Lambda}, \eta_{0}^{\Lambda})+c_{3}^{\mathrm{TL}}(a) (\log n / n)^{1 / 4}\{\mathcal{B}^{\mathrm{RMST}}(a; \widehat{\eta}^{\Lambda}, \eta_{0}^{\Lambda})\}^{1/2}
\end{aligned}
\end{equation}
}
For RA-learner, IPTW-learner, AIPTW-learner, and U-learner, we have % we define $\widehat{\psi}^{*}(\tau, X)=\widehat{E}\{Y^{*}(Y^{\mathrm{AIPCW}}(\tau; \widehat{\eta}^{\Lambda}); (\widehat{\mu}, \widehat{\pi})) \mid X\}$, the partial-oracle estimator as $\widetilde{\psi}^{*}(\tau, X; \widehat{\eta}^{\Lambda})=\widehat{E}\{Y^{*}(Y^{\mathrm{AIPCW}}(\tau; \widehat{\eta}^{\Lambda}); (\mu_{\infty}, \pi_{\infty})) \mid X\}$, and the oracle estimator as $\widetilde{\psi}^{*}(\tau, X; \eta_{\infty}^{\Lambda})=\widehat{E}\{Y^{*}(Y^{\mathrm{AIPCW}}(\tau; \eta_{\infty}^{\Lambda}); (\mu_{\infty}, \pi_{\infty})) \mid X\}$, such that % (\mu_{\infty}(\widehat{\eta}^{\Lambda})
{\footnotesize
\begin{equation} \label{eq:oracle_continuous}
\begin{aligned}
\mathcal{E}(\widehat{\psi}^{*}; \rho^{\star}(\widehat{\psi}^{*})) \leq & (1+2 \delta)^{1 / 2} \mathcal{E}(\widetilde{\psi}^{*}(\widehat{\eta}^{\Lambda}); \rho^{\star}(\widetilde{\psi}^{*}; \widehat{\eta}^{\Lambda}))+\widetilde{c}_{1}^{*}\{(1+\log n)/n\}^{1 / 2}+c_{r}^{*} r^{*}((\widehat{\mu}, \widehat{\pi}), (\mu_{0}, \pi_{0})) \\
&+c_{\mathcal{B}}^{*}(\log n / n)^{1 / 4}\mathcal{B}^{*}((\widehat{\mu}, \widehat{\pi}), (\mu_{0}, \pi_{0})) \\ % \widehat{\eta}(\widehat{\eta}^{\Lambda}, \widehat{\rho^{\star}}_{v1}), \eta_{0}(\widehat{\eta}^{\Lambda}, \widehat{\rho^{\star}}_{v1})) \\ % \{\widehat{\eta}(\widehat{\eta}^{\Lambda}, \widehat{\rho^{\star}}_{v1}), \eta_{0}(\widehat{\eta}^{\Lambda}, \widehat{\rho^{\star}}_{v1})\}
\mathcal{E}(\widetilde{\psi}^{*}(\widehat{\eta}^{\Lambda}); \rho^{\star}(\widetilde{\psi}^{*}; \widehat{\eta}^{\Lambda})) \leq & (1+2 \delta)^{1 / 2} \mathcal{E}(\widetilde{\psi}^{*}; \rho^{\star}(\widetilde{\psi}^{*}))+c_{1}^{*}\{(1+\log n)/n\}^{1 / 2}+\widetilde{c}_{2}^{*} r(\widehat{\eta}^{\Lambda}, \eta_{0}^{\Lambda}) \\
&+c_{3}^{*}(\log n / n)^{1 / 4}\mathcal{B}^{\mathrm{RMST}}(\widehat{\eta}^{\Lambda}, \eta_{0}^{\Lambda}) % \widehat{\eta}(\widehat{\eta}^{\Lambda}, \widehat{\rho^{\star}}_{v1}), \eta_{0}(\widehat{\eta}^{\Lambda}, \widehat{\rho^{\star}}_{v1})) \\
\end{aligned}
\end{equation}
}
where
{\footnotesize
$$
\begin{aligned}
r^{\mathrm{RA}}((\widehat{\mu}, \widehat{\pi}), (\mu_{0}, \pi_{0}))=&\max_{k_{2}} E_{0}|\widehat{\mu}_{k_{2}}(\tau, 1-A, X; \rho^{\star}(\widehat{\mu}))-\mu_{0}(\tau, 1-A, X)| \\
r^{\mathrm{IPTW}}((\widehat{\mu}, \widehat{\pi}), (\mu_{0}, \pi_{0}))=&\max_{k_{2}} E_{0}|\widehat{\pi}_{k_{2}}(1 \mid X; \rho^{\star}(\widehat{\pi}))-\pi_{0}(1 \mid X)| \\
r^{\mathrm{AIPTW}}((\widehat{\mu}, \widehat{\pi}), (\mu_{0}, \pi_{0}))=&\max_{k_{2}} \sum_{a \in \{0, 1\}} E_{0}|\{\widehat{\pi}_{k_{2}}(a \mid X; \rho^{\star}(\widehat{\pi}))-\pi_{0}(a \mid X)\} \{\widehat{\mu}_{k_{2}}(\tau, a, X; \rho^{\star}(\widehat{\mu}, a))-\mu_{0}(\tau, a, X)\}| \\
r^{\mathrm{UL}}((\widehat{\mu}, \widehat{\pi}), (\mu_{0}, \pi_{0}))=&\max_{k_{2}} \sum_{a \in \{0, 1\}} E_{0}|\{\widehat{\pi}_{k_{2}}(a \mid X; \rho^{\star}(\widehat{\pi}))-\pi_{0}(a \mid X)\} \{\widehat{\mu}_{k_{2}}(\tau, a, X; \rho^{\star}(\widehat{\mu}, a))-\mu_{0}(\tau, a, X)\}| \\
\mathcal{B}^{\mathrm{RA}}((\widehat{\mu}, \widehat{\pi}), (\mu_{0}, \pi_{0}))=&\max_{k_{2}} E_{0}\{I(\mu_{\infty}(\tau, 1-A, X)=\mu_{0}(\tau, 1-A, X)) \\
&\times |\widehat{\mu}_{k_{2}}(\tau, 1-A, X; \rho^{\star}(\widehat{\mu}))-\mu_{0}(\tau, 1-A, X)|\}^2 \\
\mathcal{B}^{\mathrm{IPTW}}((\widehat{\mu}, \widehat{\pi}), (\mu_{0}, \pi_{0}))=&\max_{k_{2}} E_{0} \{I(\pi_{\infty}(1 \mid X)=\pi_{0}(1 \mid X)) |\widehat{\pi}_{k_{2}}(1 \mid X; \rho^{\star}(\widehat{\pi}))-\pi_{0}(1 \mid X)|\}^2 \\
\mathcal{B}^{\mathrm{AIPTW}}((\widehat{\mu}, \widehat{\pi}), (\mu_{0}, \pi_{0}))=&\max_{k_{2}} \sum_{a \in \{0, 1\}} E_{0} \{I(\pi_{\infty}(a \mid X)=\pi_{0}(a \mid X)) |\widehat{\pi}_{k_{2}}(a \mid X; \rho^{\star}(\widehat{\pi}))-\pi_{0}(a \mid X)| \\
&+I\{\mu_{\infty}(\tau, a, X)=\mu_{0}(\tau, a, X)\}|\widehat{\mu}_{k_{2}}(\tau, a, X; \rho^{\star}(\widehat{\mu}, a))-\mu_{0}(\tau, a, X)|\}^2 \\
\mathcal{B}^{\mathrm{UL}}((\widehat{\mu}, \widehat{\pi}), (\mu_{0}, \pi_{0}))=&\max_{k_{2}} \sum_{a \in \{0, 1\}} E_{0} [I(\pi_{\infty}(a \mid X)=\pi_{0}(a \mid X), \mu_{\infty}(\tau, a, X)=\mu_{0}(\tau, a, X)) \\
&\times\{|\widehat{\pi}_{k_{2}}(a \mid X; \rho^{\star}(\widehat{\pi}))-\pi_{0}(a \mid X)|+|\widehat{\mu}_{k_{2}}(\tau, a, X; \rho^{\star}(\widehat{\mu}, a))-\mu_{0}(\tau, a, X)|\}]^2
\end{aligned}
$$
}

As X-learner is a plugin estimator with two selectors for each treatment arm, we can not apply~\eqref{eq:oracle_continuous} directly. Instead, we decompose X-learner into two as
{\footnotesize
$$
\begin{aligned}
&\mathcal{E}^{2}(\widehat{\psi}^{\mathrm{XL}}; \rho^{\star}(\widehat{\psi}^{\mathrm{XL}}, 0), \rho^{\star}(\widehat{\psi}^{\mathrm{XL}}, 1)) \\
=&E[P_{0}\{\widehat{\psi}^{\mathrm{XL}}(\tau, X)-\psi_0(\tau, X)\}^2] \\
=&E(P_{0}[\widehat{\pi}(1 \mid X; \rho^{\star}(\widehat{\pi})) \{\widehat{E}[Y^{\mathrm{AIPCW}}(\tau, A; \widehat{\eta}^{\Lambda})-\widehat{\mu}(\tau, 0, X; \rho^{\star}(\widehat{\mu}, 0)) \mid A=1, X]-\psi_{0}(\tau, X)\} \\
&+\widehat{\pi}(0 \mid X; \rho^{\star}(\widehat{\pi})) \{\widehat{E}[\widehat{\mu}(\tau, 1, X; \rho^{\star}(\widehat{\mu}, 1))-Y^{\mathrm{AIPCW}}(\tau, A; \widehat{\eta}^{\Lambda}) \mid A=0, X]-\psi_{0}(\tau, X)\}]^2) \\
\leq& 2(1-1 / \epsilon)^2 E(P_{0}[\{\widehat{E}[Y^{\mathrm{AIPCW}}(\tau, A; \widehat{\eta}^{\Lambda})-\widehat{\mu}(\tau, 0, X; \rho^{\star}(\widehat{\mu}, 0)) \mid A=1, X]-\psi_{0}(\tau, X)\}]^2) \\
&+2(1-1 / \epsilon)^2 E(P_{0}[\{\widehat{E}[\widehat{\mu}(\tau, 1, X; \rho^{\star}(\widehat{\mu}, 1))-Y^{\mathrm{AIPCW}}(\tau, A; \widehat{\eta}^{\Lambda}) \mid A=0, X]-\psi_{0}(\tau, X)\}]^2) \\
=&2(1-1 / \epsilon)^2 \{\mathcal{E}^{2}(\widehat{\psi}^{\mathrm{XL}}; \rho^{\star}(\widehat{\psi}^{\mathrm{XL}}, 0))+ \mathcal{E}^{2}(\widehat{\psi}^{\mathrm{XL}}; \rho^{\star}(\widehat{\psi}^{\mathrm{XL}}, 1))\}
\end{aligned}
$$
}
where for $a \in \{0, 1\}$
{\footnotesize
\begin{equation} \label{eq:xl_oracle}
\begin{aligned}
\mathcal{E}^{2}(\widehat{\psi}^{\mathrm{XL}}; \rho^{\star}(\widehat{\psi}^{\mathrm{XL}}, a)) \leq & (1+2 \delta)^{1 / 2} \mathcal{E}^{2}(\widetilde{\psi}^{\mathrm{XL}}(\widehat{\eta}^{\Lambda}); \rho^{\star}(\widetilde{\psi}^{\mathrm{XL}}, a; \widehat{\eta}^{\Lambda}))+\widetilde{c}_{1}^{\mathrm{XL}}(a)\{(1+\log n)/n\}^{1 / 2} \\
&+c_{r}^{\mathrm{XL}}(a) r^{\mathrm{XL}}((\widehat{\mu}, \widehat{\pi}), (\mu_{0}, \pi_{0}))+c_{\mathcal{B}}^{\mathrm{XL}}(a)(\log n / n)^{1 / 4}\mathcal{B}^{\mathrm{XL}}((\widehat{\mu}, \widehat{\pi}), (\mu_{0}, \pi_{0})) \\
\mathcal{E}^{2}(\widetilde{\psi}^{\mathrm{XL}}(\widehat{\eta}^{\Lambda}); \rho^{\star}(\widetilde{\psi}^{\mathrm{XL}}, a; \widehat{\eta}^{\Lambda})) \leq & (1+2 \delta)^{1 / 2} \mathcal{E}^{2}(\widetilde{\psi}^{\mathrm{XL}}; \rho^{\star}(\widetilde{\psi}^{\mathrm{XL}}, a))+c_{1}^{\mathrm{XL}}(a)\{(1+\log n)/n\}^{1 / 2} \\
&+c_{2}^{\mathrm{XL}}(a) r(\widehat{\eta}^{\Lambda}, \eta_{0}^{\Lambda})+c_{3}^{\mathrm{XL}}(a)(\log n / n)^{1 / 4}\mathcal{B}^{\mathrm{RMST}}(\widehat{\eta}^{\Lambda}, \eta_{0}^{\Lambda}) \\
r^{\mathrm{XL}}((\widehat{\mu}, \widehat{\pi}), (\mu_{0}, \pi_{0}))=&\max_{k_{2}} E_{0}|\widehat{\mu}_{k_{2}}(\tau, a, X; \rho^{\star}(\widehat{\mu}, a))-\mu_{0}(\tau, a, X)| \\
\mathcal{B}^{\mathrm{XL}}((\widehat{\mu}, \widehat{\pi}), (\mu_{0}, \pi_{0}))=&\max_{k_{2}} E_{0}[I\{\mu_{\infty}(\tau, a, X)=\mu_{0}(\tau, a, X)\} |\widehat{\mu}_{k_{2}}(\tau, a, X; \rho^{\star}(\widehat{\mu}, a))-\mu_{0}(\tau, a, X)|]^2
\end{aligned}
\end{equation}
}

For R-learner
{\footnotesize
$$
\begin{aligned}
&\mathcal{E}^{2}(\widehat{\psi}^{\mathrm{RL}}; \rho^{\star}(\widehat{\psi}^{\mathrm{RL}}))= E[P_{0}\{\widehat{\psi}^{\mathrm{RL}}(\tau, X; \rho^{\star}(\widehat{\psi}^{\mathrm{RL}}))-\psi_{0}(\tau, X)\}^{2}] \\
\leq& \epsilon^{2} E(P_{0}[\{A-\pi_{0}(1 \mid X)\}^{2}\{\widehat{\psi}^{\mathrm{RL}}(\tau, X; \rho^{\star}(\widehat{\psi}^{\mathrm{RL}}))-\psi_{0}(\tau, X)\}^{2}]) \\
=&\epsilon^{2}E(P_{0}[\{A-\pi_{0}(1 \mid X)\}^{2}\{\widehat{\psi}^{\mathrm{RL}, 2}(\tau, X; \rho^{\star}(\widehat{\psi}^{\mathrm{RL}}))-\psi_{0}^{2}(\tau, X)\} \\
&-2\{A-\pi_{0}(1 \mid X)\}^{2}\psi_{0}(\tau, X)\{\widehat{\psi}^{\mathrm{RL}}(\tau, X; \rho^{\star}(\widehat{\psi}^{\mathrm{RL}}))-\psi_{0}(\tau, X)\}]) \\
=&\epsilon^{2}E(P_{0}[\{A-\pi_{0}(1 \mid X)\}^{2}\{\widehat{\psi}^{\mathrm{RL}, 2}(\tau, X; \rho^{\star}(\widehat{\psi}^{\mathrm{RL}}))-\psi_{0}^{2}(\tau, X)\} \\
&-2\{A-\pi_{0}(X)\}\{Y^{\mathrm{AIPCW}}(\tau, A; \eta_{\infty}^{\Lambda})-\mu_{0}(\tau, X)\}\{\widehat{\psi}^{\mathrm{RL}}(\tau, X; \rho^{\star}(\widehat{\psi}^{\mathrm{RL}}))-\psi_{0}(\tau, X)\}]) \\
=&\epsilon^{2}E(P_{0}[\{A-\pi_{\infty}(1 \mid X)\}^{2}\{\widehat{\psi}^{\mathrm{RL}, 2}(\tau, X; \rho^{\star}(\widehat{\psi}^{\mathrm{RL}}))-\psi_{0}^{2}(\tau, X)\} \\
&-2\{A-\pi_{\infty}(1 \mid X)\}\{Y^{\mathrm{AIPCW}}(\tau, A; \eta_{\infty}^{\Lambda})-\mu_{\infty}(X)\}\{\widehat{\psi}^{\mathrm{RL}}(\tau, X; \rho^{\star}(\widehat{\psi}^{\mathrm{RL}}))-\psi_{0}(\tau, X)\}]) \\
=&\epsilon^{2}E\{P_{0}([\{Y^{\mathrm{AIPCW}}(\tau, A; \eta_{\infty}^{\Lambda})-\mu_{\infty}(\tau, X)\}-\{A-\pi_{\infty}(1 \mid X)\}\widehat{\psi}^{\mathrm{RL}}(\tau, X; \rho^{\star}(\widehat{\psi}^{\mathrm{RL}}))]^{2} \\
&-[\{Y^{\mathrm{AIPCW}}(\tau, A; \eta_{\infty}^{\Lambda})-\mu_{\infty}(\tau, X)\}-\{A-\pi_{\infty}(1 \mid X)\}\psi_{0}(\tau, X)]^{2})\} \\
=&\epsilon^{2}E\{L_{2}^{\mathrm{RL}}(\widehat{\psi}^{\mathrm{RL}}; \eta_{\infty}^{\Lambda}, (\mu_{\infty}, \pi_{\infty}), \rho^{\star}(\widehat{\psi}^{\mathrm{RL}}))-L_{2}^{\mathrm{RL}}(\psi_{0}; \eta_{\infty}^{\Lambda}, (\mu_{\infty}, \pi_{\infty}))\}
\end{aligned}
$$
}
Applying~\Cref{lemma1} while taking $Y^{\mathrm{AIPCW}}(\tau, A; \eta_{\infty}^{\Lambda})$ as the continuous outcome, we have
{\footnotesize
$$
\begin{aligned}
&[E\{L_{2}^{\mathrm{RL}}(\widehat{\psi}^{\mathrm{RL}}; \eta_{\infty}^{\Lambda}, (\mu_{\infty}, \pi_{\infty}), \rho^{\star}(\widehat{\psi}^{\mathrm{RL}}))-L_{2}^{\mathrm{RL}}(\psi_{0}; \eta_{\infty}^{\Lambda}, (\mu_{\infty}, \pi_{\infty}))\}]^{1/2} \\
\leq& \{(1+2 \delta) [E\{L_{2}^{\mathrm{RL}}(\widetilde{\psi}^{\mathrm{RL}}(\widehat{\eta}^{\Lambda}); \eta_{\infty}^{\Lambda}, (\mu_{\infty}, \pi_{\infty}), \rho^{\star}(\widetilde{\psi}^{\mathrm{RL}}; \widehat{\eta}^{\Lambda}))-L_{2}^{\mathrm{RL}}(\psi_{0}; \eta_{\infty}^{\Lambda}, (\mu_{\infty}, \pi_{\infty}))\}]^{1/2} \\
&+\widetilde{c}_{1}^{\mathrm{RL}}(\delta)\{(1+\log n) / n\}^{1 / 2}+\{c_{r1}^{\mathrm{RL}}(\delta) r_{1}^{\mathrm{RL}}((\widehat{\mu}, \widehat{\pi}), (\mu_{0}, \pi_{0}))+c_{r2}^{\mathrm{RL}}(\delta)r_{2}^{\mathrm{RL}}((\widehat{\mu}, \widehat{\pi}), (\mu_{0}, \pi_{0}))\} \\
&+(\log n/n)^{1/4} \{c_{\mathcal{B}1}^{\mathrm{RL}, 2}(\delta) \mathcal{B}_{1}^{\mathrm{RL}}((\widehat{\mu}, \widehat{\pi}), (\mu_{0}, \pi_{0}))+c_{\mathcal{B}2}^{\mathrm{RL}, 2}(\delta)\mathcal{B}_{2}^{\mathrm{RL}}((\widehat{\mu}, \widehat{\pi}), (\mu_{0}, \pi_{0}))+c_{\mathcal{B}3}^{\mathrm{RL}, 2}(\delta)\mathcal{B}_{3}^{\mathrm{RL}}((\widehat{\mu}, \widehat{\pi}), (\mu_{0}, \pi_{0}))\}^{1/2}
\end{aligned}
$$
}
where
{\footnotesize
$$
\begin{aligned}
r_{1}^{\mathrm{RL}}((\widehat{\mu}, \widehat{\pi}), (\mu_{0}, \pi_{0}))=&\max_{k}\{ E(P_{0}[\{A-\widehat{\pi}_{k}(1 \mid X; \rho^{\star}(\widehat{\pi}))\}\{Y^{\mathrm{AIPCW}}(\tau, A; \eta_{\infty}^{\Lambda})-\widehat{\mu}_{k}(\tau, X; \rho^{\star}(\widehat{\mu}))\}-\{A-\pi_{0}(1 \mid X)\} \\
&\times \{Y^{\mathrm{AIPCW}}(\tau, A; \eta_{\infty}^{\Lambda})-\mu_{0}(\tau, X)\}]^2)\}^{1/2} \\
=&\max_{k} \{E(P_{0}[\{\widehat{\mu}_{k}(\tau, X; \rho^{\star}(\widehat{\mu}))-\mu_{0}(\tau, X)\}\{\widehat{\pi}_{k}(1 \mid X; \rho^{\star}(\widehat{\pi}))-\pi_{0}(1 \mid X)\}]^2)\}^{1/2} \\
r_{2}^{\mathrm{RL}}((\widehat{\mu}, \widehat{\pi}), (\mu_{0}, \pi_{0}))=&\max_{k} \{E(P_{0}[\{A-\widehat{\pi}_{k}(1 \mid X; \rho^{\star}(\widehat{\pi}))\}^{2}-\{A-\pi_{0}(1 \mid X)\}^{2}]^{2})\}^{1/2} \\
=&\max_{k} \{E(P_{0}[\{\widehat{\pi}_{k}(1 \mid X; \rho^{\star}(\widehat{\pi}))-\pi_{0}(1 \mid X)\}\{\widehat{\pi}_{k}(1 \mid X; \rho^{\star}(\widehat{\pi}))+\pi_{0}(1 \mid X)\} \\
&-2A\{\widehat{\pi}_{k}(1 \mid X; \rho^{\star}(\widehat{\pi}))-\pi_{0}(1 \mid X)\}]^2)\}^{1/2} \\
\leq& 4 \max_{k} (E[P_{0}\{\widehat{\pi}_{k}(1 \mid X; \rho^{\star}(\widehat{\pi}))-\pi_{0}(1 \mid X)\}^2])^{1/2} \\
\mathcal{B}_{1}^{\mathrm{RL}}((\widehat{\mu}, \widehat{\pi}), (\mu_{0}, \pi_{0}))=&\max_{k} E(P_{0}[\{\widehat{\mu}_{k}(\tau, X; \rho^{\star}(\widehat{\mu}))-\mu_{\infty}(\tau, X)\}\{\widehat{\pi}_{k}(1 \mid X; \rho^{\star}(\widehat{\pi}))-\pi_{\infty}(1 \mid X)\}]^2) \\
\mathcal{B}_{2}^{\mathrm{RL}}((\widehat{\mu}, \widehat{\pi}), (\mu_{0}, \pi_{0}))=&\max_{k} E(P_{0}[\{\widehat{\mu}_{k}(\tau, X; \rho^{\star}(\widehat{\mu}))-\mu_{\infty}(\tau, X)\}\{\widehat{\pi}_{k}(1 \mid X; \rho^{\star}(\widehat{\pi}))-\pi_{\infty}(1 \mid X)\}^{2} \\
&\times\{\widehat{\pi}_{k}(1 \mid X; \rho^{\star}(\widehat{\pi}))+\pi_{\infty}(1 \mid X)-2A\}] \\
\leq& 2\max_{k} E(P_{0}[|\widehat{\mu}_{k}(\tau, X; \rho^{\star}(\widehat{\mu}))-\mu_{\infty}(\tau, X)|\{\widehat{\pi}_{k}(1 \mid X; \rho^{\star}(\widehat{\pi}))-\pi_{\infty}(1 \mid X)\}^{2}] \\
\mathcal{B}_{3}^{\mathrm{RL}}((\widehat{\mu}, \widehat{\pi}), (\mu_{0}, \pi_{0}))=&16 \max_{k} E[P_{0}\{\widehat{\pi}_{k}(1 \mid X; \rho^{\star}(\widehat{\pi}))-\pi_{\infty}(1 \mid X)\}^2]
\end{aligned}
$$
}
where $r_{1}^{\mathrm{RL}}((\widehat{\mu}, \widehat{\pi}), (\mu_{0}, \pi_{0}))$ is often referred to as the mixed bias property~\citep{rotnitzky2021characterization}. Then, since the transformed outcome for R-learner is linear in $Y^{\mathrm{AIPCW}}$, we can further bound the partial-oracle loss by the oracle loss using Theorem 1 in~\citet{diaz2018targeted} as % dukes2021doubly
{\footnotesize
$$
\begin{aligned}
& [E\{L_{2}^{\mathrm{RL}}(\widetilde{\psi}^{\mathrm{RL}}(\widehat{\eta}^{\Lambda}); \eta_{\infty}^{\Lambda}, (\mu_{\infty}, \pi_{\infty}), \rho^{\star}(\widetilde{\psi}^{\mathrm{RL}}; \widehat{\eta}^{\Lambda}))-L_{2}^{\mathrm{RL}}(\psi_{0}; \eta_{\infty}^{\Lambda}, (\mu_{\infty}, \pi_{\infty}))\}]^{1/2} \\
\leq & (1+2 \delta)^{1 / 2} [E\{L_{2}^{\mathrm{RL}}(\widetilde{\psi}^{\mathrm{RL}}; \eta_{\infty}^{\Lambda}, (\mu_{\infty}, \pi_{\infty}), \rho^{\star}(\widetilde{\psi}^{\mathrm{RL}}))-L_{2}^{\mathrm{RL}}(\psi_{0}; \eta_{\infty}^{\Lambda}, (\mu_{\infty}, \pi_{\infty}))\}]^{1/2} +c_{1}^{\mathrm{RL}}(\delta)\{(1+\log n)/n\}^{1 / 2} \\
&+c_{2}^{\mathrm{RL}}(\delta) r(\widehat{\eta}^{\Lambda}, \eta_{0}^{\Lambda}) +c_{3}^{\mathrm{RL}}(\delta)(\log n / n)^{1 / 4}\mathcal{B}^{\mathrm{RMST}}(\widehat{\eta}^{\Lambda}, \eta_{0}^{\Lambda})
\end{aligned}
$$
}
and eventually, we bound the oracle loss by the oracle risk by a constant as
{\footnotesize
$$
\begin{aligned}
E\{L_{2}^{\mathrm{RL}}(\widetilde{\psi}^{\mathrm{RL}}; \eta_{\infty}^{\Lambda}, (\mu_{\infty}, \pi_{\infty}), \rho^{\star}(\widetilde{\psi}^{\mathrm{RL}}))-L_{2}^{\mathrm{RL}}(\psi_{0}; \eta_{\infty}^{\Lambda}, (\mu_{\infty}, \pi_{\infty}))\}\leq \left(1-\frac{1}{\epsilon}\right)^{2}\mathcal{E}^{2}(\widetilde{\psi}^{\mathrm{RL}}; \rho^{\star}(\widetilde{\psi}^{\mathrm{RL}}))
\end{aligned}
$$
}

For MC-learner, we assume the the excess risk of the selector is bounded by the loss function with a constant $c^{\mathrm{MC}}(\mathcal{E})$
{\footnotesize
$$
\mathcal{E}^2(\widehat{\psi}^{\mathrm{MC}}; \rho^{\star}(\widehat{\psi}^{\mathrm{MC}}))\leq c^{\mathrm{MC}}(\mathcal{E}) E\{L_{2}^{\mathrm{MC}}(\widehat{\psi}^{\mathrm{MC}}; \eta_{\infty}^{\Lambda}, (\mu_{\infty}, \pi_{\infty}), \rho^{\star}(\widehat{\psi}^{\mathrm{MC}}))-L_{2}^{\mathrm{MC}}(\psi_{0}; \eta_{\infty}^{\Lambda}, (\mu_{\infty}, \pi_{\infty}))\}
$$
}
where $c^{\mathrm{MC}}(\mathcal{E})=(\epsilon-1)/(4\epsilon)$ if and only if $\pi(0 \mid X)=\pi(1 \mid X)=1/2$, the case of RCTs with equal units in each treatment arm.

Applying~\Cref{lemma1} while taking $Y(\tau, A; \eta_{\infty}^{\Lambda})$ as a continuous outcome, we have
{\footnotesize
$$
\begin{aligned}
&[E\{L_{2}^{\mathrm{MC}}(\widehat{\psi}^{\mathrm{MC}}; \eta_{\infty}^{\Lambda}, (\mu_{\infty}, \pi_{\infty}), \rho^{\star}(\widehat{\psi}^{\mathrm{MC}}))-L_{2}^{\mathrm{MC}}(\psi_{0}; \eta_{\infty}^{\Lambda}, (\mu_{\infty}, \pi_{\infty}))\}]^{1/2} \\
\leq& \{(1+2 \delta) [E\{L_{2}^{\mathrm{MC}}(\widetilde{\psi}^{\mathrm{MC}}(\widehat{\eta}^{\Lambda}); \eta_{\infty}^{\Lambda}, (\mu_{\infty}, \pi_{\infty}), \rho^{\star}(\widetilde{\psi}^{\mathrm{MC}}; \widehat{\eta}^{\Lambda}))-L_{2}^{\mathrm{MC}}(\psi_{0}; \eta_{\infty}^{\Lambda}, (\mu_{\infty}, \pi_{\infty}))\}]^{1/2} \\
&+\widetilde{c}_{1}^{\mathrm{MC}}(\delta)\{(1+\log n) / n\}^{1 / 2}+\{c_{r1}^{\mathrm{MC}}(\delta) r_{1}^{\mathrm{MC}}((\widehat{\mu}, \widehat{\pi}), (\mu_{0}, \pi_{0}))+c_{r2}^{\mathrm{MC}}(\delta)r_{2}^{\mathrm{MC}}((\widehat{\mu}, \widehat{\pi}), (\mu_{0}, \pi_{0}))\} \\
&+(\log n/n)^{1/4} \{c_{\mathcal{B}1}^{\mathrm{MC}, 2}(\delta) \mathcal{B}_{1}^{\mathrm{MC}}((\widehat{\mu}, \widehat{\pi}), (\mu_{0}, \pi_{0}))+c_{\mathcal{B}2}^{\mathrm{MC}, 2}(\delta)\mathcal{B}_{2}^{\mathrm{MC}}((\widehat{\mu}, \widehat{\pi}), (\mu_{0}, \pi_{0}))+c_{\mathcal{B}3}^{\mathrm{MC}, 2}(\delta)\mathcal{B}_{3}^{\mathrm{MC}}((\widehat{\mu}, \widehat{\pi}), (\mu_{0}, \pi_{0}))\}^{1/2}
\end{aligned}
$$
}
where
{\footnotesize
$$
\begin{aligned}
r_{1}^{\mathrm{MC}}((\widehat{\mu}, \widehat{\pi}), (\mu_{0}, \pi_{0}))=&\max_{k}\{ E(P_{0}[\{\frac{A-\widehat{\pi}_{k}(1 \mid X; \rho^{\star}(\widehat{\pi}))}{\widehat{\pi}_{k}(0 \mid X; \rho^{\star}(\widehat{\pi}))\widehat{\pi}_{k}(1 \mid X; \rho^{\star}(\widehat{\pi}))}-\frac{A-\pi_{0}(1 \mid X)}{\pi_{0}(0 \mid X)\pi_{0}(1 \mid X)}\} Y^{\mathrm{AIPCW}}(\tau, A; \eta_{\infty}^{\Lambda})]^2)\}^{1/2} \\
\lesssim&\max_{k} (E[P_{0}\{\widehat{\pi}_{k}(1 \mid X; \rho^{\star}(\widehat{\pi}))-\pi_{0}(1 \mid X)\}^2])^{1/2} \\
r_{2}^{\mathrm{MC}}((\widehat{\mu}, \widehat{\pi}), (\mu_{0}, \pi_{0}))=&\max_{k} \{E(P_{0}[\{\frac{(2A-1)\{A-\widehat{\pi}_{k}(1 \mid X; \rho^{\star}(\widehat{\pi}))\}}{4\widehat{\pi}_{k}(0 \mid X; \rho^{\star}(\widehat{\pi}))\widehat{\pi}_{k}(1 \mid X; \rho^{\star}(\widehat{\pi}))}-\frac{(2A-1)\{A-\pi_{0}(1 \mid X)\}}{4\pi_{0}(0 \mid X)\pi_{0}(1 \mid X)}\}^{2}]^{2})\}^{1/2} \\
\lesssim& \max_{k} (E[P_{0}\{\widehat{\pi}_{k}(1 \mid X; \rho^{\star}(\widehat{\pi}))-\pi_{0}(1 \mid X)\}^2])^{1/2} \\
\mathcal{B}_{1}^{\mathrm{MC}}((\widehat{\mu}, \widehat{\pi}), (\mu_{0}, \pi_{0}))\lesssim& \max_{k} E[P_{0}\{\widehat{\pi}_{k}(1 \mid X; \rho^{\star}(\widehat{\pi}))-\pi_{\infty}(1 \mid X)\}^2] \\
\mathcal{B}_{2}^{\mathrm{MC}}((\widehat{\mu}, \widehat{\pi}), (\mu_{0}, \pi_{0}))\lesssim& \max_{k} E[P_{0}\{\widehat{\pi}_{k}(1 \mid X; \rho^{\star}(\widehat{\pi}))-\pi_{\infty}(1 \mid X)\}^2] \\
\mathcal{B}_{3}^{\mathrm{MC}}((\widehat{\mu}, \widehat{\pi}), (\mu_{0}, \pi_{0}))\lesssim& \max_{k} E[P_{0}\{\widehat{\pi}_{k}(1 \mid X; \rho^{\star}(\widehat{\pi}))-\pi_{\infty}(1 \mid X)\}^2]
\end{aligned}
$$
}
The terms are almost the same as the MC-learner is absent of conditional outcome models. Because the transformed outcome of MC-learner is linear in $Y^{\mathrm{AIPCW}}$, we can further bound the partial-oracle loss by the oracle loss using Theorem 1 in~\citet{diaz2018targeted} as
{\footnotesize
$$
\begin{aligned}
& [E\{L_{2}^{\mathrm{MC}}(\widetilde{\psi}^{\mathrm{MC}}(\widehat{\eta}^{\Lambda}); \eta_{\infty}^{\Lambda}, (\mu_{\infty}, \pi_{\infty}), \rho^{\star}(\widetilde{\psi}^{\mathrm{MC}}; \widehat{\eta}^{\Lambda}))-L_{2}^{\mathrm{MC}}(\psi_{0}; \eta_{\infty}^{\Lambda}, (\mu_{\infty}, \pi_{\infty}))\}]^{1/2} \\
\leq & (1+2 \delta)^{1 / 2} [E\{L_{2}^{\mathrm{MC}}(\widetilde{\psi}^{\mathrm{MC}}; \eta_{\infty}^{\Lambda}, (\mu_{\infty}, \pi_{\infty}), \rho^{\star}(\widetilde{\psi}^{\mathrm{MC}}))-L_{2}^{\mathrm{MC}}(\psi_{0}; \eta_{\infty}^{\Lambda}, (\mu_{\infty}, \pi_{\infty}))\}]^{1/2} +c_{1}^{\mathrm{MC}}(\delta)\{(1+\log n)/n\}^{1 / 2} \\
&+c_{2}^{\mathrm{MC}}(\delta) r^{\mathrm{RMST}}(\widehat{\eta}^{\Lambda}, \eta_{0}^{\Lambda}) +c_{3}^{\mathrm{MC}}(\delta) (\log n / n)^{1 / 4}\mathcal{B}^{\mathrm{RMST}}(\widehat{\eta}^{\Lambda}, \eta_{0}^{\Lambda})
\end{aligned}
$$
}
and eventually, we bound the oracle loss by the oracle risk with a constant as
{\footnotesize
$$
\begin{aligned}
E\{L_{2}^{\mathrm{MC}}(\widetilde{\psi}^{\mathrm{MC}}; \eta_{\infty}^{\Lambda}, (\mu_{\infty}, \pi_{\infty}), \rho^{\star}(\widetilde{\psi}^{\mathrm{MC}}))-L_{2}^{\mathrm{MC}}(\psi_{0}; \eta_{\infty}^{\Lambda}, (\mu_{\infty}, \pi_{\infty}))\}\leq c^{\mathrm{MC}}(L) \mathcal{E}^{2}(\widetilde{\psi}^{\mathrm{MC}}; \rho^{\star}(\widetilde{\psi}^{\mathrm{MC}}))
\end{aligned}
$$
}
where $c^{\mathrm{MC}}(L)=4 \epsilon$ if and only if $\pi(0 \mid X)=\pi(1 \mid X)=1/2$.

For the MCEA-learner
{\footnotesize
$$
\begin{aligned}
&E\{P_{0}(\frac{(2A-1)\{A-\pi_0(1 \mid X)\}}{4\pi_0(0 \mid X)\pi_0(1 \mid X)}[2(2A-1)\{Y-\mu_{0}(X)\}-\widehat{\psi}(X)]^{2}-\frac{(2A-1)\{A-\pi_0(1 \mid X)\}}{4\pi_0(0 \mid X)\pi_0(1 \mid X)} \\
&\times[2(2A-1)\{Y-\mu_{0}(X)\}-\psi_{0}(X)]^{2})\} \\
=&E\{P_{0}(\frac{(2A-1)\{A-\pi_0(1 \mid X)\}}{4\pi_0(0 \mid X)\pi_0(1 \mid X)} [\{\widehat{\psi}(X)+\psi_{0}(X)\}\{\widehat{\psi}(X)-\psi_{0}(X)\}-4(2A-1)\{Y-\mu_{0}(X)\} \\
&\times\{\widehat{\psi}(X)-\psi_{0}(X)\}])\} \\
=&E\{P_{0}(\frac{(2A-1)\{A-\pi_0(1 \mid X)\}}{4\pi_0(0 \mid X)\pi_0(1 \mid X)}[\{\widehat{\psi}(X)-\psi_{0}(X)\}^{2}+[2\psi_{0}(X)-4(2A-1)\{Y-\mu_{0}(X)\}]\{\widehat{\psi}(X)-\psi_{0}(X)\}])\} \\
=&E(P_{0}[\frac{(2A-1)\{A-\pi_0(1 \mid X)\}}{4\pi_0(0 \mid X)\pi_0(1 \mid X)}\{\widehat{\psi}(X)-\psi_{0}(X)\}^{2}])
\end{aligned}
$$
}
One can simply replace the superscript MC by MCEA, except for
{\footnotesize
$$
\begin{aligned}
r_{1}^{\mathrm{MCEA}}((\widehat{\mu}, \widehat{\pi}), (\mu_{0}, \pi_{0}))\lesssim&\max_{k} \{E[P_{0}\{\widehat{\pi}_{k}(1 \mid X; \rho^{\star}(\widehat{\pi}))-\pi_{0}(1 \mid X)\}^2]+E[P_{0}\{\widehat{\mu}_{k}(\tau, X; \rho^{\star}(\widehat{\mu}))-\mu_{0}(\tau, X)\}^2] \\ 
&+E(P_{0}[\{\widehat{\mu}_{k}(\tau, X; \rho^{\star}(\widehat{\mu}))-\mu_{0}(\tau, X)\}\{\widehat{\pi}_{k}(1 \mid X; \rho^{\star}(\widehat{\pi}))-\pi_{0}(1 \mid X)\}]^{2})\}^{1/2} \\
\mathcal{B}_{1}^{\mathrm{MCEA}}((\widehat{\mu}, \widehat{\pi}), (\mu_{\infty}, \pi_{\infty}))\lesssim& \max_{k} E[P_{\infty}\{\widehat{\pi}_{k}(1 \mid X; \rho^{\star}(\widehat{\pi}))-\pi_{\infty}(1 \mid X)\}^2]+E[P_{\infty}\{\widehat{\mu}_{k}(\tau, X; \rho^{\star}(\widehat{\mu}))-\mu_{\infty}(\tau, X)\}^2 \\ 
&+E(P_{0}[\{\widehat{\mu}_{k}(\tau, X; \rho^{\star}(\widehat{\mu}))-\mu_{\infty}(\tau, X)\}\{\widehat{\pi}_{k}(1 \mid X; \rho^{\star}(\widehat{\pi}))-\pi_{\infty}(1 \mid X)\}])^{2} \\
\mathcal{B}_{2}^{\mathrm{MCEA}}((\widehat{\mu}, \widehat{\pi}), (\mu_{0}, \pi_{0}))\lesssim& \max_{k} E[P_{0}\{\widehat{\pi}_{k}(1 \mid X; \rho^{\star}(\widehat{\pi}))-\pi_{\infty}(1 \mid X)\}]+E[P_{0}\{\widehat{\pi}_{k}(1 \mid X; \rho^{\star}(\widehat{\pi}))-\pi_{\infty}(1 \mid X)\}^2] \\
&+E[P_{0}\{\widehat{\mu}_{k}(\tau, X; \rho^{\star}(\widehat{\mu}))-\mu_{\infty}(\tau, X)\}]+E[P_{0}\{\widehat{\mu}_{k}(\tau, X; \rho^{\star}(\widehat{\mu}))-\mu_{\infty}(\tau, X)\}^2] \\
&+E(P_{0}[\{\widehat{\mu}_{k}(\tau, X; \rho^{\star}(\widehat{\mu}))-\mu_{\infty}(\tau, X)\}\{\widehat{\pi}_{k}(1 \mid X; \rho^{\star}(\widehat{\pi}))-\pi_{\infty}(1 \mid X)\}])
\end{aligned}
$$
}

\clearpage

\section{Additional Simulation Results} \label{app:sim}

The code implementing all estimators is available at \url{https://github.com/xushenbo}. \\

Overlap weighted MSE~\citep{morzywolek2023general}
{\footnotesize
$$
\begin{aligned}
\epsilon_{\mathrm{ATE}}=&P_{n} \{ \psi_{0}(X)-\widehat{\psi}(X) \} \\
\epsilon_{\mathrm{ATE}}^{h}=&\frac{P_{n} \{ h_{0}(X) \psi_{0}(X)\}}{P_{n} h_{0}(X)}-\frac{P_{n} \{\widehat{h}(X)\widehat{\psi}(X)\}}{P_{n} \widehat{h}(X)} \\
\mathrm{PEHE}^{h}=& \frac{P_{n} [h_{0}(X)\{\psi_{0}(X)-\widehat{\psi}(X)\}^2]}{P_{n} h_{0}(X)} \\
% \mathrm{PEWHE}=& P_{n} \left[\frac{h_{0}(X)\psi_{0}(X)}{P_{n}h_{0}(X)}-\frac{\widehat{h}(X)\widehat{\psi}(X)}{P_{n}\widehat{h}(X)}\right]^2 \\
\mathrm{prevalence}=& P_{n}I(\psi_{0}(X)>0) \\ % \max\{P_{n}I(\psi_{0}(X)>0), P_{n}I(\psi_{0}(X)<0)\} \\
\mathrm{accuracy}=&P_n I(\psi_{0}(X)\widehat{\psi}(X)>0) \\
\mathrm{gain}^{h}=& \frac{P_{n} \{h_{0}(X)I(\psi_{0}(X)\widehat{\psi}(X)>0) \left|\psi_{0}(X) \right|\}}{P_{n} h_{0}(X)} \\
\mathrm{regret}^{h}=& \frac{P_{n} \{h_{0}(X)I(\psi_{0}(X)\widehat{\psi}(X)<0) \left|\psi_{0}(X) \right|\}}{P_{n} h_{0}(X)} (\text{baseline}: \widehat{\psi}(X)=\frac{P_{n} \{ h_{0}(X) \psi_{0}(X)\}}{P_{n} h_{0}(X)}) \\
\mathrm{GRR}=& \frac{\mathrm{gain}}{\mathrm{regret}} \\
\mathrm{GRR}^{h}=& \frac{\mathrm{gain}^{h}}{\mathrm{regret}^{h}}
\end{aligned}
$$
}

\newpage

\begin{table}[h]
\small
\centering
\caption{Abbreviation mapping of HTE learners in figures}
\begin{tabular}{l l}
\hline Abbreviation & Stands for \\
\hline
single.learner & Single learner \\
two.learners & Two learners \\
csf & Causal survival forest \\
dml & Double/debiased machine learning of the IF transformation \\
bj & Buckley-James transformation \\
ipcw & Inverse probability of censoring weighted transformation \\
drcut & Doubly robust censoring unbiased transformation \\
iptw & Inverse probability treatment weighted learner \\
ra & Regression adjusted learner \\
aiptw & Augmented inverse probability treatment weighted learner \\
xlearner & X-learner \\
mc & Modified covariate learner \\
eamc & Efficiency augmented modified covariate learner \\
ulearner & U-learner \\
rlearner & R-learner \\
causal.forest & Causal forest \\
\hline
\end{tabular}
\label{tab:abbreviation}
\end{table}

\newpage

\begin{figure}[H]
\centering
\includegraphics[scale=0.55]{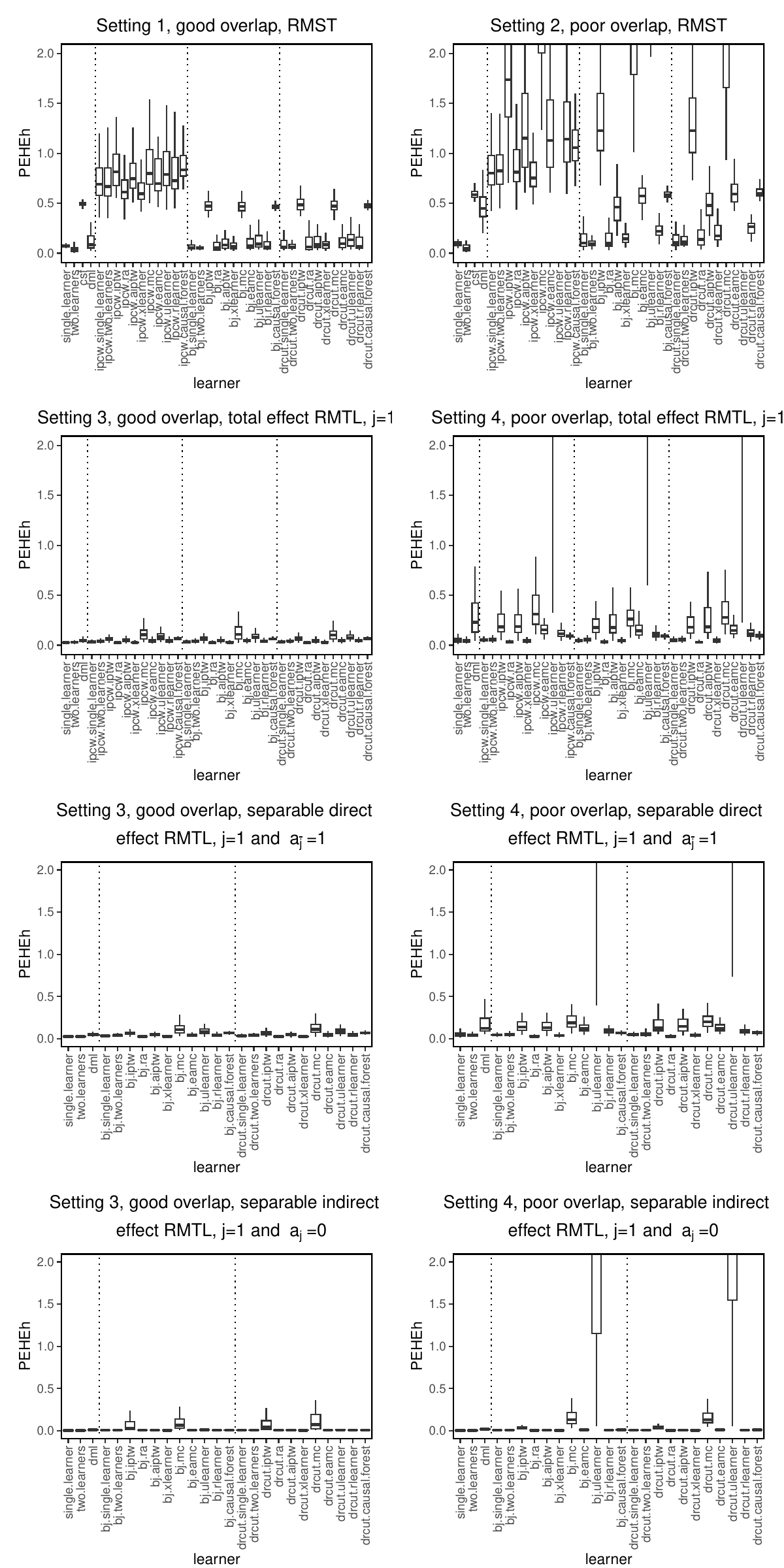}
\caption{PEHEh of estimated HTE.}
\label{fig:sim_peheh}
\end{figure}

\begin{figure}[H]
\centering
\includegraphics[scale=0.55]{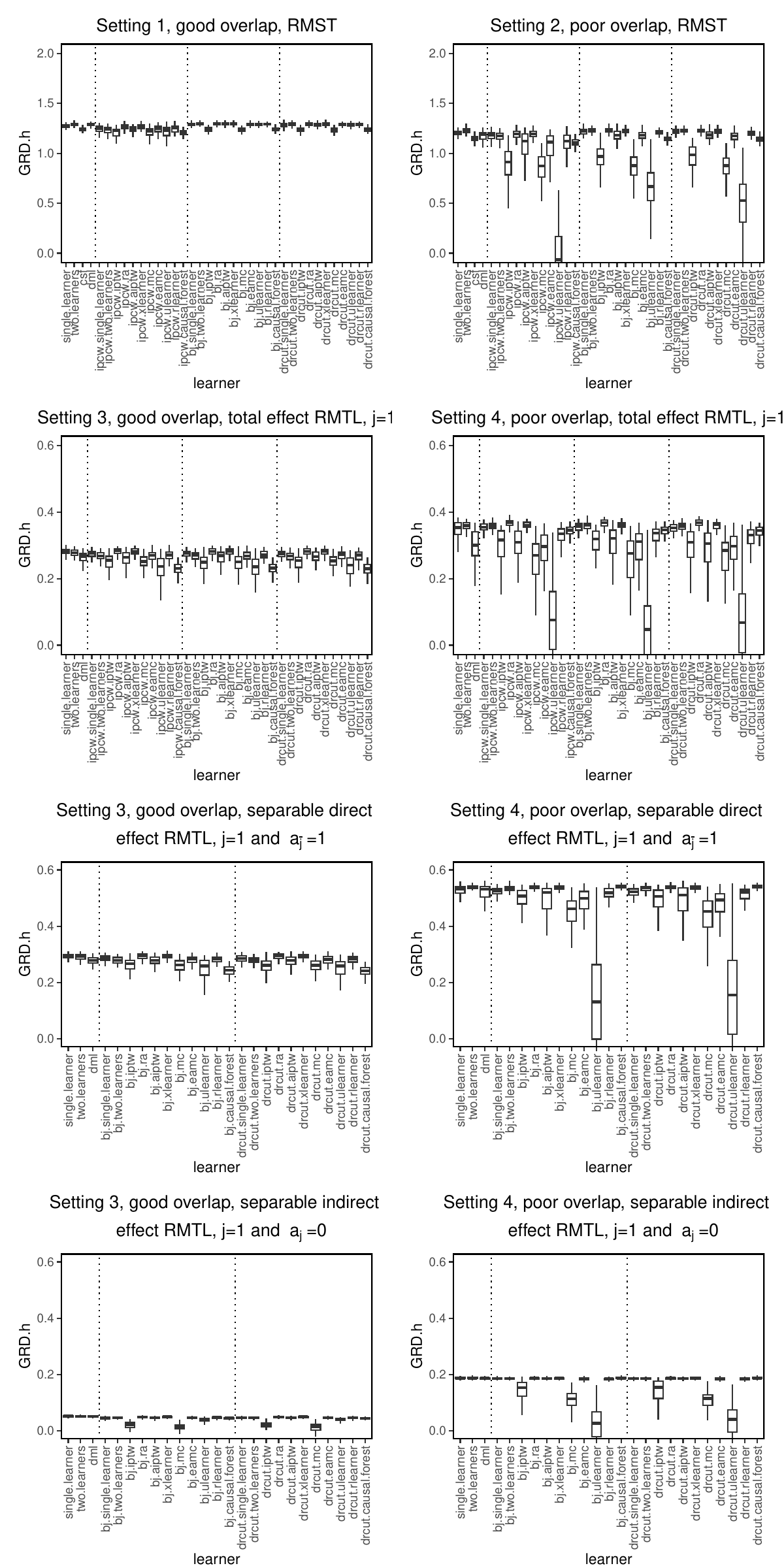}
\caption{GRDh of estimated HTE.}
\label{fig:sim_grdh}
\end{figure}

\end{document}